\documentclass[tight,twocolumn]{aastex62}
\usepackage{apjfonts}
\usepackage{graphicx}
\usepackage{float}
\usepackage[caption=false]{subfig}
\usepackage{graphics} 
\usepackage{amssymb,amsmath}
\usepackage{color}

\turnoffedit

\newcommand{\jz}{\hbox{ASASSN-17jz}}
\newcommand{\fullhost}{\hbox{SDSS J171955.84+414049.4}}
\newcommand{\host}{\hbox{SDSS J171955}}

\def\lsim{\lower0.3em\hbox{$\,\buildrel <\over\sim\,$}}
\def\gsim{\lower0.3em\hbox{$\,\buildrel >\over\sim\,$}}

\newcommand{\msun}{\hbox{M$_{\odot}$}}

\newcommand{\halpha}{\hbox{H$\alpha$}}
\newcommand{\hbeta}{\hbox{H$\beta$}}
\newcommand{\llambda}{\hbox{$\lambda$$\lambda$}}

\newcommand{\HST}{\textit{HST}}

\newcommand{\swift}{\textit{Swift}}

\newcommand{\mujyb}{\,\mbox{\ensuremath{\mathrm{\mu Jy/beam}}}}     
\newcommand{\mjyb}{\,\mbox{\ensuremath{\mathrm{mJy/beam}}}}         
\newcommand{\jyb}{\,\mbox{\ensuremath{\mathrm{Jy/beam}}}}         
\newcommand{\lunits}{\,\ensuremath{\mathrm{erg~s^{-1}~Hz^{-1}}}}  

\begin{document}

\submitjournal{ApJ}

\title{Investigating the Nature of the Luminous Ambiguous Nuclear Transient ASASSN-17jz}
\shorttitle{The Unusual Nuclear Transient ASASSN-17jz} 
\shortauthors{Holoien, et al. }

\author[0000-0001-9206-3460]{Thomas~W.-S.~Holoien}
\altaffiliation{NHFP Einstein Fellow}
\affiliation{The Observatories of the Carnegie Institution for Science, 813 Santa Barbara St., Pasadena, CA 91101, USA}

\author[0000-0001-7351-2531]{Jack~M.~M.~Neustadt}
\affiliation{Department of Astronomy, The Ohio State University, 140 West 18th Avenue, Columbus, OH 43210, USA}

\author{Patrick~J.~Vallely}
\affiliation{Department of Astronomy, The Ohio State University, 140 West 18th Avenue, Columbus, OH 43210, USA}

\author[0000-0002-4449-9152]{Katie~Auchettl}
\affiliation{School of Physics, The University of Melbourne, Parkville, VIC 3010, Australia}
\affiliation{ARC Centre of Excellence for All Sky Astrophysics in 3 Dimensions (ASTRO 3D)}
\affiliation{Department of Astronomy and Astrophysics, University of California, Santa Cruz, CA 95064, USA}

\author[0000-0001-9668-2920]{Jason~T.~Hinkle}
\affiliation{Institute for Astronomy, University of Hawai`i, 2680 Woodlawn Drive, Honolulu, HI 96822, USA}

\author[0000-0001-6301-9073]{Cristina Romero-Ca\~{n}izales}
\affiliation{Institute of Astronomy and Astrophysics, Academia Sinica, 11F of Astronomy-Mathematics Building, AS/NTU No. 1, Sec. 4, Roosevelt Rd, Taipei 10617, Taiwan, R.O.C}

\author[0000-0003-4631-1149]{Benjamin.~J.~Shappee}
\affiliation{Institute for Astronomy, University of Hawai`i, 2680 Woodlawn Drive, Honolulu, HI 96822, USA}

\author{Christopher~S.~Kochanek}
\altaffiliation{Radcliffe Fellow}
\affiliation{Department of Astronomy, The Ohio State University, 140 West 18th Avenue, Columbus, OH 43210, USA}
\affiliation{Center for Cosmology and AstroParticle Physics (CCAPP), The Ohio State University, 191 W.\ Woodruff Ave., Columbus, OH 43210, USA}

\author{K.~Z.~Stanek}
\affiliation{Department of Astronomy, The Ohio State University, 140 West 18th Avenue, Columbus, OH 43210, USA}
\affiliation{Center for Cosmology and AstroParticle Physics (CCAPP), The Ohio State University, 191 W.\ Woodruff Ave., Columbus, OH 43210, USA}

\author{Ping~Chen}
\affiliation{Kavli Institute for Astronomy and Astrophysics, Peking University, Yi He Yuan Road 5, Hai Dian District, Beijing 100871, China}

\author{Subo~Dong}
\affiliation{Kavli Institute for Astronomy and Astrophysics, Peking University, Yi He Yuan Road 5, Hai Dian District, Beijing 100871, China}

\author{Jose~L.~Prieto}
\affiliation{N\'ucleo de Astronom\'ia de la Facultad de Ingenier\'ia y Ciencias, Universidad Diego Portales, Av. Ej\'ercito 441, Santiago, Chile}
\affiliation{Millennium Institute of Astrophysics, Santiago, Chile}

\author{Todd~A.~Thompson}
\affiliation{Center for Cosmology and AstroParticle Physics (CCAPP), The Ohio State University, 191 W.\ Woodruff Ave., Columbus, OH 43210, USA}
\affiliation{Department of Astronomy, The Ohio State University, 140 West 18th Avenue, Columbus, OH 43210, USA}

\author[0000-0001-5955-2502]{Thomas G. Brink}
\affiliation{Department of Astronomy, University of California, Berkeley, CA 94720-3411, USA}

\author[0000-0003-3460-0103]{Alexei V. Filippenko}
\affiliation{Department of Astronomy, University of California, Berkeley, CA 94720-3411, USA}
\affiliation{Miller Institute for Basic Research in Science, University of California, Berkeley, CA 94720, USA}

\author{WeiKang Zheng}
\affiliation{Department of Astronomy, University of California, Berkeley, CA 94720-3411, USA}

\author{David~Bersier}
\affiliation{Astrophysics Research Institute, Liverpool John Moores University, 146 Brownlow Hill, Liverpool L3 5RF, UK}

\author{Subhash~Bose}
\affiliation{Department of Astronomy, The Ohio State University, 140 West 18th Avenue, Columbus, OH 43210, USA}
\affiliation{Center for Cosmology and AstroParticle Physics (CCAPP), The Ohio State University, 191 W.\ Woodruff Ave., Columbus, OH 43210, USA}

\author[0000-0002-6523-9536]{Adam J. Burgasser}
\affiliation{Center for Astrophysics and Space Sciences,
University of California, San Diego, La Jolla, CA 92093, USA}

\author{Sanyum Channa}
\affiliation{Department of Astronomy, University of California, Berkeley, CA 94720-3411, USA}

\author{Thomas~de~Jaeger}
\affiliation{Department of Astronomy, University of California, Berkeley, CA 94720-3411, USA}
\affiliation{Institute for Astronomy, University of Hawai`i, 2680 Woodlawn Drive, Honolulu, HI 96822, USA}

\author{Julia Hestenes}
\affiliation{Department of Astronomy, University of California, Berkeley, CA 94720-3411, USA}
\affiliation{Department of Applied Physics and Applied Mathematics, Columbia University, New York, NY 10027-6623, USA}

\author[0000-0002-8537-6714]{Myungshin Im}
\affiliation{SNU Astronomy Research Center, Astronomy Program, Dept. of Physics and Astronomy, Seoul National University, Seoul 08826, Republic of Korea}

\author{Benjamin Jeffers}
\affiliation{Department of Astronomy, University of California, Berkeley, CA 94720-3411, USA}

\author[0000-0003-1470-5901]{Hyunsung~D.~Jun}
\affiliation{School of Physics, Korea Institute for Advanced Study, 85 Hoegiro, Dongdaemun-gu, Seoul 02455, Korea}

\author[0000-0002-5328-9827]{George~Lansbury}
\affiliation{European Southern Observatory, Karl-Schwarzschild-Strasse 2, D-85748 Garching, Germany}

\author{Richard~S.~Post}
\affiliation{Post Observatory, Lexington, MA 02421, USA}

\author{Timothy W. Ross}
\affiliation{Department of Astronomy, University of California, Berkeley, CA 94720-3411, USA}

\author[0000-0003-2686-9241]{Daniel~Stern}
\affiliation{Jet Propulsion Laboratory, California Institute of Technology, 4800 Oak Grove Drive, Pasadena, CA 91109, USA}

\author{Kevin Tang}
\affiliation{Department of Astronomy, University of California, Berkeley, CA 94720-3411, USA}

\author[0000-0002-2471-8442]{Michael~A.~Tucker}
\affiliation{Institute for Astronomy, University of Hawai`i, 2680 Woodlawn Drive, Honolulu, HI 96822, USA}
\altaffiliation{DOE CSGF Fellow}

\author{Stefano~Valenti}
\affiliation{Department of Physics and Astronomy, University of California, 1 Shields Avenue, Davis, CA 95616-5270, USA}

\author{Sameen Yunus}
\affiliation{Department of Astronomy, University of California, Berkeley, CA 94720-3411, USA}

\author{Keto D. Zhang}
\affiliation{Department of Astronomy, University of California, Berkeley, CA 94720-3411, USA}
\affiliation{IPAC, Mail Code 100-22, California Institute of Technology,
1200 E. California Blvd.,
Pasadena, CA 91125, USA}

\correspondingauthor{T.~W.-S.~Holoien}
\email{tholoien@carnegiescience.edu}

\date{\today}

\begin{abstract}
We present observations of the extremely luminous but ambiguous nuclear transient (ANT) ASASSN-17jz, spanning roughly 1200 days of the object's evolution. ASASSN-17jz was discovered by the All-Sky Automated Survey for Supernovae (ASAS-SN) in the galaxy \fullhost{} on UT 2017 July 27 \edit1{at a redshift of $z=0.1641$}. The transient peaked at an absolute $B$-band magnitude of $M_{B,{\rm peak}}=-22.81$, corresponding to a bolometric luminosity of $L_{\rm bol,peak}=8.3\times10^{44}$~erg~s$^{-1}$, and exhibited late-time ultraviolet emission \edit1{that was still ongoing in our latest observations.} \edit1{Integrating the full light curve gives a total emitted energy of $E_{\rm tot}=(1.36\pm0.08)\times10^{52}$~erg, with $(0.80\pm0.02)\times10^{52}$~erg of this emitted within 200 days of peak light.} This late-time \edit1{ultraviolet emission} is accompanied by increasing X-ray emission that becomes softer as it brightens. ASASSN-17jz exhibited a large number of spectral emission lines most commonly seen in active galactic nuclei (AGNs) with little evidence of evolution. \edit1{It also showed transient Balmer features which became fainter and broader over time, and are still being detected $>1000$ days after peak brightness}. We consider various physical scenarios for the origin of the transient, including supernovae (SNe), tidal disruption events (TDEs), AGN outbursts, and ANTs. We find that the most likely explanation is that ASASSN-17jz was an SN~IIn occurring in or near the disk of an existing AGN, and that the late-time emission is caused by the AGN transitioning to a more active state.
\end{abstract}

\section{Introduction}

In recent years, the proliferation of large-area, untargeted sky surveys has led to the discovery of new types of transients associated with the nuclei of their host galaxies. These have included tidal disruption events \edit1{(TDEs; see \citealt{saxton20} and \citealt{velzen20} for recent reviews)}, unusual accretion events in active galactic nuclei \citep[AGNs; e.g.,][]{shappee14,wyrzykowski17,trakhtenbrot19a,frederick20}, and other outbursts whose origins cannot be definitively determined \citep[e.g.,][]{kankare17,neustadt20,hinkle21c}. \edit1{These events, particularly the last two groups, can be difficult to classify and interpret. They often involve unusual transient phenomena; atypical, large-scale changes to formerly stable accretion systems; or even a combination of these, such as a supernova (SN) or TDE occurring in or around an existing AGN and affecting the accretion flow. With these being rarer phenomena and a lack of prioritization of nuclear sources until recently, there can be few observations or simulations to use for comparison to new observations, making the classification  of each individual event difficult. Regardless of the specific origin of each event, however, nuclear outbursts usually} involve accretion onto a supermassive black hole (SMBH). \edit1{They thus} allow us to study BHs and accretion physics in novel ways and in galaxies that are otherwise quiescent.

\edit1{These transients are typically flagged for detailed study by TDE searches, and because of this} they often have several observational characteristics of TDEs. These include a strong blue ultraviolet (UV) continuum with broad Balmer and/or helium emission lines in their spectra, relatively smooth light curves that remain bright for a year or more, and, in some cases, hard or soft X-ray emission \citep[e.g.,][]{saxton20,velzen20}. As only a few dozen of these outbursts have been identified, each additional discovery allows for new observations that can refine our physical understanding of these events and make them better probes of SMBHs and accretion physics.

Here we study the nuclear outburst ASASSN-17jz (AT 2017fro\footnote{\url{https://wis-tns.weizmann.ac.il/object/2017fro}}) spanning from 88 days prior to peak light through 1081 days after peak. ASASSN-17jz was discovered by the All-Sky Automated Survey for Supernovae \citep[ASAS-SN;][]{shappee14} on UT 2017 July 27 in the galaxy \fullhost{} \citep[hereafter \host;][]{asassn17jz_disc_atel}, \edit1{at a redshift of $z=0.1641$}. \edit1{Because} an optical spectrum obtained on 2017 July 29 showed that the transient exhibited a blue continuum and broad Balmer emission features, consistent with a TDE \citep[e.g.,][]{arcavi14}, we requested target-of-opportunity \edit1{(TOO)} observations from the \textit{Neil Gehrels Swift Observatory} \citep[\swift;][]{gehrels04} UltraViolet and Optical Telescope \citep[UVOT;][]{roming05} and X-ray Telescope \citep[XRT;][]{burrows05} (Target ID: 10218). \edit1{These initial \swift{} observations, taken near peak light,} showed that ASASSN-17jz was UV-bright but did not exhibit any strong soft X-ray emission, which again is seen in many TDEs \citep[e.g.,][]{auchettl17}. The transient was also very luminous, with $M_{B,{\rm peak}}=-22.81$ mag, making it one of the most luminous supernovae (SNe) or TDEs discovered thus far. With this potential TDE or \edit1{superluminous} SN classification, we began a long-term, multiwavelength monitoring campaign to characterize the object.

 As ASASSN-17jz evolved past peak light, its spectra remained relatively constant, and the transient was classified as an AGN outburst by the Global Supernova Project \citep{asassn17jz_spec_tns}. However, the light curve of ASASSN-17jz showed a smooth rise and fall, which is much more commonly seen in TDEs or SNe than in AGN outbursts, and the spectral energy distribution (SED) was well-fit by an evolving blackbody, also typical of these transients. ASASSN-17jz thus joins a growing group of ambiguous nuclear transients (ANTs) --- luminous nuclear outbursts with an unclear origin. We continued to observe this transient for several years, finding that some of the features are still visible nearly \edit1{1200} days after peak light, and we compare our observations to those of AGNs, TDEs, and SNe to determine the nature of this unusual event.
 
Section~\ref{sec:obs} describes the archival data available for the host galaxy and its physical properties, discusses our follow-up observations of the transient, and describes the available prediscovery variability data from ASAS-SN and ATLAS. In Section~\ref{sec:phot_anal} we analyze the photometry, fit the light curves with several transient models, and model the blackbody emission of the transient to compare it with other sources. We analyze the evolution of the optical and UV spectroscopic features in Section~\ref{sec:spec_anal} and compare these to AGNs, TDEs, and SNe.  Section~\ref{sec:disc} summarizes our findings and discusses the likelihoods of different origin scenarios. We adopt a distance of $d=792.9$~Mpc (H$_0=69.6$~km~s$^{-1}$~Mpc$^{-1}$, $\Omega_M=0.296$, and $\Omega_{\Lambda}=0.714$; see Sec.~\ref{sec:params}), a Galactic extinction of $A_V = 0.06$~mag \citep{schlafly11}, converted to other filters using a \citet{cardelli89} extinction law, and use UT dates throughout this paper.
 

\section{Observations and Survey Data}
\label{sec:obs}


\subsection{Archival Data and Host-Galaxy Fits}
\label{sec:archival}

We retrieved archival $ugriz$ images of the host galaxy {\host} from the Sloan Digital Sky Survey (SDSS) Data Release 16 \citep[DR16;][]{ahumada20} and $JHK_S$ images from the Two Micron All-Sky Survey \citep[2MASS;][]{skrutskie06,2massDOI} and measured aperture magnitudes using the \textsc{Iraf} \citep{tody86,tody93} task \texttt{phot}. \edit1{We used a 5\farcs{0} (radius) aperture as this is large enough to contain the majority of the host galaxy light while also not sacrificing S/N on the nuclear region. It is also the default aperture size for the {\swift} photometric pipeline.} We also obtained UV $NUV$ and $FUV$ magnitudes from the {\it Galaxy Evolution Explorer (GALEX)} All-Sky Imaging Survey (AIS) catalog and infrared $W1$ and $W2$ magnitudes from the {\it Wide-field Infrared Survey Explorer} \citep[{\it WISE};][]{wright10,allwiseDOI} AllWISE catalog. These archival host magnitudes are listed in Table~\ref{tab:host_mags}.


\begin{deluxetable}{ccc}
\tabletypesize{\footnotesize}
\tablecaption{Archival Photometry of \host}
\tablehead{
\colhead{Filter} &
\colhead{Magnitude} &
\colhead{Uncertainty} }
\startdata
$FUV$ & 20.41 & 0.21 \\
$NUV$ & 19.57 & 0.11 \\
$u$ & 19.62 & 0.22 \\
$g$ & 18.15 & 0.15 \\
$r$ & 17.61 & 0.11 \\
$i$ & 17.23 & 0.10 \\
$z$ & 17.14 & 0.09 \\
$J$ & 16.74 & 0.11 \\
$H$ & 16.67 & 0.11 \\
$K_S$ & 16.08 & 0.09 \\
$W1$ & 16.88 & 0.03 \\
$W2$ & 16.92 & 0.03
\enddata 
\tablecomments{5\farcs{0} aperture magnitudes of {\host} measured from SDSS DR16 ($ugriz$) and 2MASS ($JHK_S$) data, $FUV$ and $NUV$ point-spread-function (PSF) magnitudes from the {\it GALEX} AIS, and $W1$ and $W2$ infrared (IR) magnitudes from the AllWISE catalog. All magnitudes are presented in the AB system \citep{oke83}.}
\label{tab:host_mags}
\vspace{-0.8cm}
\end{deluxetable}

The mid-infrared (MIR) {\it WISE} colors of ($W1-W2$) = $0.60 \pm 0.04$~mag and ($W3-W4$) = $2.34 \pm 0.19$~mag imply that \host{} does not host a strong AGN based on the criteria of \citet{assef13}. This, however, does not rule out the presence of a low-luminosity AGN whose light is dominated by stars in the host. The ($W1-W2$) color of \host{} is redder than the majority of hosts of TDEs and other ANTs \citep[e.g.,][]{hinkle21b}, and there is likely some AGN activity in the galaxy based on our follow-up data \edit1{(see Sections~\ref{sec:xray_anal} and \ref{sec:spec_anal})}.

To further constrain the presence of AGN activity associated with the host galaxy, we analysed ROSAT All-Sky Survey \citep{voges99} and pointed observations of the source (ObsID: RP201238N00). No source was detected nearby in either of these observations. As the exposure time of the pointed observation is much longer, we use this observation to constrain the X-ray flux to a 3$\sigma$ upper limit of 0.007~counts~s$^{-1}$ in the 0.3--2.0~keV energy range. Assuming a photon index typical of known AGNs \citep[$\Gamma=1.75$;][]{ricci17} and a Galactic column density of $1.67 \times 10^{20}$~cm$^{-2}$ \citep{hi4pi16}, we derive a 3$\sigma$ upper limit to the absorbed flux of $2.3 \times 10^{-13}$~erg~cm$^{-2}$~s$^{-1}$ in the 0.3--10.0~keV energy range. This corresponds to an absorbed luminosity of $2.0 \times 10^{43}$~erg~s$^{-1}$, \edit1{again} implying that the host does not harbor a strong AGN.

\edit1{We also checked archival survey data from the Catalina Real-Time Transient Survey \citep[CRTS;][]{drake09} to search for signs of previous variability from \host. CRTS obtained observations of the galaxy beginning on UT 2005 July 01, roughly 12~yr prior to our first detection of \jz{} in ASAS-SN $V$-band data, and there is no evidence of prior flaring in these data. This provides further evidence that there is no strong AGN in the host galaxy, and gives constraints on the presence of an SMBH binary (see Section~\ref{sec:disc_single}).}

\edit1{We fit the SED of the host galaxy using the archival host magnitudes and} the Fitting and Assessment of Synthetic Templates \citep[\textsc{fast};~][]{kriek09} software. We assumed a \citet{cardelli89} extinction law with $R_V=3.1$ and a Galactic extinction of $A_V = 0.06$~mag \citep{schlafly11}, and adopted an exponentially declining star-formation history, a Salpeter initial mass function, and the \citet{bruzual03} stellar population models for the fit. In order to robustly estimate the uncertainties, we generated 1000 realizations of the archival fluxes perturbed by their respective uncertainties, assuming Gaussian errors. We then used \textsc{fast} to model the host SED for each set of input fluxes. Our resulting median and 68\% confidence intervals on the host parameters are a stellar mass of $M_{\star}=5.5^{+1.6}_{-1.6} \times 10^{10}$~M$_{\odot}$, a stellar age of $2.3^{+1.2}_{-1.0}$~Gyr, and a star-formation rate (SFR) of $2.9^{+0.4}_{-0.5}$~M$_{\odot}$~yr$^{-1}$. Scaling the stellar mass using the average stellar-mass-to-bulge-mass ratio of several ASAS-SN TDE hosts, as we have done in the past \citep[e.g.,][]{holoien20a}, we estimate a bulge mass of $M_B \approx 10^{10.1}$~\msun. This gives an estimated black hole mass of $M_{\rm BH}=10^{7.5}$~{\msun}, using the $M_B-M_{\rm BH}$ relation from \citet{mcconnell13}. This is well within the range of $10^6~\msun\leq M_{\rm BH} \leq 10^8~\msun$ expected for TDE host galaxies \citep[e.g,][]{rees88,kochanek16}, and is consistent with the SMBH masses of several TDE hosts in the literature \citep[e.g.,][]{holoien14b,holoien16a,holoien16b,brown17a,wevers17,mockler19}.

For several of the filters we used for our follow-up photometry ($BVRI$ and the {\swift} UVOT filters), there are no archival images of the host that can be used as image-subtraction templates. To estimate the host flux in these filters, we computed synthetic 5\farcs{0} aperture magnitudes for each photometric band for each of the 1000 bootstrapped host SED fits. This yielded a distribution of synthetic magnitudes for each filter, and we used the median and 68\% confidence intervals on the host magnitude in each filter as our host magnitudes and uncertainties. We then subtracted our synthetic host fluxes from our follow-up flux measurements to isolate the transient flux, as described in the following sections. The synthetic host magnitudes and uncertainties for each follow-up filter are shown in Table~\ref{tab:synth_mags}.


\begin{deluxetable}{cc}
\tabletypesize{\footnotesize}
\tablecaption{5\farcs{0} Synthetic Host Aperture Magnitudes}
\tablehead{
\colhead{Filter} & \colhead{Magnitude and Uncertainty}}
\startdata
$UVW2$ & $19.97\pm0.19$\\
$UVM2$ & $19.85\pm0.18$ \\
$UVW1$ & $19.70\pm0.17$ \\ 
$U_{UVOT}$ & $19.33\pm0.14$ \\
$B$ & $18.65\pm0.12$ \\
$V$ & $17.80\pm0.11$ \\
$R$ & $17.55\pm0.09$ \\
$I$ & $17.26\pm0.07$ \\
\enddata 
\tablecomments{5\farcs{0} aperture magnitudes of {\host} synthesized for the {\swift} UV$+U$ and $BVRI$ filters as described in Section~\ref{sec:archival}. All magnitudes are in the AB system \citep{oke83}.}
\label{tab:synth_mags} 
\vspace{-0.8cm}
\end{deluxetable}

\subsection{ASAS-SN Light Curve}
\label{sec:ASASSN_LC}

ASAS-SN monitors the visible sky nightly to find bright, nearby transients using units of four 14~cm telescopes \citep{shappee14,kochanek17}. Currently, ASAS-SN consists of five units in Hawaii, Chile, Texas, and South Africa hosted by the Las Cumbres Observatory global telescope network \citep{brown13} using $g$-band filters, but at the time of the discovery of ASASSN-17jz, it was composed of only single units in Chile and in Hawaii using $V$-band filters. Roughly two months after ASASSN-17jz was discovered our Texas unit began survey operations, and also started to observe ASASSN-17jz in the $g$ band. In order to remove host-galaxy flux from the ASASSN-17jz photometry, we constructed a reference image of the host galaxy and surrounding sky for the two cameras that could detect it. In order to ensure that no flux from the transient was present in the reference image, we constructed the reference images using only data taken more than two months prior to our discovery date for the $V$-band data and more than two years after discovery for the $g$-band data. 

We performed image subtraction on the science images using these reference images, and aperture photometry on each subtracted image using the {\sc Iraf} {\tt apphot} package. We calibrated the magnitudes using several stars in the vicinity of the transient with known magnitudes in the AAVSO Photometric All-Sky Survey \citep[APASS;][]{henden15}. For some prediscovery epochs we stacked several science images to improve the signal-to-noise ratio (S/N) of our detections and obtain deeper limits prior to our first detection. Some ASAS-SN $V$- and $g$-band data were also stacked with other ground-based and {\swift} photometric observations to improve the \edit1{S/N} of our detections, as described in Section~\ref{sec:other_phot}. The ASAS-SN photometry is presented in Table~\ref{tab:phot} and shown in Figure~\ref{fig:lc}. The temporal error bars indicate the time span of epochs that were combined to obtain higher S/N.


\begin{deluxetable}{ccc}
\tabletypesize{\footnotesize}
\tablecaption{Stacked Host-Subtracted Photometry of ASASSN-17jz}
\tablehead{
\colhead{MJD Range} &
\colhead{Filter} &
\colhead{Magnitude} }
\startdata
57968.28  &  $I$ & $17.03 \pm 0.11$  \\
57970.29$-$57972.28  &  $I$ & $16.90 \pm 0.04$  \\
57973.27$-$57974.28  &  $I$ & $16.88 \pm 0.05$  \\
... & &  \\
58181.57  &  $UVW2$ & $19.49 \pm 0.17$  \\
58377.30  &  $UVW2$ & $20.45 \pm 0.40$  \\
59051.15$-$59052.67 & $UVW2$ & $20.02 \pm 0.22$ \\
\enddata 
\tablecomments{Host-subtracted AB magnitudes and $3\sigma$ upper limits for ASASSN-17jz. All $BVRI$, $gri$, and ATLAS $o$ data have been stacked in bins of 1~day, as described in the text. {\swift} UV+$U$ data are not stacked except for the final two epochs. The ``MJD Range'' column contains the range of dates that were combined to obtain a given measurement; a single MJD without a range indicates that the magnitude was measured from a single epoch. All magnitudes are corrected for Galactic extinction and are presented in the AB system. The entire table is published in machine-readable format in the online journal, with only a selection shown here for guidance regarding its form and content.}
\label{tab:phot} 
\vspace{-0.8cm}
\end{deluxetable}

\subsection{ATLAS Light Curve}
\label{sec:ATLAS_LC}

The Asteroid Terrestrial-impact Last Alert System \citep[ATLAS;][]{tonry18} is an ongoing survey that uses fully robotic 0.5~m telescopes located on the summit of Haleakal\=a and Mauna Loa Observatory to monitor the sky visible from Hawai'i with a primary goal of detecting small asteroids on a collision course with Earth. In normal survey operations, each telescope obtains four 30~s exposures of 200--250 fields per night, covering roughly a quarter of the visible sky. ATLAS uses two broad filters for its survey operations: ``cyan" ($c$), covering 420--650~nm, and ``orange'' ($o$) covering 560--820~nm \citep{tonry18}.


\begin{figure*}
\begin{minipage}{\textwidth}
\centering
{\includegraphics[width=0.95\textwidth]{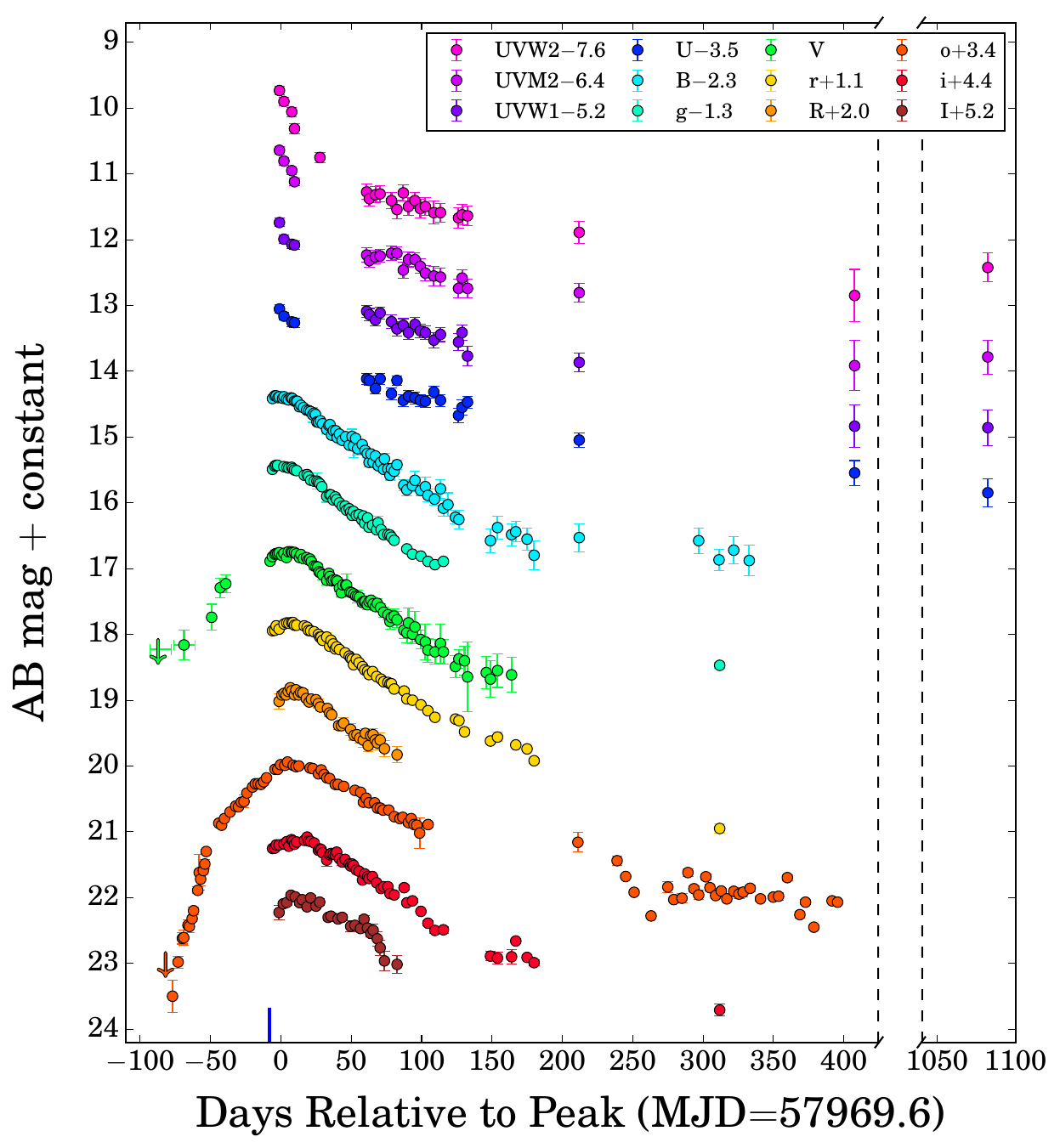}}
\caption{Host-subtracted UV and optical light curves of ASASSN-17jz spanning from 87 days prior to peak $B$-band brightness (MJD = 57969.6; see Sec.~\ref{sec:params}) to 1083 days after peak. The date of discovery (MJD = 57961.4) is shown with a blue line at the bottom. The gap in the abscissa indicates a period of roughly 675~days where no new observations were taken. The $BVRI$, $gri$, and ATLAS $o$ data, as well as the final two epochs of {\swift} UVOT+$U$ data, have been stacked, generally in bins of 1~day except where noted otherwise in Section~\ref{sec:obs}. Error bars in time are used to indicate the date ranges of data that have been combined, but in general these are smaller than the data points. Upper limits ($3\sigma$) are shown with downward arrows. We converted {\swift} $B$- and $V$-band data to the Johnson $B$ and $V$ filters to enable direct comparison with the ground-based data. All data have been corrected for Galactic extinction and are presented in the AB system.}
\label{fig:lc}
\end{minipage}
\end{figure*}

ATLAS images are processed by an automated pipeline that performs flat fielding, astrometric calibration, and photometric calibration. To isolate transient flux, a low-noise reference image of the host field of ASASSN-17jz was constructed by stacking multiple images taken under good conditions, and this reference was then subtracted from each science image of ASASSN-17jz. We performed forced photometry on the host-subtracted ATLAS images of ASASSN-17jz as described by \citet{tonry18} and computed a single weighted average flux from the images obtained in each night of observation. We present the stacked $o$-band photometry and $3\sigma$ limits in Table~\ref{tab:phot} and Figure~\ref{fig:lc}. As there were comparatively few $c$-band observations taken of ASASSN-17jz owing to weather and the design of the ATLAS survey, we do not include the $c$-band photometry in Figure~\ref{fig:lc} or use it in the analysis presented in Section~\ref{sec:phot_anal}, but we do include these data in Table~\ref{tab:phot} for completeness.

\subsection{Swift Observations}
\label{sec:swift}

During our initial follow-up campaign of ASASSN-17jz, we obtained 23 epochs of {\swift} \edit1{TOO} observations spanning from 1~day before peak light through 212~days after peak. We later obtained one epoch of observations on 2018 September 16, and two epochs on 2020 July 21 and 2020 July 22 to monitor the long-term UV and X-ray evolution of this transient. UVOT observations were obtained in the $V$ (5468~\AA), $B$ (4392~\AA), $U$ (3465~\AA), $UVW1$ (2600~\AA), $UVM2$ (2246~\AA), and $UVW2$ (1928~\AA) filters \citep{poole08} at most epochs, with later epochs only using the UV and $U$ filters, as the transient had faded in the optical. Each epoch of UVOT observations consisted of two images for each filter, and we first combined these two images using the HEAsoft \citep{heasarc14} software task {\tt uvotimsum}. We then used the task {\tt uvotsource} to extract counts from a 5\farcs{0} radius region around the transient, using a nearby $\sim$~40\farcs{0} sky region to estimate and subtract the sky background. We then calculated magnitudes and fluxes from the UVOT count rates using the most recent UVOT calibration \citep{poole08,breeveld10}. In the process of preparing this manuscript, the {\swift} team announced an update to the UVOT calibration to correct for the loss of sensitivity over time that affected observations in the UV filters taken after 2017 by up to 0.3~mag. Our {\swift} magnitudes include this update.

We subtracted the 5\farcs{0} host fluxes from each follow-up observation and corrected for Galactic extinction to isolate the transient flux. To directly compare the {\swift} $B$- and $V$-band data to our ground-based observations, we also converted the UVOT $BV$ magnitudes to the Johnson-Cousins system using the standard color corrections\footnote{\url{https://heasarc.gsfc.nasa.gov/docs/heasarc/caldb/swift/docs/uvot/uvot_caldb_coltrans_02b.pdf}}. As the final two epochs of {\swift} UV and $U$ data were taken within a day and a half of each other, we combined the flux from these two epochs using a weighted average to improve the S/N of our late-time flux measurements. Some epochs of {\swift} $B$- and $V$-band data were also stacked with ground-based photometric observations as described in Section~\ref{sec:other_phot}.

ASASSN-17jz was also observed with the \swift{} X-Ray Telescope (XRT) in photon-counting mode concurrent with the UVOT observations. All XRT observations were reduced following the standard \swift{} XRT analysis procedures\footnote{\url{http://swift.gsfc.nasa.gov/analysis/xrt_swguide_v1_2.pdf}}. Level-one XRT data are reprocessed using \textsc{xrtpipeline} version 0.13.2 and standard filters and screening, along with the most up-to-date calibration files. To place constraints on the presence of X-ray emission, we used a source region with a radius of 50\farcs{0} centered on the position of ASASSN-17jz and a source-free background region with a radius of 150\farcs{0} centered at $(\alpha,\delta) = (17^{\rm h}19^{\rm m}28.8975^{\rm s}, +41^\circ 38'27\farcs{00})$. All extracted count rates were aperture corrected as this source radius contains only $\sim90$\% of the 1.5~keV counts \citep{moretti04}. 

At most epochs, we do not detect X-ray emission from ASASSN-17jz, so we calculate $3\sigma$ flux limits. Owing to the detection of bright X-ray emission at late times, we combined our late-time XRT observations using \textsc{XSELECT} version 2.4k to increase the S/N. We then extracted a spectrum from our merged observations using \textsc{xrtproducts} version 0.4.2 and the regions defined above. The ancillary response file (ARF) was derived by merging individual exposure maps using \textsc{XIMAGE} version 4.5.1 and the task \textsc{xrtmkarf}. The response matrix file (RMF) was taken from the \swift\, calibration database. Using the FTOOLs command \textsc{grppha} we grouped the stacked spectrum using a minimum of 10 counts per energy bin. We give further details on the X-ray flux measurements and investigate the nature of the late-time X-ray brightening in Section~\ref{sec:xray_anal}.

\subsection{Other Photometric Observations}
\label{sec:other_phot}

We obtained $BVgri$ observations from the Las Cumbres Observatory 1~m telescopes located at McDonald Observatory, Texas \citep{brown13} and from the 24~inch Post Observatory robotic telescopes located in Mayhill, New Mexico, and Sierra Remote Observatory in California. We further obtained $BVRI$ images from the 0.76~m Katzman Automatic Imaging Telescope \citep[KAIT;][]{filippenko01} and the 1~m Anna L. Nickel telescope at Lick Observatory. 

For the $BVRI$ data, we measured 5\farcs{0} aperture magnitudes using the {\sc Iraf} {\tt apphot} package, with a 10\farcs{0}--15\farcs{0} annulus to estimate and subtract the background. We calibrated the magnitudes using several stars in the field with well-defined magnitudes in SDSS DR16 \citep{ahumada20}, using the color corrections from \citet{lupton05} to calculate $BVRI$ magnitudes for the comparison stars from the SDSS $ugriz$ magnitudes. We then subtracted the flux of the host galaxy and corrected all the $BVRI$ ground-based aperture magnitudes for Galactic extinction.

For the $gri$ data, we aligned archival SDSS $gri$ images with our follow-up images and used the software {\tt HOTPANTS} \citep{becker15} to produce host-subtracted images of the transient. We then measured 5\farcs{0} magnitudes of the transient from the subtracted images using {\tt apphot}, calibrating the magnitudes to $gri$ magnitudes of several stars in the field from SDSS DR16. We then corrected the magnitudes for Galactic extinction, as with the $BVRI$ and UVOT data.


\begin{figure*}
\begin{minipage}{\textwidth}
\centering
{\includegraphics[width=0.9\textwidth]{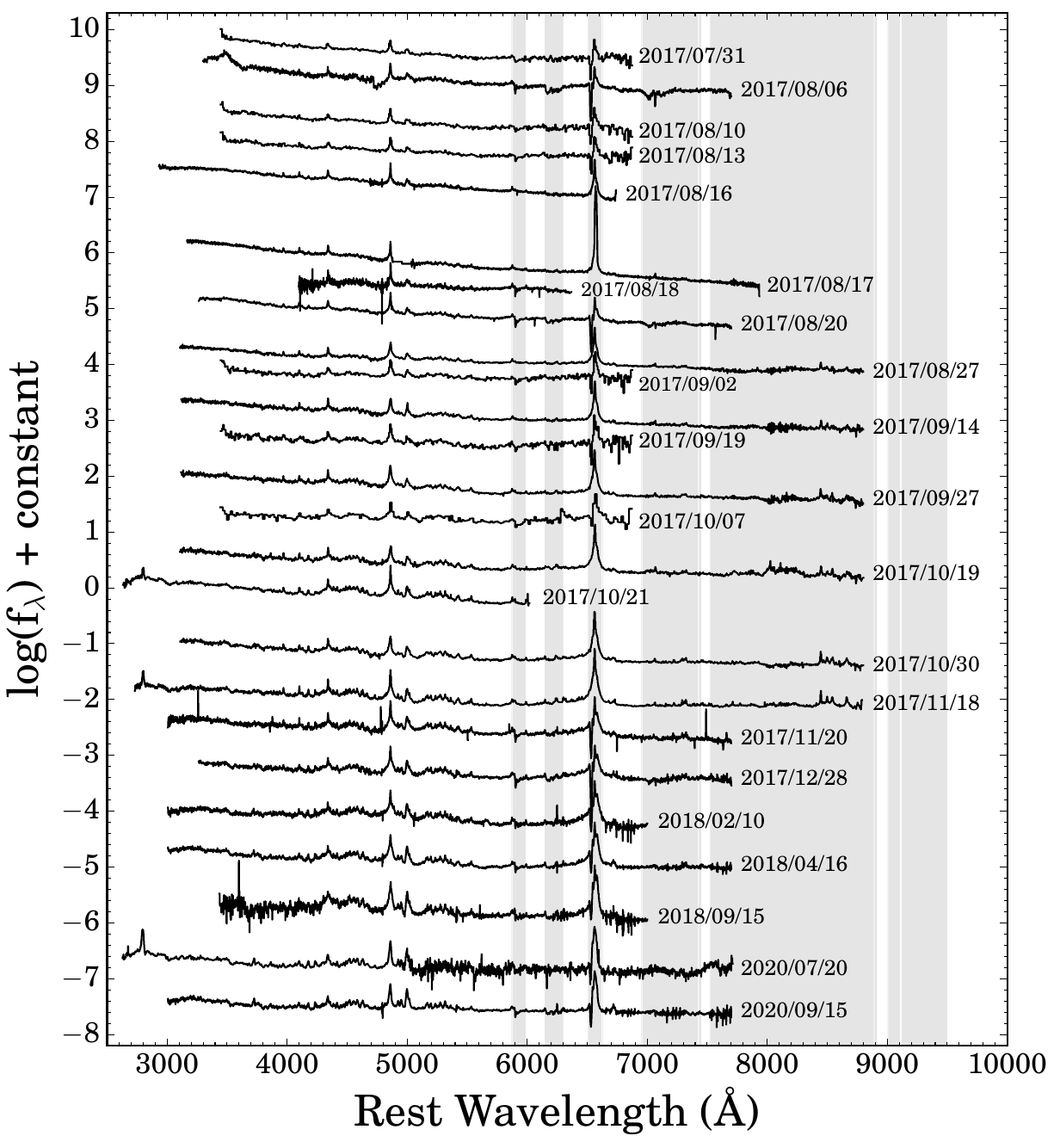}}
\caption{Optical spectroscopic evolution of ASASSN-17jz spanning from 5~days before peak brightness (2017 August 04) through 1137~days after peak. We have flux calibrated the spectra using our follow-up photometry, as described in Section~\ref{sec:spec}. Some spectra have been binned to reduce noise and improve clarity. Regions of sky contamination are shown in light gray.}
\label{fig:spec_evol}
\end{minipage}
\end{figure*}

In many cases we obtained subdaily-cadence photometry. To obtain better S/N photometric measurements and to see the overall trends more clearly, we stacked our follow-up $BVRI$ and $gri$ data in time intervals of 1~day using a weighted average of all the flux measurements taken within that interval. For $B$ and $V$, we included {\swift} and ASAS-SN data when stacking the photometry. The host-subtracted, stacked photometry is presented in Table~\ref{tab:phot} and shown in Figure~\ref{fig:lc}.

\subsection{Optical Spectroscopic Observations}
\label{sec:spec}

After confirming the discovery of ASASSN-17jz and classifying it as a potential TDE candidate, we began a program of optical spectroscopic monitoring to complement our photometric follow-up campaign. The telescopes and instruments used to obtain follow-up spectra included the Spectrograph for the Rapid Acquisition of Transients (SPRAT) on the 2~m Liverpool Telescope, the Kast Spectrograph on the 3~m Shane telescope at Lick Observatory, the Double Spectrograph (DBSP) on the 5.1~m Hale telescope at Palomar Observatory, the Multi-Object Double Spectrograph \citep[MODS;][]{Pogge2010} on the dual 8.4~m Large Binocular Telescope (LBT), the Low-Resolution Imaging Spectrometer \citep[LRIS;][]{oke95} on the Keck~I 10~m telescope, and the Deep Imaging Multi-Object Spectrograph \citep[DEIMOS;][]{faber03} on the Keck~II 10~m telescope. 
Most of the  spectra were acquired at or near the parallactic angle \citep{filippenko82} to minimize slit losses caused by atmospheric dispersion.

The majority of our spectra were reduced and calibrated using {\sc Iraf} following standard procedures, including bias subtraction, flat-fielding, one-dimensional spectral extraction, and wavelength calibration by comparison to a lamp spectrum. We reduced the MODS spectra using the MODS spectroscopic pipeline\footnote{\url{http://www.astronomy.ohio-state.edu/MODS/Software/modsIDL/}} \edit1{and the LRIS spectra using the {\sc LPipe} pipeline software \citep{perley19}}. The spectra were flux calibrated using \edit1{observations of standard stars} obtained on the same nights as the science spectra. For some cases we also performed telluric corrections using the standard-star spectra, but in most cases telluric features are left uncorrected. 

We used our follow-up photometric data to refine the flux calibration of the spectra. For each filter completely contained in the wavelength range covered by each spectrum and for which we could interpolate the photometric light curves to the spectroscopic epoch, we extracted synthetic photometric magnitudes. We then fit a line to the difference between the observed photometric flux and the synthetic flux as a function of the central wavelength of the filter, and scaled the spectrum by this fit. Finally, we corrected the spectra for Galactic extinction.

Figure~\ref{fig:spec_evol} shows the spectroscopic evolution of ASASSN-17jz. Only one spectrum is shown for dates where multiple \edit1{spectroscopic} observations were obtained on the same night. \edit1{Instrument setups and exposure times for each spectrum are listed in Table~\ref{tab:spec_details}.} We analyze the spectra and compare them to those of TDEs, SNe, AGNs, and other nuclear outbursts in Section~\ref{sec:spec_anal}.

\subsection{HST Spectroscopic Observations}
\label{sec:hst_spec}

\begin{figure}
\centering
{\includegraphics[width=0.49\textwidth]{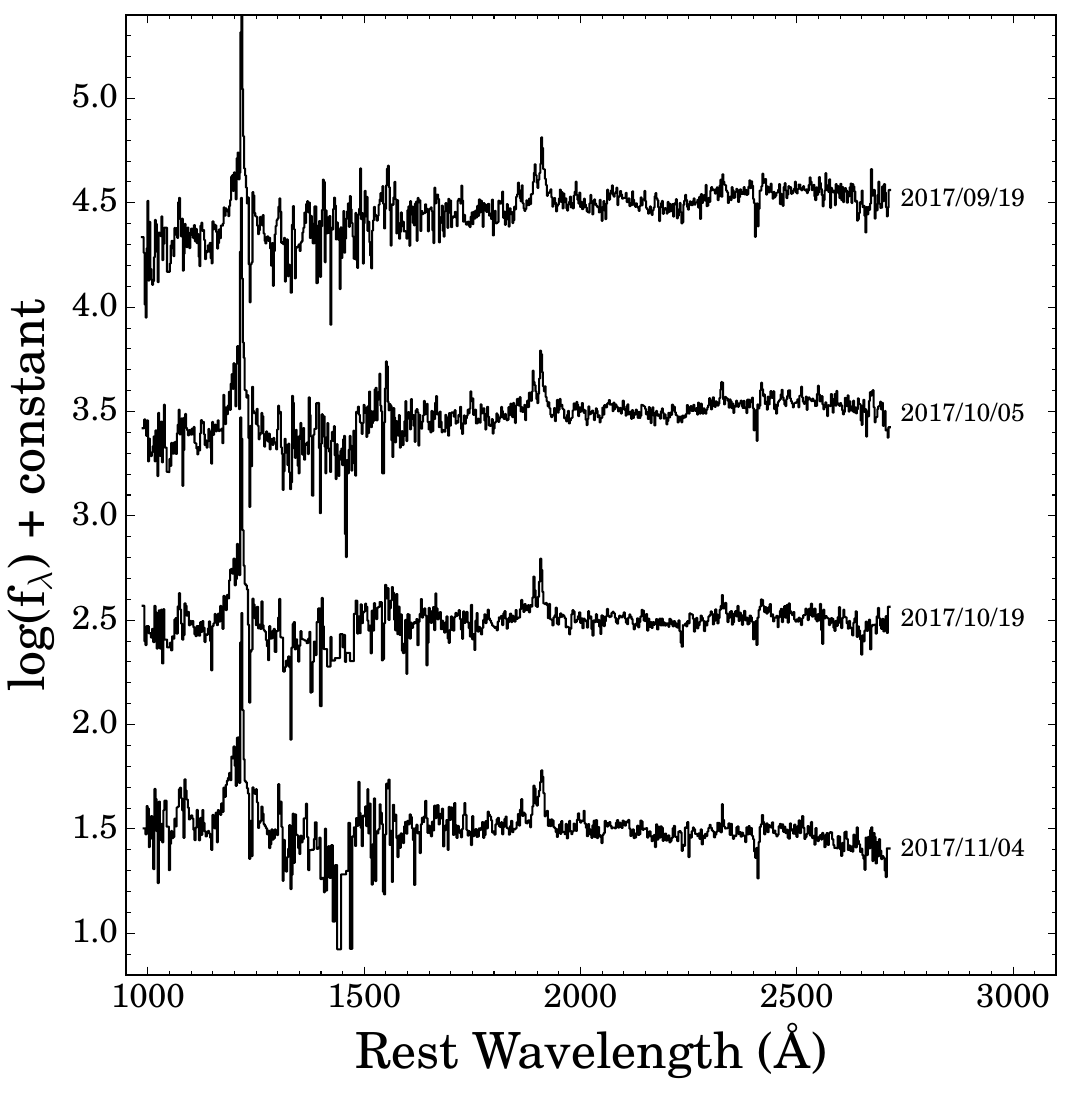}}
\caption{UV spectroscopic evolution of ASASSN-17jz from \emph{HST}/STIS (GO-14781; PI C. Kochanek) spanning from 46~days after peak brightness through 92~days after peak. The spectra have been flux calibrated using {\swift} photometry as described in Section~\ref{sec:hst_spec} and binned by a factor of 2--3 to reduce noise.}
\label{fig:spec_evol_stis}
\end{figure}

We obtained 4 observations (GO-14781; PI C. Kochanek) using the Space Telescope Imaging Spectrograph 
\citep[STIS;][]{woodgate98} on the \emph{Hubble Space Telescope} (\HST). We used the FUV/NUV MAMA detectors with the G140L (1150--1730~\AA, FUV-MAMA) and G230L (1570--3180~\AA, NUV-MAMA) gratings and 52\farcs{0}$\times$0\farcs{2} slit. \edit1{Full details and exposure times for each spectrum are listed in Table~\ref{tab:uv_spec_details}}. The source was clearly detected and spatially unresolved in the two-dimensional frames, so we used the standard \HST{} pipeline for producing one-dimensional spectra. We performed inverse-variance-weighted combinations of the individual exposures and merged the FUV and NUV channels to produce the spectra seen in Figure \ref{fig:spec_evol_stis}. We analyze the \HST{} spectra and compare them to UV spectra of other transients in Section~\ref{sec:spec_anal}.

\subsection{Radio Observations}
\label{sec:radio_obs}

We obtained radio observations of ASASSN-17jz using Director's Discretionary Time observations at 10~GHz with the Karl G. Jansky VLA (legacy code AR981) in its BnA configuration on 23 February 2018, and \edit1{TOO} observations at 5~GHz with the electronic European very long baseline interferometry Network (e-EVN) on 11 April 2018 (project code RR011). 

\begin{figure}
\centering
{\includegraphics[width=0.49\textwidth]{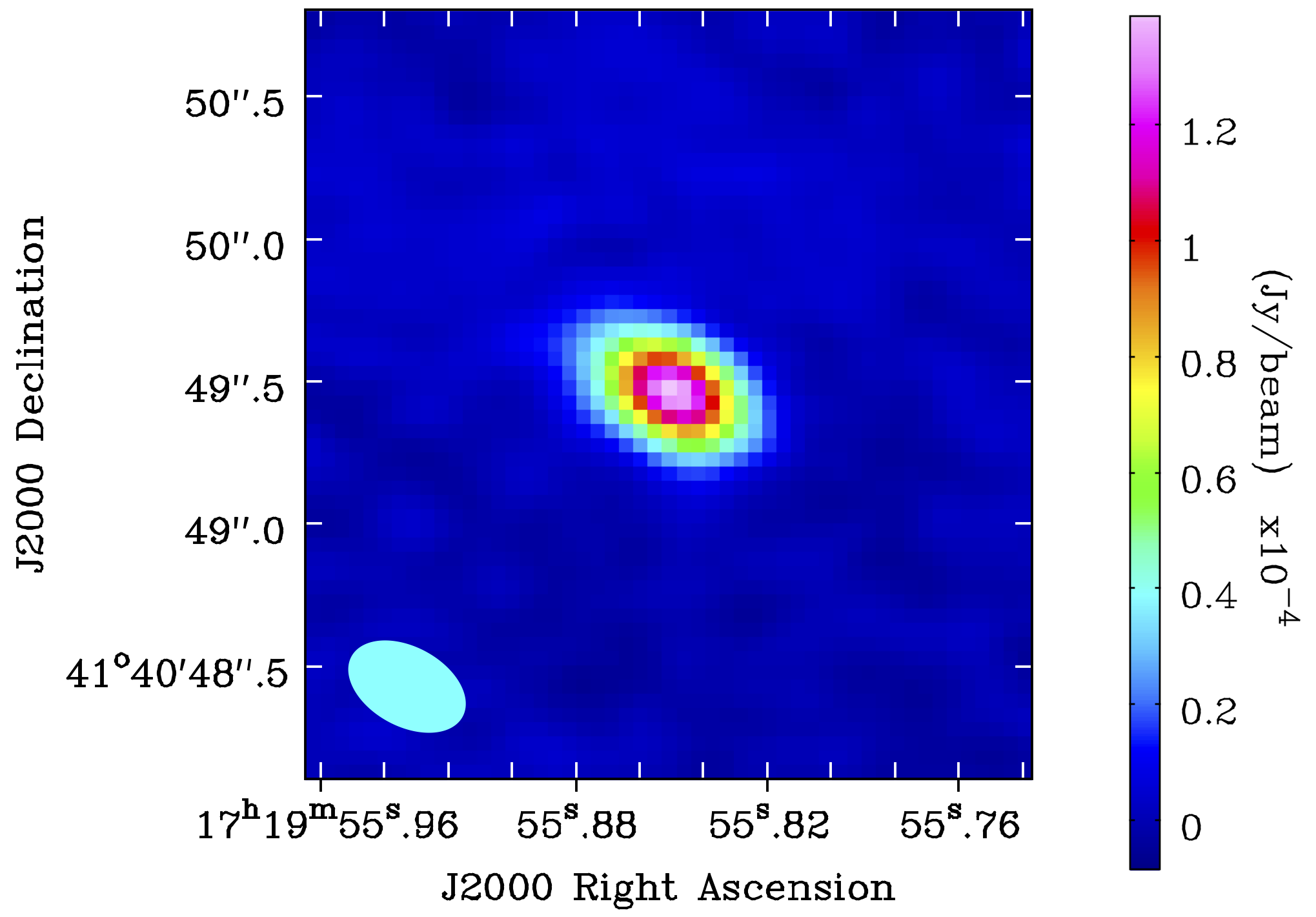}}
\caption{VLA image \edit1{at the position of} ASASSN-17jz at a central frequency of 10~GHz (FWHM $=0\farcs43\times0\farcs27$
at PA $=61.5\degr{}$), from observations obtained on 23 February 2018.}
\label{fig:radio_im}
\end{figure}

The VLA observations consisted of 2.2~hr on source using the full bandwidth ($2\times2$~GHz) X-band receiver in full polarization. We reduced the data with the VLA recipe for total intensity continuum observations within the Common Astronomy Software Applications package \citep[{\sc casa};][]{mcmullin07} version 5.1.2-4. We reran the pipeline with additional data flagging after inspecting the first pipeline products. Using the full bandwidth we attained a root-mean square (RMS) of 2.1~\mujyb{}, improving the pipeline product by 5\%. We used J1331+3030 as a flux-density calibrator and J1734+3857 as a phase calibrator. Both calibrators appear as point sources with 10~GHz peak intensities of $4.39\pm0.22$~\jyb{} and $0.88\pm0.04$~\jyb{}, respectively. We detected a source at the position of ASASSN-17jz with an S/N of $\sim 65$ when using the full bandwidth. Figure~\ref{fig:radio_im} shows the VLA image \edit1{at the position of the} transient.

The e-EVN observations were carried out with the antennas Effelsberg (100~m, DE), Jodrell Bank ($38\times25$~m, UK), Westerbork (25~m, NL), Medicina (32~m, IT), Noto (32~m, IT), Onsala (25~m, SE), Tianma (65~m, CN), Torun (32~m, PL), Yebes (40~m, ES), Hartebeesthoek (26~m, SA), and Irbene (32~m, LV). From the half-day e-EVN run, 7.35~hr were spent on source, resulting in an RMS of $\sim 9.69$~\mujyb{} at the position of ASASSN-17jz. We reduced the data following standard procedures within the NRAO Astronomical Image Processing System ({\sc aips}) and using some products corresponding to the initial steps of the EVN pipeline. We used J1724+4004 as the phase calibrator. \edit1{The calibrator has a core-jet morphology with a peak} intensity of $2.78\pm0.67$~\mjyb{} \edit1{for the brightest component and we took this structure into account when fringe-fitting the data.} The resulting image has a resolution of $4.1 \times 3.2$~mas at a central frequency of 5~GHz. Taking into account the zero-level emission and a 5\% uncertainty in the calibration, we obtained an observed peak intensity upper limit for ASASSN-17jz of $<34$~\mujyb{}. 


In our VLA observations we detected a slightly resolved source with an observed flux density of about $148\pm16$\mujyb{} in the 8--12~GHz X-band beam (full width at half-maximum intensity [FWHM] $=0\farcs4\times0\farcs3$ at position angle [PA] $=61.5\degr{}$). The \edit1{e-EVN} observations made only 45~days after the VLA observations resulted in an upper limit of $<34$~\mujyb{}. Observations made at 3~GHz within the VLA Sky Survey \citep[VLASS;][]{lacy20} one month after the \edit1{e-EVN} observations resulted in an even less constraining upper limit of $<378$~\mjyb{}.

\edit1{We analyze these radio data and discuss their possible physical origins in Section~\ref{sec:radio_analysis}.}


\section{Photometric Analysis}
\label{sec:phot_anal}

\subsection{Position, Redshift, and $t_{\rm peak}$ Measurements}
\label{sec:params}

The position and host offset of ASASSN-17jz was measured using a host-subtracted $g$-band image from the Las Cumbres Observatory 1~m telescopes taken on 2017 September 5. We used the \textsc{Iraf} task \texttt{imcentroid} to measure a centroid position of the transient in the subtracted image and a centroid position of the nucleus of the host galaxy in the archival $g$-band image from SDSS. The resulting position of ASASSN-17jz is $(\alpha,\delta) = (17^{\rm h}19^{\rm m}55.85^{\rm s}, +41^\circ40'49\farcs{45})$, which is offset by 0\farcs{12} from the measured position of the host nucleus. This offset is likely dominated by a systematic offset in the astrometric solutions of the two images. To determine the likely systematic offset, we also measured the centroid positions of several stars in both the nonsubtracted follow-up image and the archival host image, and calculated an average offset for the positions of these comparison stars. The resulting average is 0\farcs{27}, and thus the transient is offset by $0\farcs{12}\pm0\farcs{27}$, corresponding to a physical offset of $460\pm1040$~pc, consistent with the nucleus of \host. 

\begin{figure}
\centering
\includegraphics[width=0.49\textwidth]{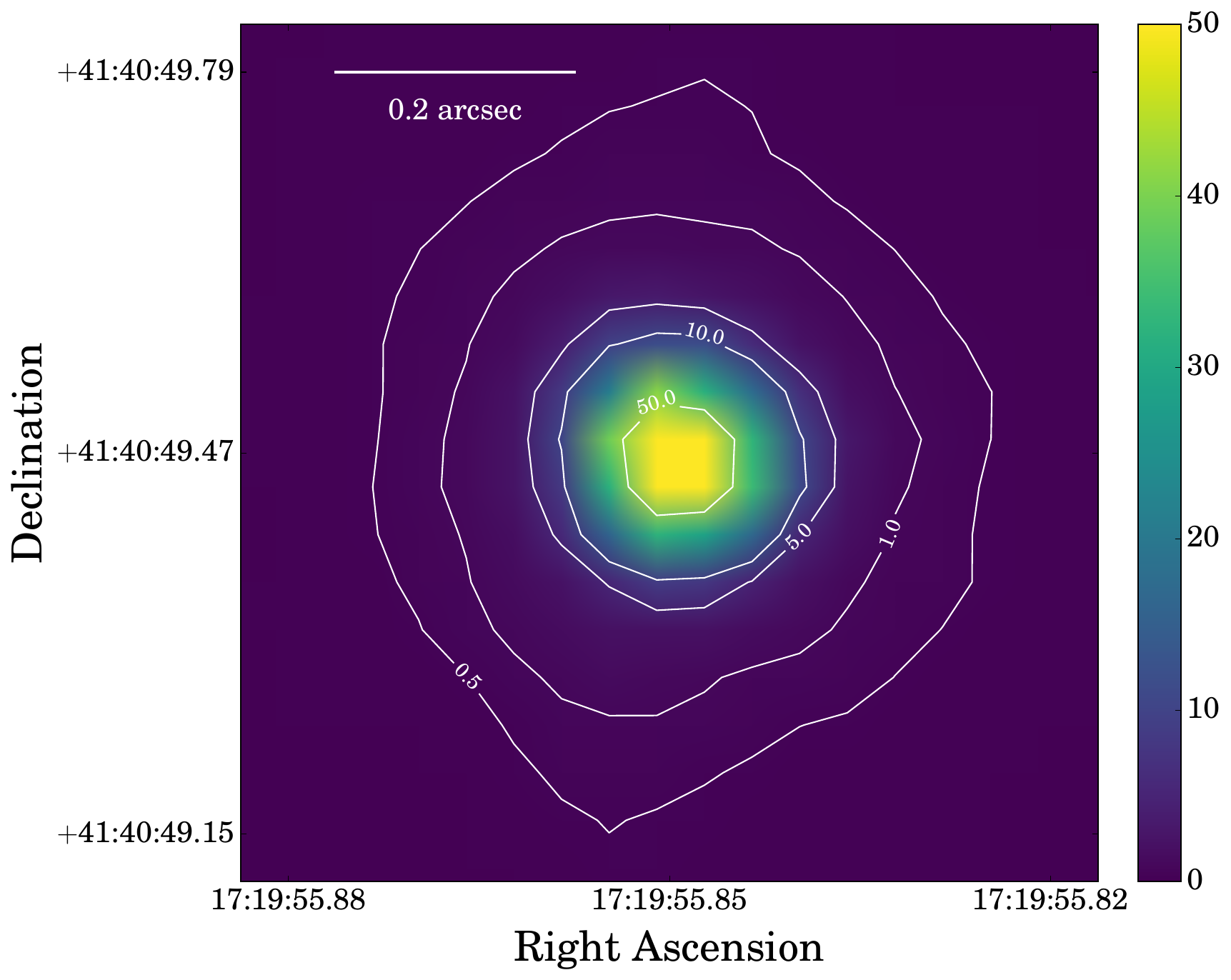}
\caption{Contour plot of the nucleus of \host{} using the \HST{} WFC3/UVIS $F555W$ image of ASASSN-17jz taken on 2018 March 22 (GO-15166; PI A. Filippenko). The contours indicate regions of increasing signal and the scale is given in the top-left.}
\label{fig:hst_contour}
\end{figure}

We further examined the position of ASASSN-17jz using \HST{} WFC3/UVIS data obtained on 2018 March 22, 230 days after peak light (GO-15166; PI A. Filippenko). Using both $F555W$ and $F814W$ images, we measured a centroid position of the transient, finding $(\alpha,\delta) = (17^{\rm h}19^{\rm m}55.85^{\rm s}, +41^\circ40'49\farcs{47})$, consistent with the position we measured from the host-subtracted LCOGT data. We also used the significantly higher resolution of \HST{} to examine whether two distinct sources are visible. Figure~\ref{fig:hst_contour} shows a contour plot of the nuclear region of \host{} in the $F555W$ image. To the resolution limit of WFC3/UVIS, no second emission source can be seen outside of the nucleus of the host galaxy. Since the transient was still roughly as \edit1{optically} bright as the host nucleus at this time, this strongly suggests that ASASSN-17jz occurred in the nucleus of its host.


There is no previously measured redshift for the host galaxy. The initial classification of ASASSN-17jz reported a transient redshift of $z=0.164$ based on the broad Balmer lines \citep{asassn17jz_spec_tns}. We measured the redshift of the transient using the narrow \ion{O}{1}~8446~\AA\ line that is visible in the 2017 November 18 LRIS spectrum, finding $z=0.1641$. We adopt this slightly more precise measurement throughout the manuscript, corresponding to a luminosity distance of $d=792.9$~Mpc (H$_0=69.6$~km~s$^{-1}$~Mpc$^{-1}$, $\Omega_M=0.296$, and $\Omega_{\Lambda}=0.714$).

We estimated the time of peak light using the stacked host-subtracted $B$-band light curve by fitting a parabola to the $B$ data taken prior to MJD = 57985. In order to estimate the uncertainty in the peak time, we generated 10,000 $B$-band light curves for our specified date range, perturbing each magnitude by its uncertainty and assuming Gaussian errors. We then fit each of these 10,000 light curves using the same parabolic model and calculated the 68\% confidence interval in the peak date and magnitude, finding $t_{B,{\rm peak}}=57969.6\pm0.9$ and $m_{B,{\rm peak}}=16.69\pm0.01$~mag. We adopt this time, corresponding to 2017 August 4.6, as our reference time throughout our analysis \edit1{when referring to the phase of the transient}. We used the same method to calculate the peak times for each of the filters redder than $B$ and find that there is some evidence that redder filters peaked later than bluer filters, with $t_{I,{\rm peak}}=57978.2\pm0.9$ being the latest. We do not have enough data points near peak in the {\swift} UVOT+$U$ filters to make such fits. However, our first epoch of {\swift} data was obtained on MJD = 57968.6, prior to our calculated $B$-band peak, and the {\swift} light curves are clearly declining at all epochs, implying that the transient peaked at an earlier time in the UV. This behavior is similar to what has been seen in several TDEs \citep[e.g.,][]{holoien18a,holoien19b,hinkle21b,hinkle21a}. 

At the host distance of $d=792.9$~Mpc, the $B$-band peak magnitude of ASASSN-17jz corresponds to an absolute magnitude of $M_{B,{\rm peak}}=-22.81$. This is significantly more luminous than most TDEs, which tend to peak at absolute magnitudes of $M \approx -20$ \citep[e.g.,][]{hinkle21b}, and would be at the upper end of the typical luminosity range for \edit1{superluminous supernovae} (SLSNe) \citep[e.g.,][]{galyam19}. ASASSN-17jz is thus one of the most luminous transients ever discovered, and we performed several types of analysis to investigate the nature of this event.

\subsection{\texttt{MOSFiT} Light-Curve Fits}
\label{sec:lc_anal}

We fit the multiband host-subtracted light curves of the transient using the Modular Open-Source Fitter for Transients \citep[\texttt{MOSFiT};][]{guillochon18}. \texttt{MOSFiT} uses various models to generate bolometric and single-filter light curves and then fits these to the observed data. We use it here to fit the light curves of ASASSN-17jz, as it is the only software that has models for both SNe and TDEs, and is one of the only fitting tools available for generalized fitting of TDE emission. We used the nested sampling mode to fit the model parameters and estimate their uncertainties. See \citet{guillochon18} for more details about \texttt{MOSFiT} and its built-in models.

\begin{figure*}
\begin{minipage}{\textwidth}
\centering
\subfloat[][]{
\includegraphics[width=0.495\textwidth]{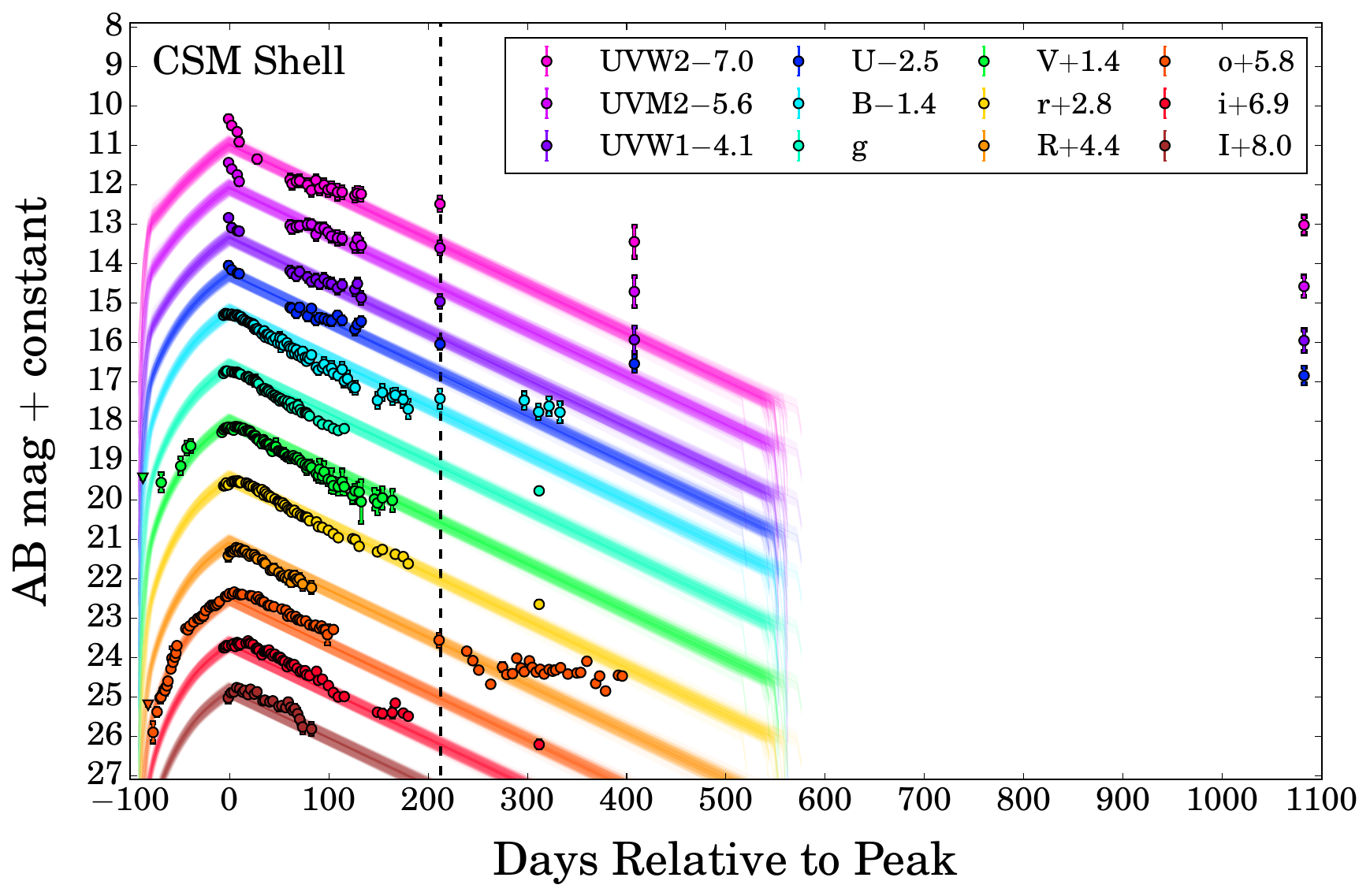}}
\subfloat[][]{
\includegraphics[width=0.495\textwidth]{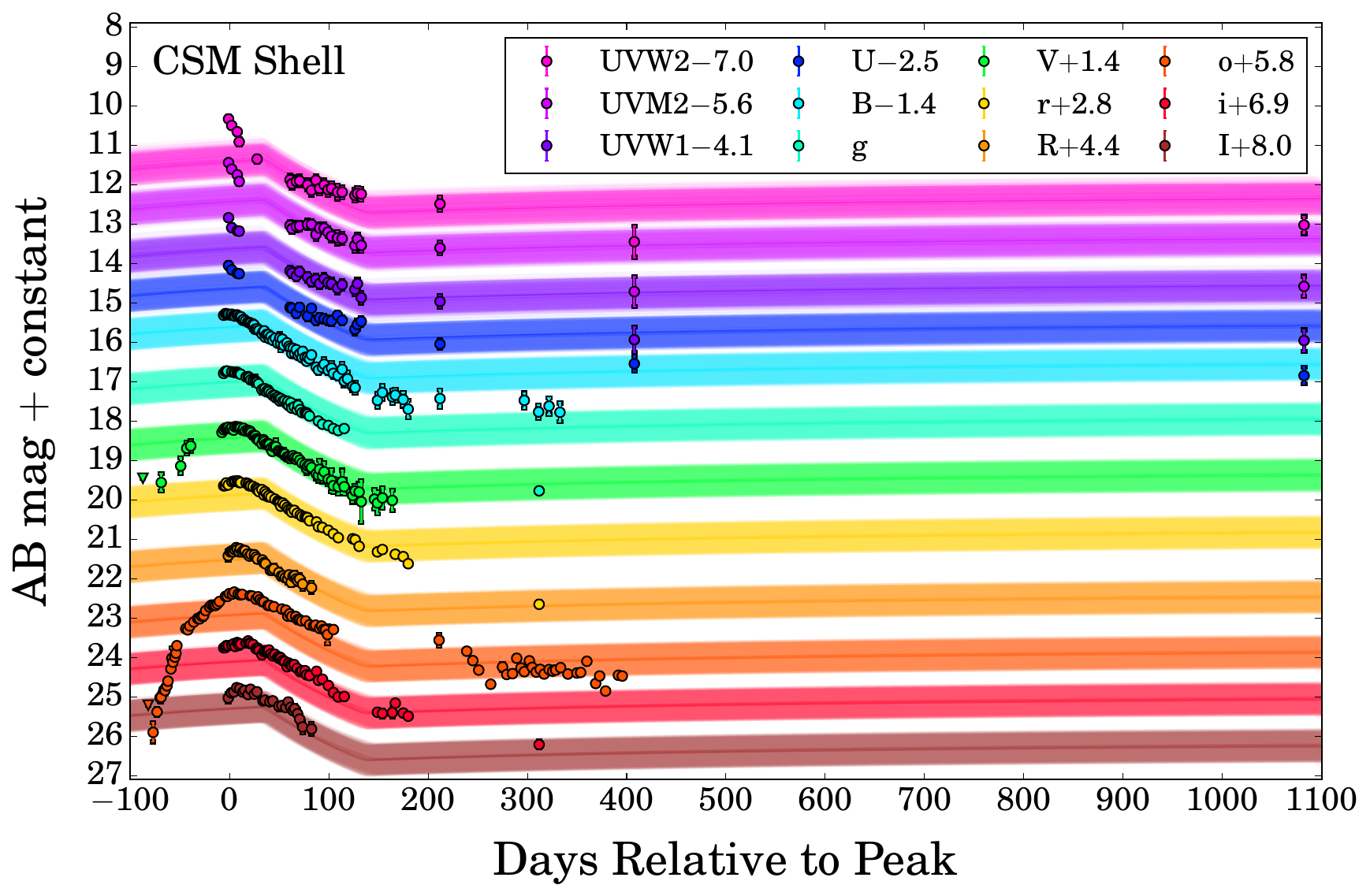}}
\\
\subfloat[][]{
\includegraphics[width=0.495\textwidth]{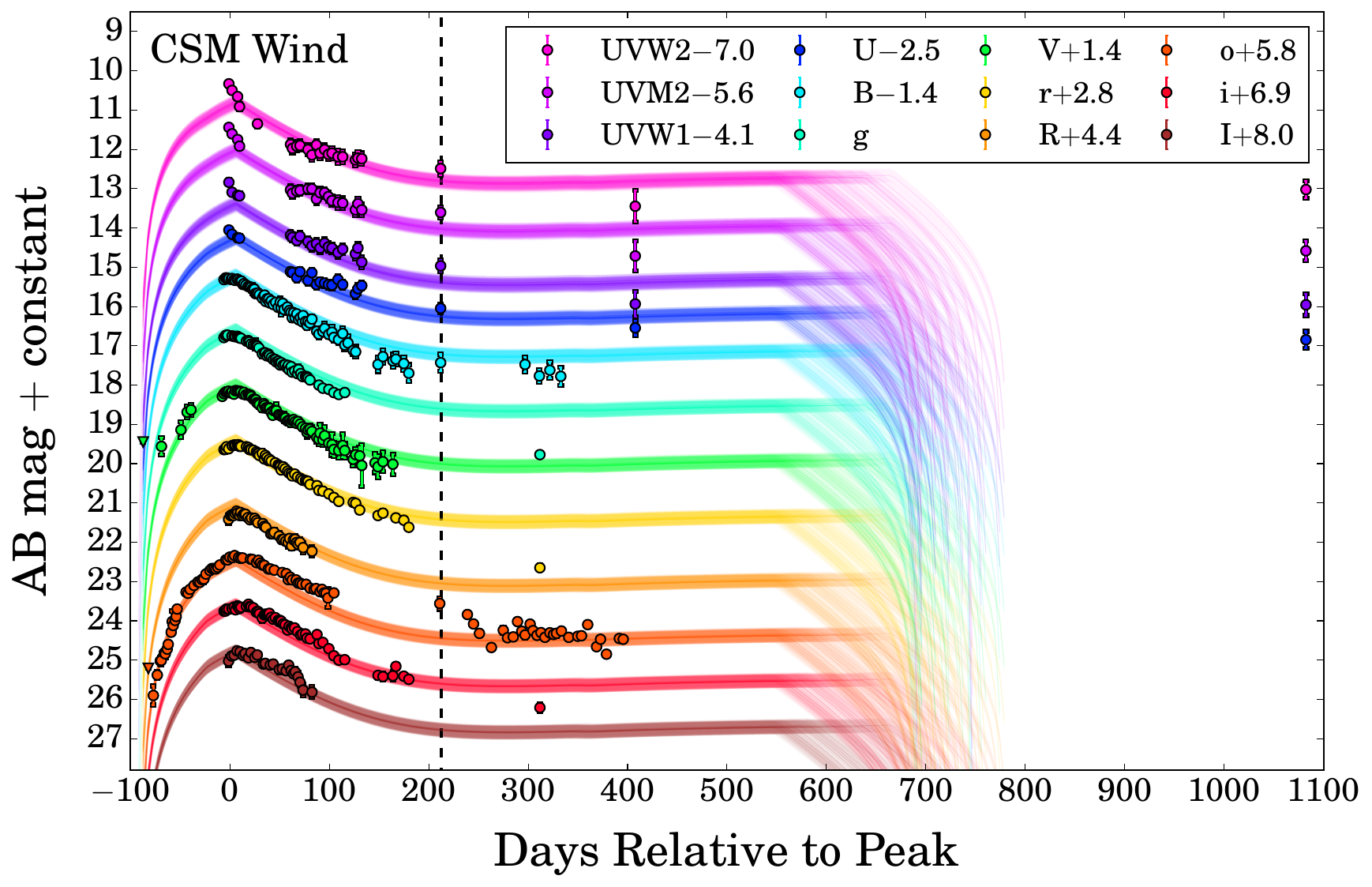}}
\subfloat[][]{
\includegraphics[width=0.495\textwidth]{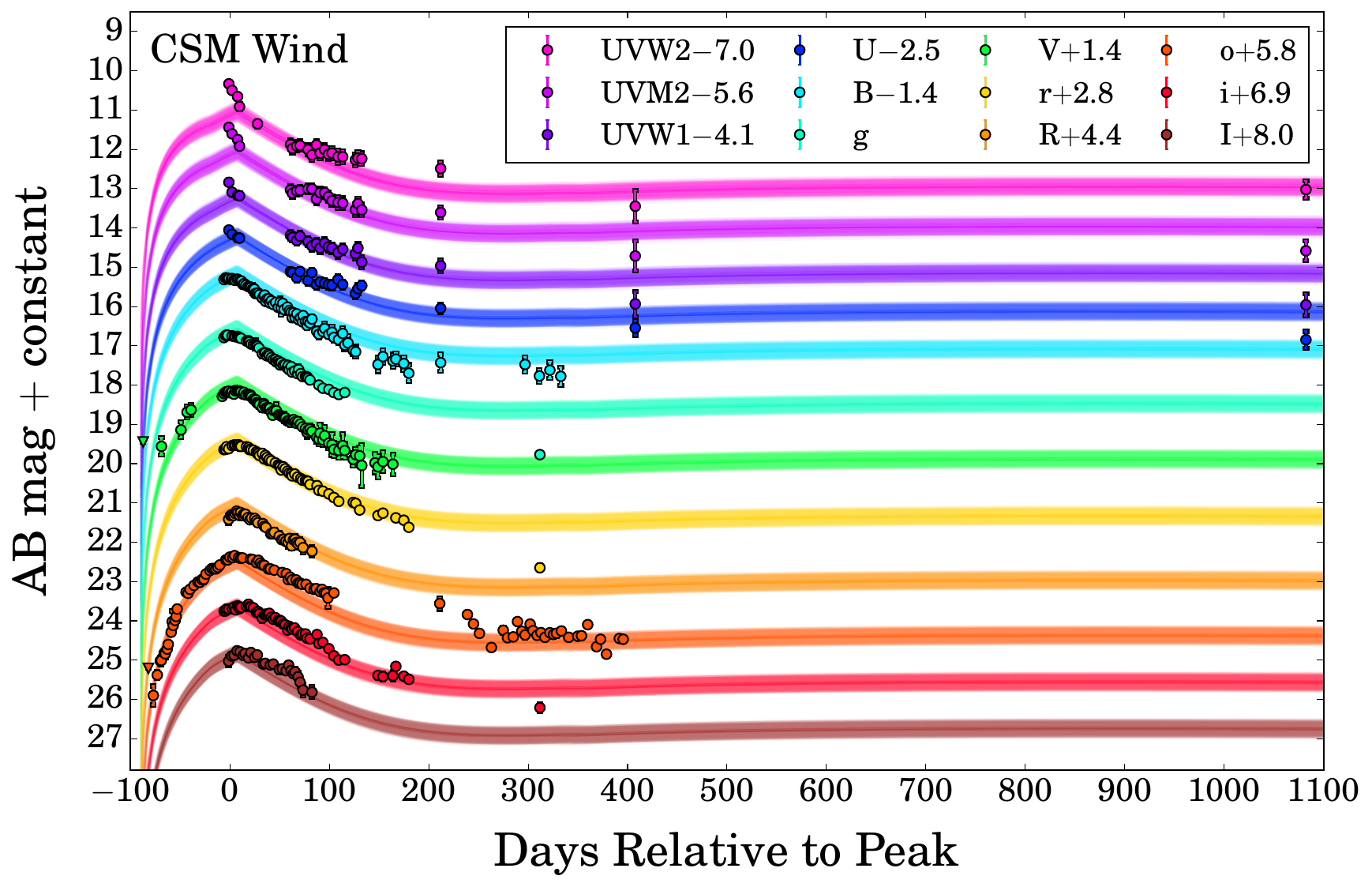}}
\\
\subfloat[][]{
\includegraphics[width=0.495\textwidth]{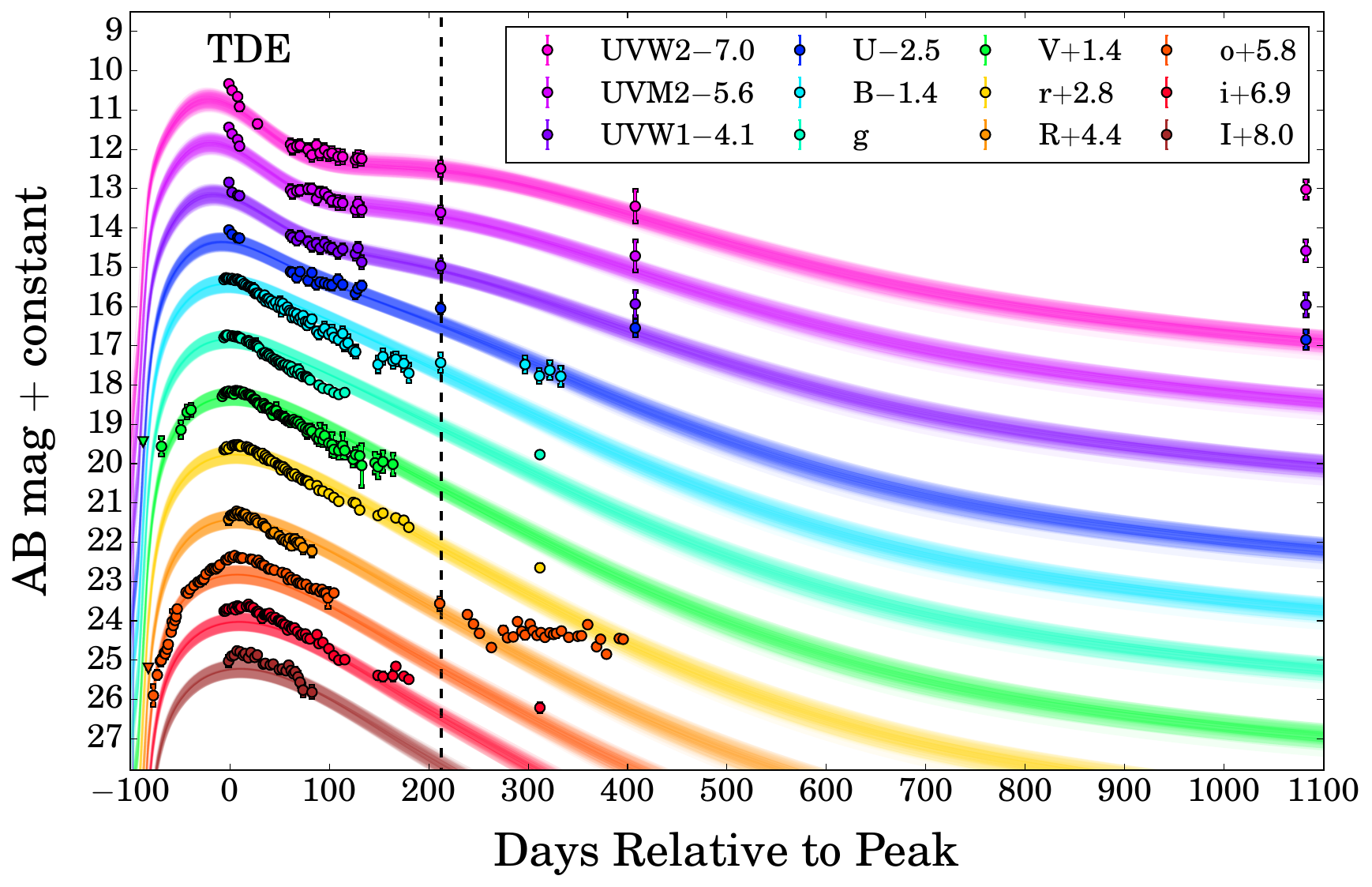}}
\subfloat[][]{
\includegraphics[width=0.495\textwidth]{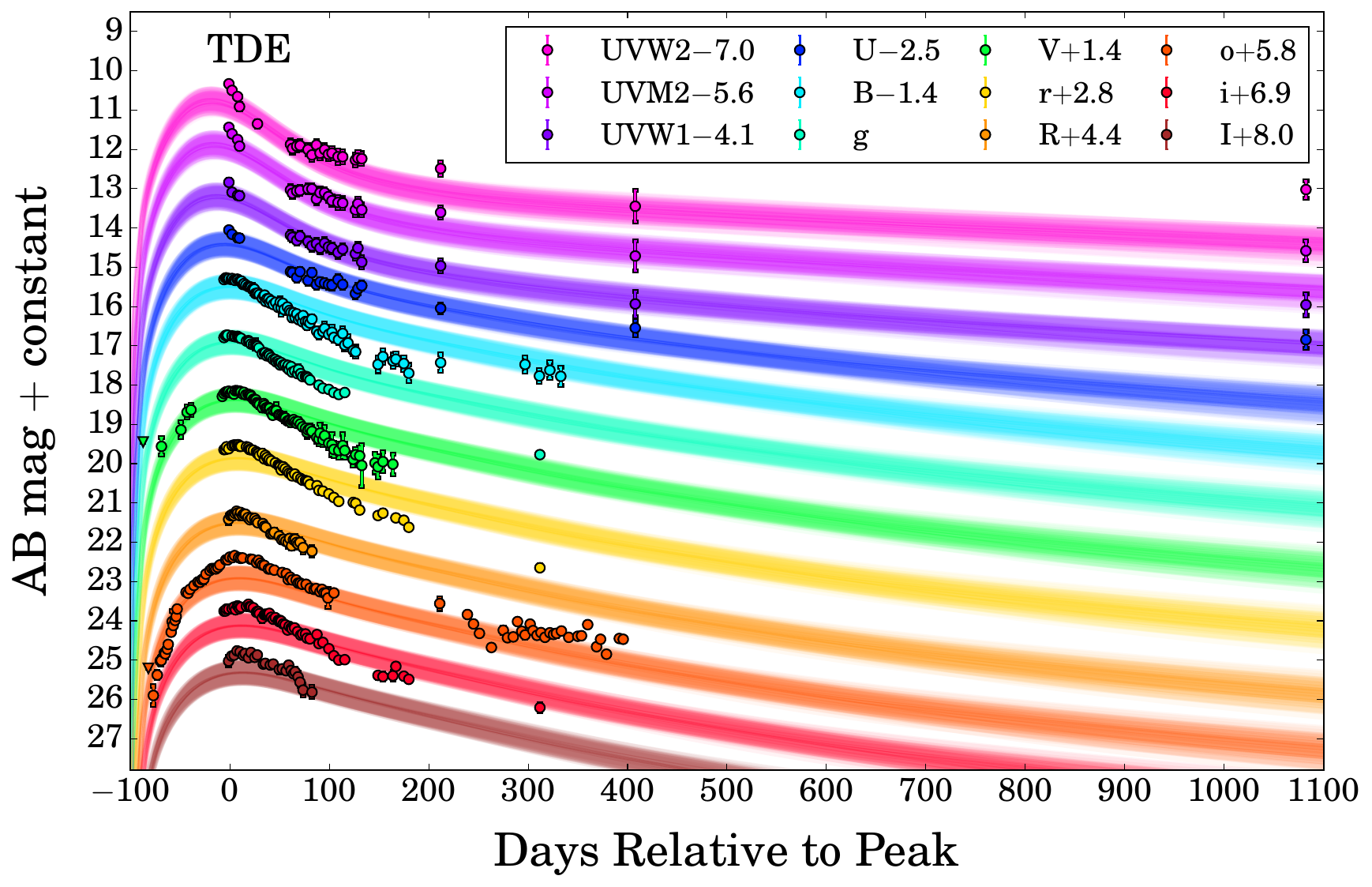}}
\caption{Host-subtracted light curves and light-curve fits from \texttt{MOSFiT} \citep{guillochon17}. The colors of the different filters match those of Figure~\ref{fig:lc}. The top row shows fits using an SN+CSM interaction model with the CSM modeled as a shell, the middle row shows fits using an SN+CSM interaction model with the CSM modeled as a wind, and the bottom row shows fits using a TDE model. The left column displays fits using only the data obtained prior to MJD = 58182, indicated by the vertical dashed black line, while the right column gives fits using the full dataset.}
\label{fig:mosfit}
\end{minipage}
\end{figure*}

We used three different models to fit the light curves of ASASSN-17jz with \texttt{MOSFiT}: a supernova interacting with circumstellar medium (SN+CSM), with two different CSM density profiles (see below), and the TDE model. As the UV light curves of ASASSN-17jz flatten at late times, possibly indicative of a secondary power source becoming dominant later, we performed two sets of fits for each model, one using only data obtained prior to MJD = 58182, which excludes the late-time flat UV emission, and one using the entire dataset. The results of these fits are shown in Figure~\ref{fig:mosfit}.

The model for SN emission with CSM interaction used by \texttt{MOSFiT} is described by \citet{chatzopoulos13}. For our fits, we varied the slope of the CSM density profile, $\rho \propto r^{-s}$, to obtain fits for both a shell CSM ($s=0$) and a wind CSM ($s=2$), and set the ejecta density index $\rho \propto r^{-n}$ to $n=7$. The free parameters of the model are the ejecta mass $M_{\textrm{ej}}$, the CSM mass $M_{\textrm{CSM}}$, the inner radius of the CSM $R_{0}$, the density $\rho$ of the CSM at $R_0$, the velocity of the ejecta $v_{\textrm{ej}}$, and the time of explosion $t_{\textrm{exp}}$. 

The top row of Figure~\ref{fig:mosfit} shows the fits using an SN+CSM-shell. This model is able to fit the early-time data fairly well (top-left plot in the Figure), though it slightly underestimates the UV emission, particularly at peak and at later times, and prefers a steeper decline than we observe. However, this model is clearly unable to fit the entire dataset. In order to match the late-time flattening it requires a much slower rise than we observe. If the transient is powered by an SN with interaction with a constant-density shell of CSM, the late-time UV emission must be powered by another source, such as \edit1{an} underlying AGN. Moreover, for both the fits to the early-time data and the fit to the full dataset, the ejecta masses are unreasonably high ($M_{\rm ej} \approx 250$~\msun). Based on these results, we disfavor an SN+CSM model with a constant-density CSM shell.


\begin{deluxetable}{cccc}
\tabletypesize{\footnotesize}
\tablecaption{\texttt{MOSFiT} Model Parameter Fits --- SN+Constant-Density Shell CSM}
\tablehead{
\colhead{Parameter} &
\colhead{Early} &
\colhead{Full} &
\colhead{Units} }
\startdata
$M_{\textrm{ej}}$ & $256.3^{+2.6}_{-5.0}$ & $246.9^{+8.7}_{-13.0}$ & \msun \\
$M_{\textrm{CSM}}$ & $17.8^{+3.9}_{-3.4}$ & $100.3^{+7.1}_{-8.6}$ & \msun \\
$R_{0}$ & $12.7^{+16.2}_{-7.6}$ & $21.7^{+20.0}_{-13.3}$ & au \\
$\log{\rho}$ & $-11.7^{+0.0}_{-0.0}$ & $-13.9^{+0.1}_{-0.0}$ & g~cm$^{-3}$ \\
$v_{\textrm{ej}}$ & $9935.9^{+43.7}_{-80.7}$ & $2961.3^{+136.3}_{-93.8}$ & km~s$^{-1}$ \\
$t_{\textrm{exp}}$ & $-15.5^{+1.1}_{-1.2}$ & $-468.7^{+36.3}_{-20.8}$ & days \\
\enddata 
\tablecomments{Best-fit parameter values and 16--84\% ranges in the uncertainties from \texttt{MOSFiT} for an SN+CSM interaction model, with the CSM described as a constant-density shell ($s=0$). The ``Early'' column shows the results for fitting only the data obtained prior to MJD = 58182 (top-left panel of Fig.~\ref{fig:mosfit}), while the ``Full'' column shows the results for fitting the entire dataset (top-right panel of Fig.~\ref{fig:mosfit}). The listed uncertainties are from the model fit, and do not include any systematic uncertainties.}
\label{tab:mosfit_csm_shell} 
\vspace{-0.8cm}
\end{deluxetable}

In the middle row of Figure~\ref{fig:mosfit} we show the fits using an SN+CSM-wind model. In this case we find that the fits to just the early-time data (middle-left panel of the Figure) and the fits to the full dataset (middle-right panel of the Figure) look very similar, with the model able to fit the leveling off of the light curves that begins around 200~days after peak, and with the early data generally being fit well with the exception of the peak UV emission again being underestimated. The fit to the early-time data nearly represents the data points at $t \approx 400$~days after peak despite these not being included in the fit. When the late-time data are included, the fit is able to reproduce the late-time UV emission without significantly changing the fits to the rise and peak, implying that in this case a secondary power source would not be required. However, while other parameters of this model are reasonable, it requires both extremely large ejecta and CSM masses for both the early-time and complete fits (Table~\ref{tab:mosfit_csm_wind}).


\begin{deluxetable}{cccc}
\tabletypesize{\footnotesize}
\tablecaption{\texttt{MOSFiT} Model Parameter Fits --- 
SN+Steady-State Wind CSM}
\tablehead{
\colhead{Parameter} &
\colhead{Early} &
\colhead{Full} &
\colhead{Units} }
\startdata
$M_{\textrm{ej}}$ & $252.4^{+5.0}_{-9.2}$ & $253.0^{+4.5}_{-6.8}$ & \msun \\
$M_{\textrm{CSM}}$ & $81.0^{+2.8}_{-3.8}$ & $129.3^{+8.7}_{-9.6}$ & \msun \\
$R_{0}$ & $18.6^{+6.7}_{-2.7}$ & $23.6^{+3.8}_{-5.0}$ & au \\
$\log{\rho}$ & $-10.7^{+0.1}_{-0.3}$ & $-12.1^{+0.2}_{-0.1}$ & g~cm$^{-3}$ \\
$v_{\textrm{ej}}$ & $9925.5^{+49.8}_{-90.5}$ & $9922.8^{+51.0}_{-75.9}$ & km~s$^{-1}$ \\
$t_{\textrm{exp}}$ & $-13.7^{+0.7}_{-0.7}$ & $-14.4^{+0.9}_{-0.8}$ & days \\
\enddata 
\tablecomments{Best-fit parameter values and 16--84\% ranges in the uncertainties from \texttt{MOSFiT} for an SN+CSM interaction model, with the CSM described as a steady-state wind ($s=2$). The ``Early'' column shows the results for fitting only the data obtained prior to MJD = 58182 (middle-left panel of Fig.~\ref{fig:mosfit}), while the ``Full'' column shows the results for fitting the entire dataset (middle-right panel of Fig.~\ref{fig:mosfit}). The listed uncertainties are from the model fit, and do not include any systematic uncertainties.}
\label{tab:mosfit_csm_wind} 
\vspace{-0.8cm}
\end{deluxetable}

We note that in both SN+CSM cases the uncertainties listed in Tables~\ref{tab:mosfit_csm_shell} and \ref{tab:mosfit_csm_wind} do not include any systematic uncertainties, as these have not been quantified for the \texttt{MOSFiT} model for SN+CSM interaction. The listed uncertainties are likely underestimating the total uncertainty and should be taken as lower limits in the uncertainties of the various model parameters.

Finally, Figure~\ref{fig:mosfit} shows the \texttt{MOSFiT} TDE model fits to the early-time data (bottom-left panel) and full dataset (bottom-right panel). Notably, the fits to the early-time data alone fit the next UV epochs well but underpredict the optical emission. The early-time fits match the rise and early decline fairly well, though the early-time SN+CSM models fit better. The fit to the full dataset exchanges matching the early rise and fall of the transient to fit the late-time data, but still underestimates the emission in the latest epoch. This implies that a TDE alone is not able to replicate the observed emission. However, \texttt{MOSFiT} has had difficulty with fitting the late-time UV emission seen in other TDEs in the past \citep[e.g.,][]{holoien20a}. The late-time flattening seen in UV emission from TDEs has been attributed to a transition from fallback-dominated emission to disk-dominated emission \citep[e.g.,][]{holoien18a,velzen19}, and the \texttt{MOSFiT} TDE model was built to predict TDE emission when the bolometric luminosity closely follows the fallback rate, so it is perhaps unsurprising that \texttt{MOSFiT} does not fit the latest data well.


\begin{deluxetable}{cccc}
\tabletypesize{\footnotesize}
\tablecaption{\texttt{MOSFiT} Model Parameter Fits --- TDE}
\tablehead{
\colhead{Parameter} &
\colhead{Early} &
\colhead{Full} &
\colhead{Units} }
\startdata
$\log{R_{\textrm{ph0}}}$ & $2.9^{+0.4}_{-0.4}$ & $1.86^{+0.4}_{-0.4}$ & --- \\
$l$ (photosphere exponent) & $2.1^{+0.2}_{-0.2}$ & $1.2^{+0.2}_{-0.2}$ & --- \\
$t_\textrm{fb}$ & $-26.7^{+15.1}_{-15.0}$ & $-27.7^{+15.2}_{-15.2}$ & days \\
$\log{T_{\textrm{viscous}}}$ & $0.6^{+0.2}_{-0.3}$ & $0.3^{+0.4}_{-1.4}$ & days \\
$b$ (scaled $\beta$) & $0.7^{+0.3}_{-0.3}$ & $0.9^{+0.3}_{-0.3}$ & --- \\
$\log{M_{\rm BH}}$ & $7.3^{+0.2}_{-0.2}$ & $7.3^{+0.2}_{-0.2}$ & \msun \\
$M_\star$ & $1.0^{+3.6}_{-0.8}$ & $1.0^{+3.6}_{-0.8}$ & \msun \\
$\log{\epsilon}$ (efficiency) & $-1.7^{+0.7}_{-0.7}$ & $-1.7^{+0.7}_{-0.7}$ & --- \\
$\log{n_{\textrm{H,host}}}$ & $17.8^{+1.2}_{-1.1}$ & $17.3^{+1.1}_{-0.9}$ & cm$^{-2}$ \\
$\log{\sigma}$ & $-0.6^{+0.0}_{-0.0}$ & $-0.5^{+0.0}_{-0.0}$ & --- \\
\enddata 
\tablecomments{Best-fit parameter values and 16--84\% ranges in the uncertainties from \texttt{MOSFiT} for the TDE model. The ``Early'' column shows the results for fitting only the data obtained prior to MJD = 58182 (bottom-left panel of Fig.~\ref{fig:mosfit}), while the ``Full'' column shows the results for fitting the entire dataset (bottom-right panel of Fig.~\ref{fig:mosfit}). The uncertainties include both from the fit and the systematic uncertainties from Table 3 of \citet{mockler19}. From top to bottom, the {\texttt{MOSFiT}} TDE model parameters are the photosphere power-law normalization, the photosphere power-law exponent, the time of first fallback, the viscous timescale, the scaled impact parameter, the black hole mass, the mass of the disrupted star, the efficiency at which material falling onto the black hole is converted to bolometric flux, the column density of the host, and the model variance parameter. These are described in more detail by \citet{mockler19}.}
\label{tab:mosfit_tde}
\vspace{-0.8cm}
\end{deluxetable}

While the \texttt{MOSFiT} TDE model is crude, it remains the only available tool for generalized fitting of TDE emission. The model parameters in Table~\ref{tab:mosfit_tde} are typical of \texttt{MOSFiT} models of TDEs \citep[e.g.,][]{mockler19,holoien20a}. Both fits prefer a star of roughly $M_\star \approx 1.0$~\msun, a black hole mass of $M_{\rm BH} \approx 10^{7.3}$~\msun, and are consistent with a full disruption of the star, though the early-time fit is only marginally consistent with a full disruption. The BH mass in both cases is consistent with our estimate based on the bulge mass, and the star and BH masses are consistent with those of several other TDEs from \citet{mockler19}. We note that as \citet{mockler19} included an estimate of the systematic uncertainties of the \texttt{MOSFiT} TDE model, we have included these in the uncertainties shown in Table~\ref{tab:mosfit_tde}.

In all cases, the light-curve fits from \texttt{MOSFiT} indicate that none of the models we tested can replicate both the peak and tail of the light curve. The \texttt{MOSFiT} results thus disfavor a single transient origin for ASASSN-17jz. However, all three models are able to fit the data prior to MJD = 58182 well, particularly the two SN+CSM models, and thus all three are consistent with a scenario where a transient powers the early-time light curve while the late-time emission is powered by an underlying AGN or other secondary power source. 

\edit1{While we have used \texttt{MOSFiT} to test various scenarios, its models are limited in their scope. In particular, the TDE model does not have the capability to test more ``exotic'' scenarios (e.g., incorporating the effects of black hole spin on the TDE emission, or pushing the components of the models to values beyond those used for the simulations on which the models are based). Thus, we do not make any conclusions from these fits alone, and we incorporate additional fitting of the light curves and other observations in the conclusions made in the discussion (Section~\ref{sec:disc}).}

\subsection{Blackbody SED Fits and Luminosity Models}
\label{sec:sed_anal}

We fit a blackbody model to the UV and optical SED of ASASSN-17jz for epochs where {\swift} data were available, as we have done with previous nuclear transients \citep[e.g.,][]{hinkle21b}. We used Markov Chain Monte Carlo methods to find the best-fit blackbody parameters for the SED at each epoch, using flat priors of $1000$~K~$\leq T \leq55000$~K and $10^{11}$~cm~$\leq R \leq10^{17}$~cm so as not to overly influence the fits.

\begin{figure*}
\begin{minipage}{\textwidth}
\centering
\subfloat[][]{
\includegraphics[width=0.325\textwidth]{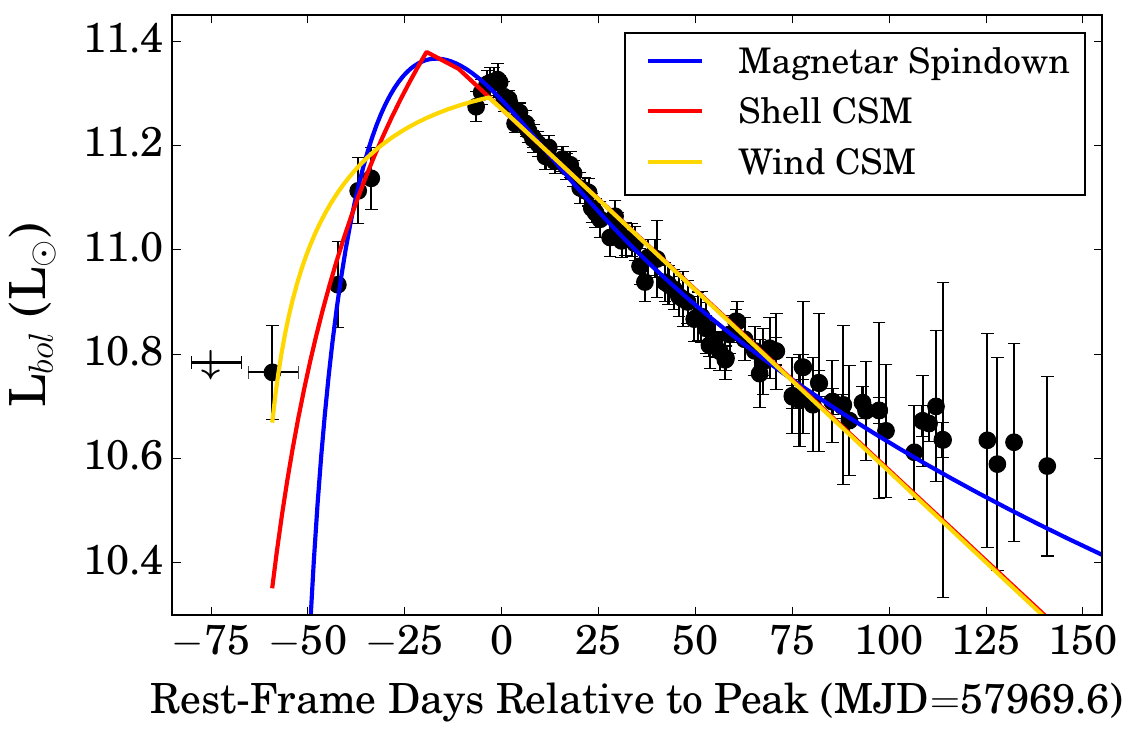}}
\subfloat[][]{
\includegraphics[width=0.325\textwidth]{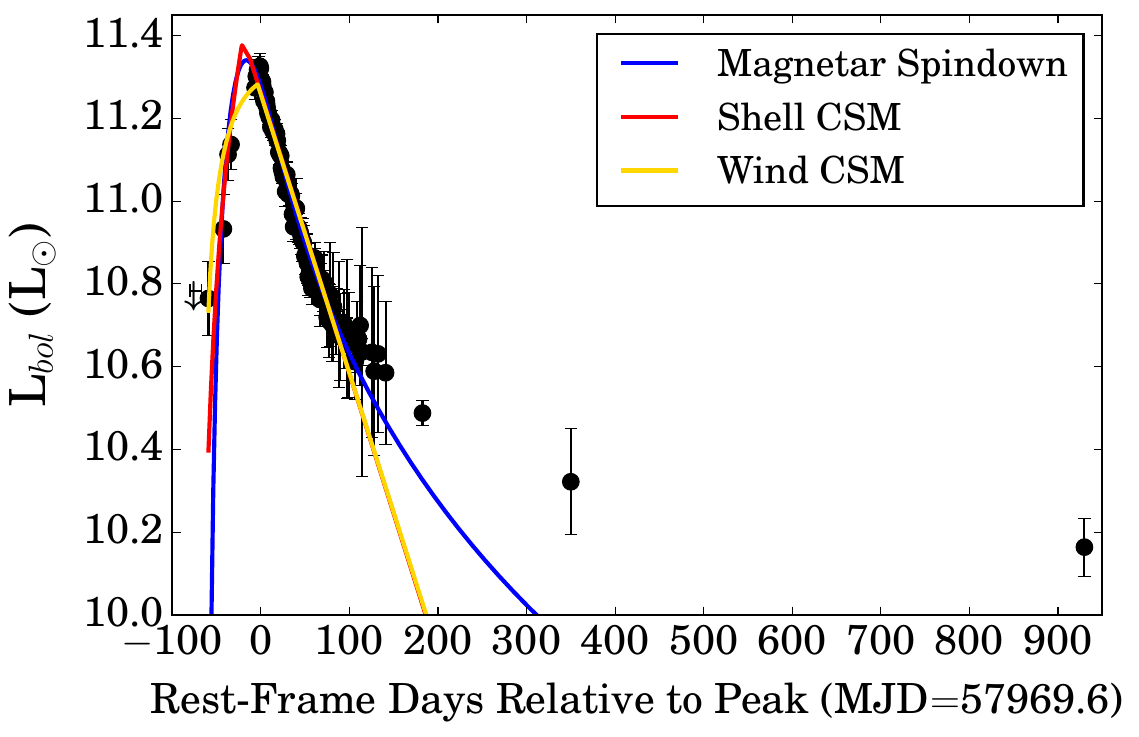}}
\subfloat[][]{
\includegraphics[width=0.325\textwidth]{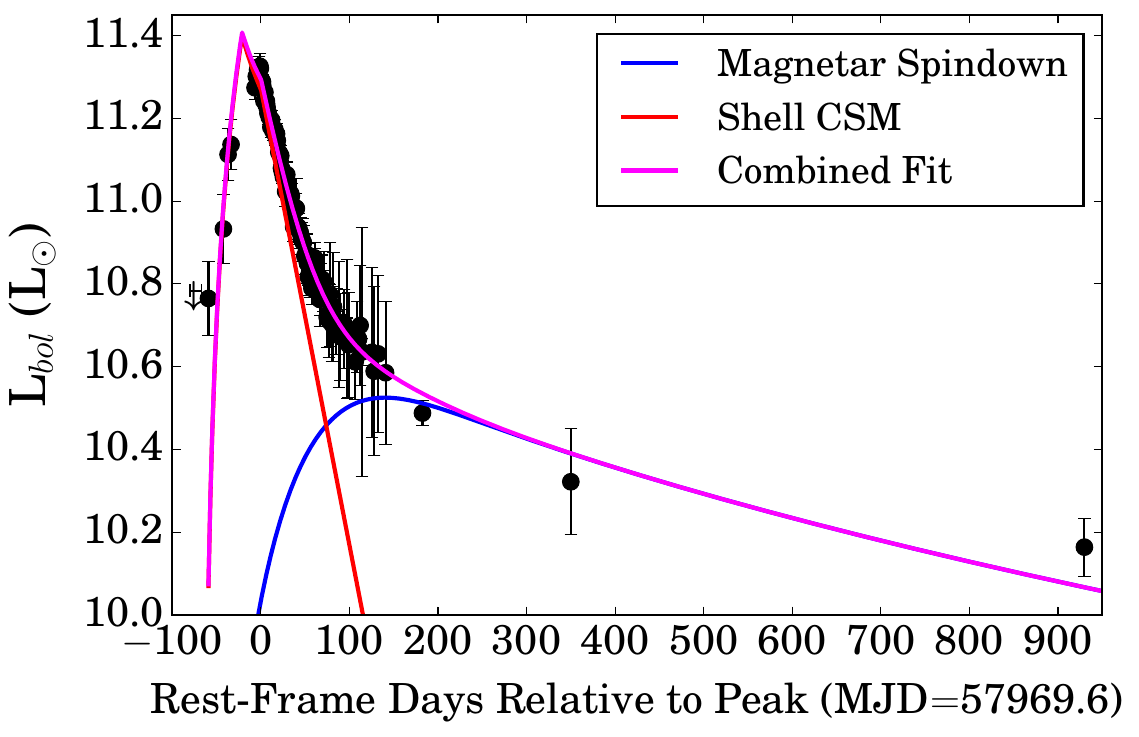}}
\\
\subfloat[][]{
\includegraphics[width=0.325\textwidth]{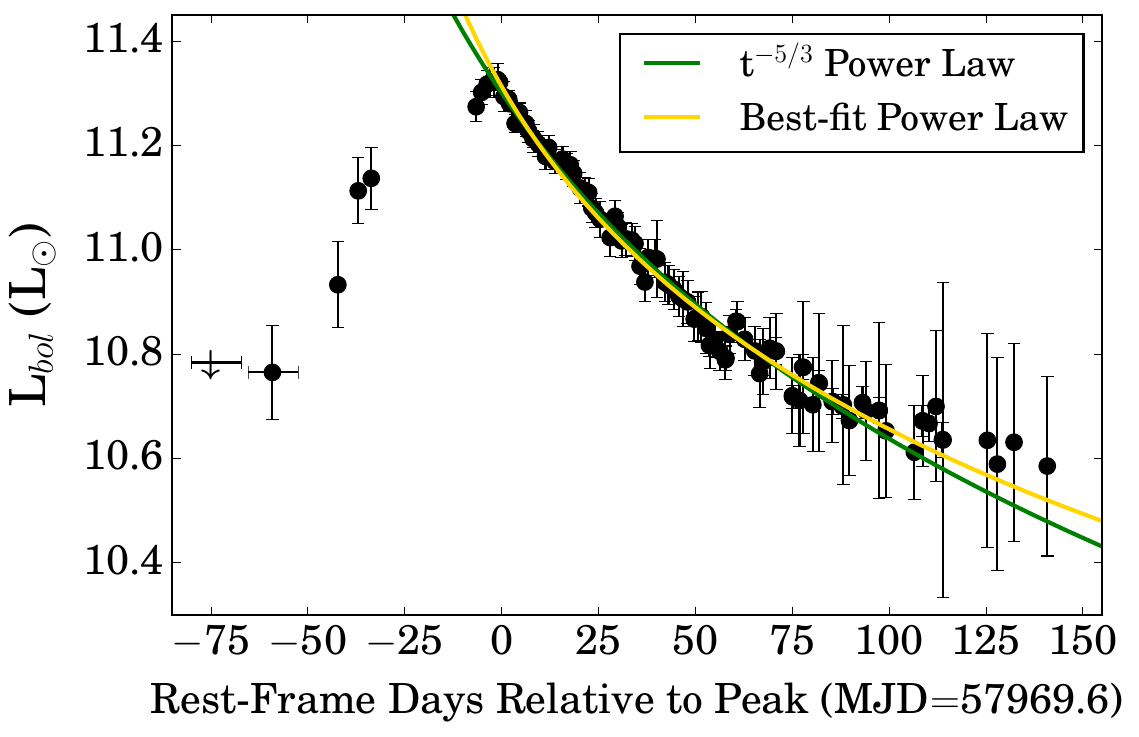}}
\subfloat[][]{
\includegraphics[width=0.325\textwidth]{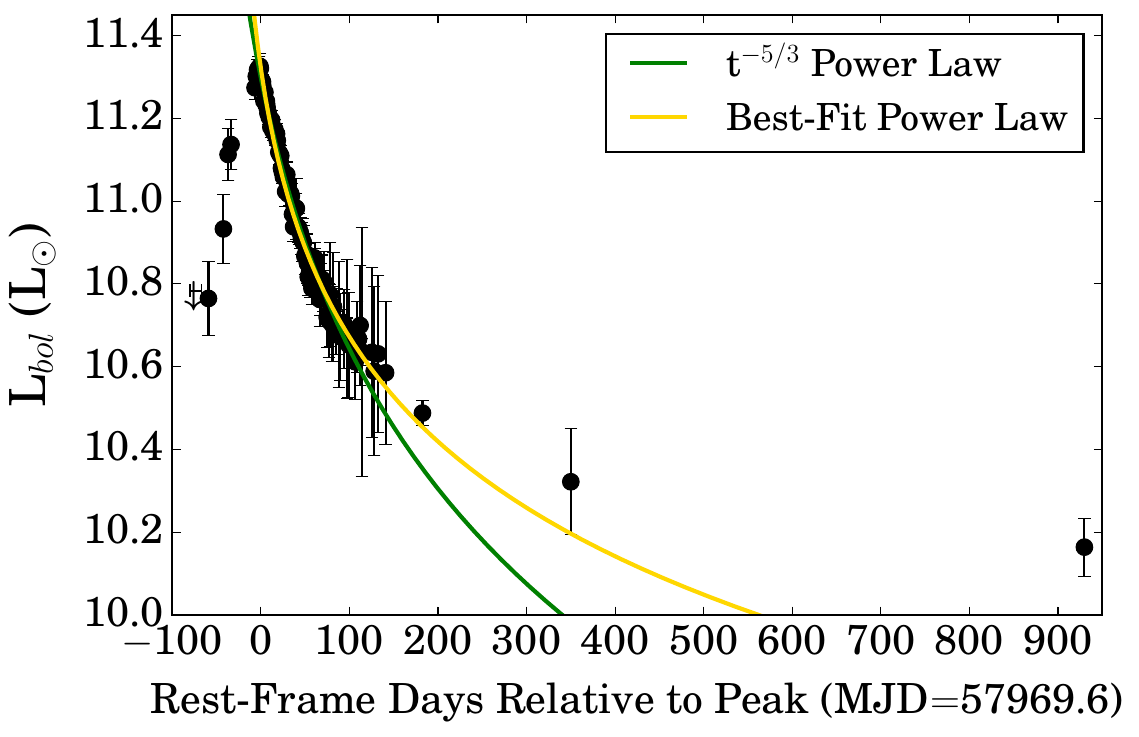}}
\caption{Bolometric luminosity \edit1{evolution} of ASASSN-17jz with various SN (top row) and TDE (bottom row) fits. \emph{Top row}: The top-left panel shows the bolometric evolution for the first 200 observer-frame days compared to fits for magnetar spindown (blue), interaction with a shell of CSM (red), and interaction with a wind of CSM (gold). The top-middle panel displays the full luminosity light curve compared to the same models used to fit the full dataset. The top-right panel gives a combined CSM-shell+magnetar spindown fit, with the two components shown in red and blue, respectively, and the combined fit shown in magenta. \emph{Bottom row}: The bottom-left panel shows $t^{-5/3}$ power-law (green) and best-fit ($t^{-\alpha}$; gold) power-law fits to the first 200 observer-frame days of the luminosity decline. The bottom-right panel shows the same models fitted to the full dataset. The parameters for the SN fits are shown in Table~\ref{tab:SNmodels} and the parameters for the TDE fits are shown in Table~\ref{tab:TDEmodels}.}
\label{fig:lum_fits}
\end{minipage}
\end{figure*}

In general, a single blackbody provides a good fit to the data at all epochs. In order to leverage our high-cadence light curves from ASAS-SN and other observatories to get an estimate of the bolometric rise, we calculated bolometric corrections to the host-subtracted and stacked $V$-band light curve by linearly interpolating between the previous and next $V$-band observations bracketing each \swift{} observation. We then calculated a $V$-band bolometric correction for each epoch, and interpolated between the previous and next bolometric correction for each $V$-band observation to obtain a bolometric luminosity. For $V$-band observations prior to the first epoch of \swift{} SED fits, we used the first \swift{} bolometric correction. 

Integrating the emission over the full light curve gives a total emitted energy of $E_{\rm tot} \approx (1.36\pm0.08)\times10^{52}$~erg over the $\sim1200$~days spanned by our observations. Of this, $(0.80\pm0.02)\times10^{52}$~erg is released during the rise and first 200~days after peak. Even if the late-time emission is not associated with the transient (a possibility discussed further below), ASASSN-17jz was an incredibly energetic transient.

We used several models for \edit1{SLSNe} and TDEs to fit the bolometric luminosity evolution of the transient. First, we modeled the luminosity of ASASSN-17jz \edit1{with several models commonly adopted to fit emission from SLSNe of all types, following the methods used by} \citet{vallely18} for the superluminous Type Ibn SN ASASSN-14ms. Here we focus primarily on two potential power sources, the interaction of SN ejecta with surrounding CSM, and the spindown of a magnetar produced during an SN explosion. These models are described in detail by \cite{chatzopoulos12}, \cite{inserra13}, and \cite{chatzopoulos13}. We also fit a model based solely on the radioactive decay of $^{56}$Ni \citep[see, e.g.,][]{arnett82}, but found that the only way to even crudely match the observed peak luminosity was to invoke a $^{56}$Ni mass larger than the SN ejecta mass. As this is an unphysical condition, we discarded $^{56}$Ni decay as a likely power source for the transient.

As with the \texttt{MOSFiT} light-curve fits, we fit both the full dataset and the early-time dataset, with the data cut off at 200~days post-peak in the observed frame (roughly 170~days in the rest frame of the transient). For the CSM interaction models, we also fit models with the CSM described both as a constant-density shell and as a steady-state wind. Throughout this work we generally adopt the same assumptions as \cite{vallely18}, except that we allow for the progenitor radius, $R_p$, to vary in our CSM-interaction models and we assume an ejecta opacity of $\kappa=0.34$~cm$^2$~g$^{-1}$, consistent with ionized H, \edit1{as the transient exhibits hydrogen emission features in its spectra}. For the magnetar spindown models, we limit the velocity of the ejecta to $v_{\rm ej}\leq10,000$~km~s$^{-1}$ to be consistent with the velocities measured from the spectroscopic emission lines (see Sec.~\ref{sec:spec_anal}).

\begin{deluxetable*}{cccccc}
\tabletypesize{\footnotesize}
\tablecaption{Physical fit parameters for SN bolometric light curve models.}
\label{tab:SNmodels} 
\tablehead{
\colhead{Parameter} &
\colhead{Shell CSM} &
\colhead{Wind CSM} &
\colhead{Magnetar Spin-Down} &
\colhead{Shell CSM+Magnetar} &
\colhead{Units} }
\startdata
$t_0$                &$-13.6\pm1.5$ & $-5.2\pm1.6$ &  $5.0\pm2.0$  & $-7.2\pm7.3$ & days \\
$M_{\rm ej}$             &  $38.7\pm7.5$  & $53.5\pm4.0$ &   $2.4\pm0.7$   & $40.2\pm8.1$ & \msun \\
$R_{p}$              &$365.1\pm187.1$ & $340.9\pm114.6$&       ---        & $384.1\pm228.7$ & R$_\odot$\\
$M_{\rm CSM}$            & $16.9\pm0.8$  & $17.8\pm0.5$  &       ---        & $10.1\pm1.7$ & \msun \\
$\rho_{{\rm CSM},1}$       &$(9.8\pm1.1)\times10^{-14}$&$(1.1\pm0.5)\times10^{-9}$&       ---        & $(6.9\pm4.7)\times10^{-14}$ &  g~cm$^{-3}$\\
$E_{\rm SN}$             & $13.3\pm1.9$  & $13.4\pm0.9$  &       ---        & $14.8\pm3.2$ & ($10^{51}$~erg) \\
$B_{14}$             &       ---      &      ---       & $0.4\pm0.0$ & $0.1\pm0.0$ & $10^{14}$~G\\
$P_{ms}$             &       ---      &      ---       & $1.2\pm0.0$  & $1.1\pm0.1$ & ms \\
$v_{\rm ej}$             &       ---      &      ---       &  $7766\pm1897$   & $8224\pm1185$ & km s$^{-1}$\\
\enddata 
\tablecomments{Parameters for the various SN bolometric light curve models. The Shell CSM+Magnetar model is fit to the full set of bolometric luminosities, while the other models are restricted to data within 200~days of peak luminosity.}
\vspace{-0.8cm}
\end{deluxetable*}

The top row of Figure~\ref{fig:lum_fits} shows the bolometric light curve of ASASSN-17jz compared to the various SN model fits. In the top-left panel we show the fits to the early-time data only, and in the top-middle panel we display the fits to the full dataset. The two CSM models provide very similar fits, particularly post-peak, with the only differences being in the shape of the rising light curves and the peak luminosities. Ultimately, models that include just magnetar spindown or just CSM interaction are both capable of fitting the first $\sim200$~days, but neither is capable of matching the observed luminosity at late times. All three models significantly underestimate the late-time luminosity.

The parameters for the best-fit SN models are given in Table~\ref{tab:SNmodels}. As none of the fits to the full dataset are able to replicate the observed light curve, we show only the parameters for the fits to the early-time data. As can be seen in the Table, the physical parameters necessary to match the early-time data in the pure CSM interaction model are rather extreme, including an ejecta mass of $M_{\rm ej}\gtrsim40$~\msun{} and an explosion energy of $E_{\rm SN} \approx 13\times10^{51}$~erg. (We note, however, that these ejecta masses are much more reasonable than those from \texttt{MOSFiT}.) Although the 1.2~ms initial rotation period required for the magnetar spindown model is quite rapid, it \edit1{is not below} the expected $\lesssim1$~ms breakup limit of a neutron star \citep{haensel95}, and the remaining physical parameters of the magnetar model have plausible values.

It is clear that a second power source component is necessary to fit the late-time data. Motivated by this, we also fit the data with a model dominated by interaction with a CSM shell at early times and magnetar spindown at late times. This fit is shown in the top-right panel of Figure~\ref{fig:lum_fits} and the parameters of the model are given in the right column of Table~\ref{tab:SNmodels}. We find that this combined model is able to replicate the full observed light curve fairly well, as shown in the top-right panel of Figure~\ref{fig:lum_fits}, though the fit does not match the rise as well as it matches the decline. The required ejecta mass for the CSM interaction part of this model is even larger, at $M_{\rm ej}=86.5$~\msun, though the mass of the CSM is lower. Ultimately, it is possible to obtain a reasonable fit to the full bolometric luminosity evolution of ASASSN-17jz using only SN emission models. However, a scenario where the early-time light curve is powered by an SN and the late-time light curve is powered by a second, non-SN power source, such as activity from the underlying AGN, could also provide a reasonable fit to the data. 

We also fit models for TDE emission to the declining light curve of ASASSN-17jz. The luminosity evolution of a TDE after peak is canonically expected to follow a $t^{-5/3}$ profile, assuming the emission from the TDE is in a fallback-dominated regime \citep[e.g.,][]{rees88,phinney89,evans89}. However, as TDE discoveries have increased, a variety of decline rates have been seen, and several models, including exponential models and power-law models with the power-law index being a free parameter. As the decline after peak is clearly not consistent with an exponential model after $\sim100$~days, we fit the declining part of the bolometric light curve with both a $L=L_0 (t-t_0)^{-5/3}$ power-law profile and a power law where the power-law index is allowed to vary, $L=L_0 (t-t_0)^{-\alpha}$. As with the SN models, we fit the models to both the early-time ($t<200$~days after peak) and the full post-peak dataset. The parameters of the TDE fits are given in Table~\ref{tab:TDEmodels}.

\begin{deluxetable}{cccccc}
\tabletypesize{\footnotesize}
\tablecaption{Physical fit parameters for TDE bolometric light curve models.}
\tablehead{
\colhead{} &
\colhead{$t^{-5/3}$,} &
\colhead{$t^{-5/3}$,} &
\colhead{$t^{-\alpha}$,} &
\colhead{$t^{-\alpha}$,} &
\colhead{} \vspace{-6pt} \\ 
\colhead{Parameter} &
\colhead{Early} &
\colhead{Full} &
\colhead{Early} &
\colhead{Full} &
\colhead{Units}}
\startdata
$t_0$  & 57891.9 & 57890.4 & 57921.1 & 57934.6 & JD\\
$\log{L_0}$   & 48.0 & 48.0 & 47.0 & 46.5 & erg s$^{-1}$\\
$\alpha$  & --- & --- & 1.23 & 1.02 & --- \\
\enddata 
\tablecomments{Parameters for the TDE power-law models fit to the declining bolometric light curve of ASASSN-17jz. The first and third columns give results for fits to the data within 200~days of peak luminosity, while the second and fourth columns provide fits to the full dataset.}
\label{tab:TDEmodels} 
\vspace{-0.8cm}
\end{deluxetable}

The results of the fits are shown in the bottom row of Figure~\ref{fig:lum_fits}. For the fit to the early-time data, the best-fit power-law index is $\alpha=1.23$, somewhat close to $5/3$, and the two fits match the data quite well. For the full dataset, there is a large difference between the two power-law models, with the model preferring a shallower decline in order to fit the late-time data better. Neither model is able to fit the emission at very late times without additional emission components.

Recent theoretical work \citep[e.g.,][]{velzen19b} has shown that the emission from TDEs transitions from a fallback-dominated regime to a disk-dominated regime at late times. However, this was only true for TDEs in host galaxies with BH masses of $M_{\rm BH}\lesssim10^{6.5}$~\msun. \edit1{\citet{velzen19b}} found that the light curves are generally consistent with an extrapolation of the early-time decline for more massive BHs. As \host{} has a BH mass of $M_{\rm BH} \approx 10^{7.5}$~\msun, the late-time emission from ASASSN-17jz is not likely to result from solely a TDE. 

Regardless of whether ASASSN-17jz involves an SN or a TDE, based on the bolometric luminosity it is highly likely that a secondary power source is required to power the full light curve. We compare ASASSN-17jz to several other luminous transients to further investigate its nature in the following section.

\subsection{Blackbody Comparison to Other Transients}
\label{sec:bbcomp}

\begin{figure}
\centering
\includegraphics[width=0.48\textwidth]{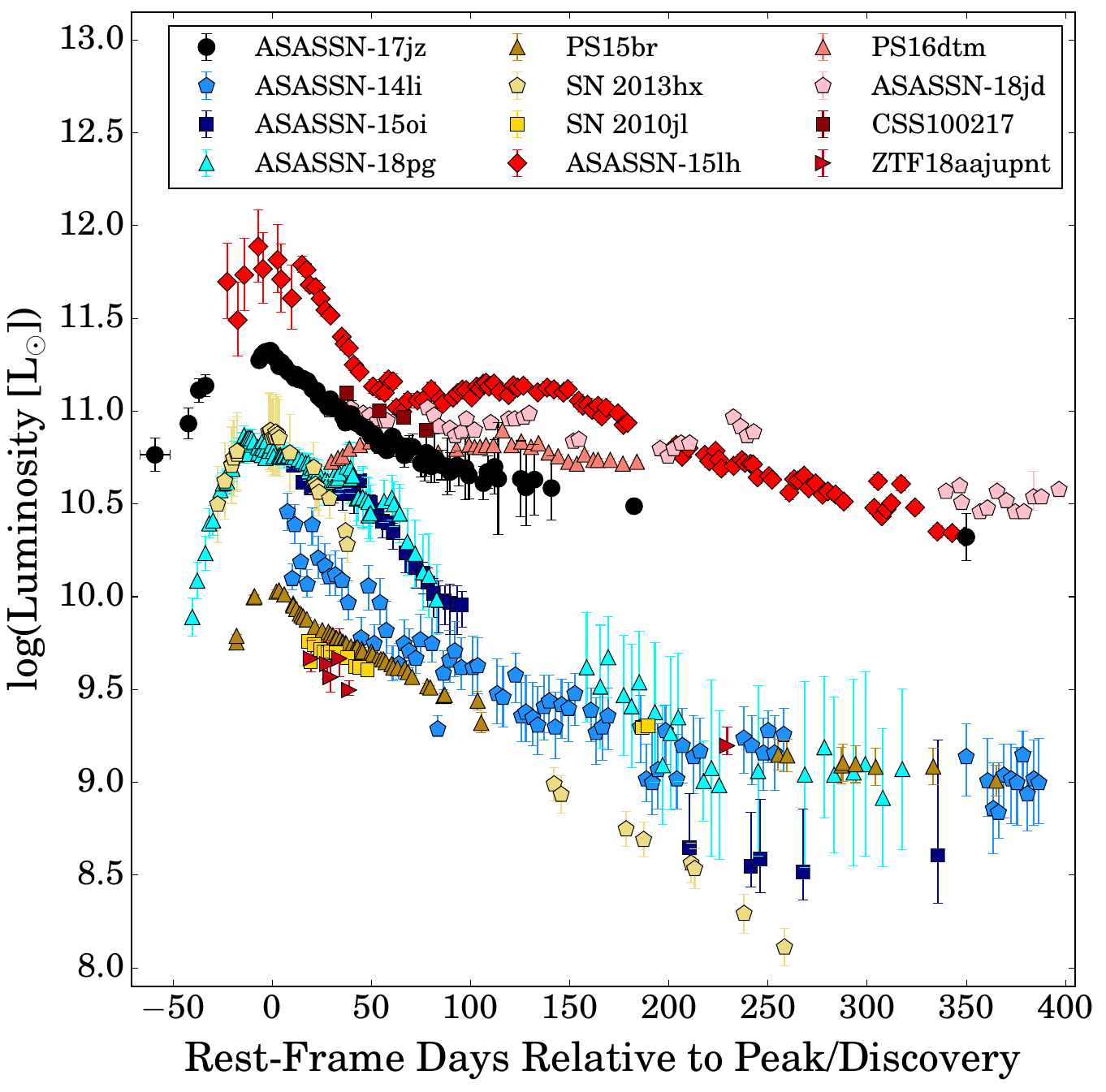}
\caption{Comparison of the luminosity evolution of ASASSN-17jz to several other nuclear and/or luminous transients. TDEs are shown in shades of blue, SLSNe in shades of gold, and ANTs in shades of red. The light curve shown for ASASSN-17jz includes both the luminosity blackbody fits calculated from the \edit1{host-subtracted} \swift{} photometry and the bolometrically corrected $V$-band data, as described in Section~\ref{sec:sed_anal}. Phase is shown in rest-frame days relative to the date of discovery for ASASSN-14li and ASASSN-15oi, and relative to the time of peak light for all other objects. We limit the figure to the first 400~days after peak/discovery.}
\label{fig:lum_comp}
\end{figure}

\begin{figure}
\centering
\includegraphics[width=0.48\textwidth]{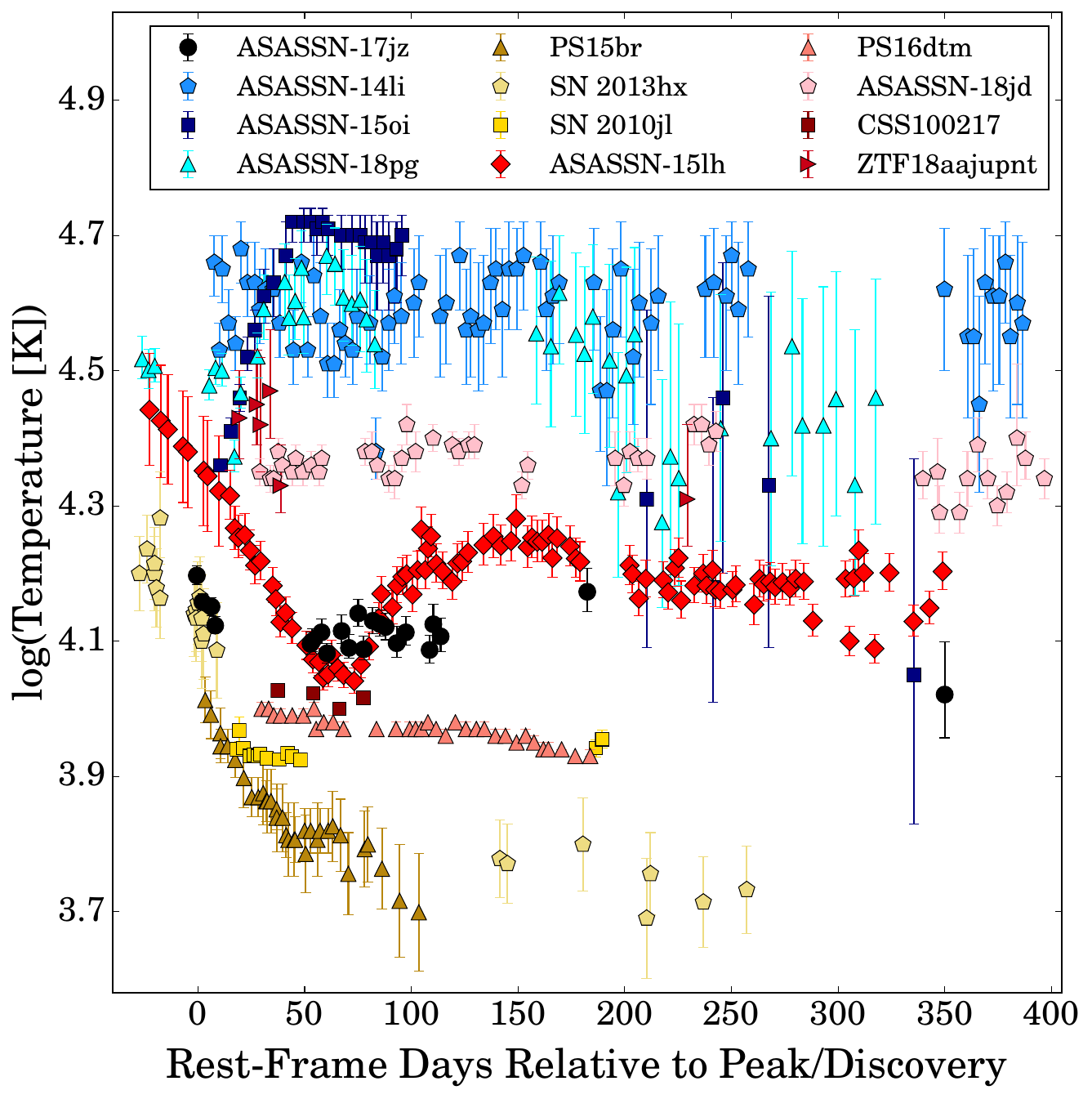}
\caption{Comparison of the blackbody temperature evolution of ASASSN-17jz to other nuclear and/or luminous transients. Colors and symbols match those of Figure~\ref{fig:lum_comp}. Phase is shown in rest-frame days relative to peak/discovery, as described in Figure~\ref{fig:lum_comp}.}
\label{fig:temp_comp}
\end{figure}

\begin{figure}
\centering
\includegraphics[width=0.48\textwidth]{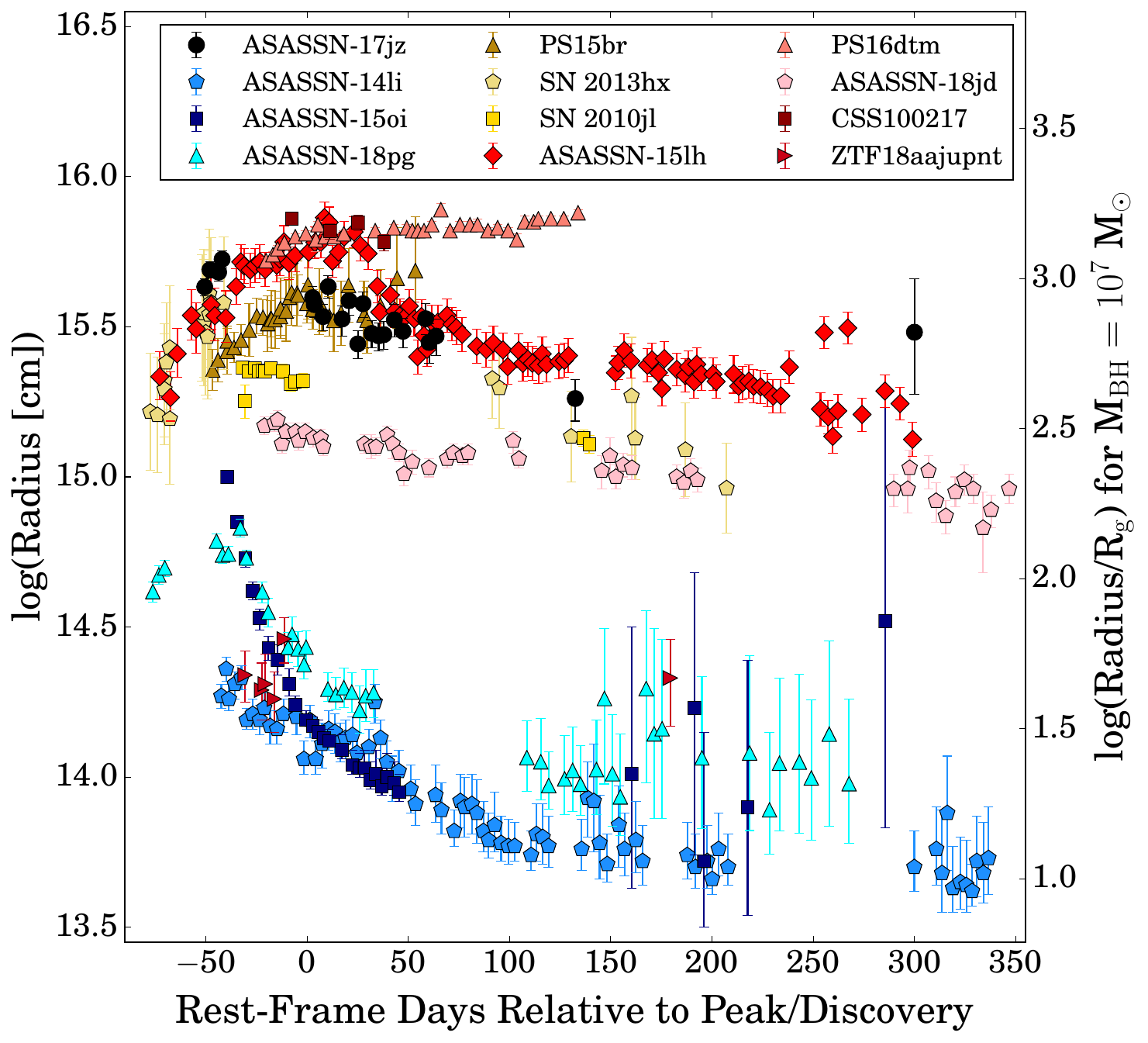}
\caption{Comparison of the blackbody radius evolution of ASASSN-17jz to that of several other nuclear and/or luminous transients. Colors and symbols match those of Figure~\ref{fig:lum_comp}. Phase is shown in rest-frame days relative to peak/discovery, as described in Figure~\ref{fig:lum_comp}. The left-hand radius scale shows the radius in units of cm, while the right-hand scale shows the corresponding radii in units of $R_g$ for a $10^7$~\msun{} black hole.}
\label{fig:rad_comp}
\end{figure}

We show the blackbody luminosity, temperature, and radius evolution of ASASSN-17jz compared to several other luminous and/or nuclear transients in Figures~\ref{fig:lum_comp}, \ref{fig:temp_comp}, and \ref{fig:rad_comp}. Our comparison sample includes the supernovae SN~2010jl \citep[an SN~IIn;][]{stoll11}, SN~2013hx \citep[an SLSN-II;][]{inserra18}, and PS15br \citep[a normal SLSN-II;][]{inserra18}; the TDEs ASASSN-14li \citep{holoien16a,brown17a}, ASASSN-15oi \citep{holoien16b,holoien18a}, and ASASSN-18pg \citep{holoien20a}; and the ANTs CSS100217:102913+404220 \citep[CSS100217, claimed to be an SN around an existing AGN;][]{drake11}, ASASSN-15lh \citep[an SLSN-I that has also been claimed to be a TDE;][]{dong16,godoy-rivera17,leloudas16}, PS16dtm \citep[claimed to be a TDE around an existing AGN;][]{blanchard17}, ZTF18aajupnt \citep[a new type of changing-look LINER;][]{frederick19}, and ASASSN-18jd \citep[either a TDE or an AGN flare;][]{neustadt20}. \edit1{For cases where a group of similar events have been identified (e.g., CSS100217 being part of a class of transients similar to the transient PS1-10adi, as noted by \citet{kankare17}), we have chosen the event with observations most similar to those of \jz{} as a representative for comparison.} For CSS100217 and SN~2010jl we downloaded archival \swift{} UVOT data, computed host-subtracted photometry using late-time \swift{} images to calculate the host flux, and fit a blackbody SED in the same way that we did for ASASSN-17jz. For ASASSN-14li, ASASSN-15oi, ASASSN-18jd, PS16dtm, and ZTF18aajupnt, we used the blackbody fits from \citet{hinkle21b}, who recently recomputed \swift{} UVOT photometry for a large number of nuclear outbursts and calculated blackbody fits in a similar way to how we fit the SED of ASASSN-17jz. For the remaining objects we took the blackbody fits from the cited papers. \edit1{Because} no peak date was reported for ZTF18aajupnt, we downloaded the public $g$-band light curve from ZTF \citep{masci19} and computed the peak date as we did for ASASSN-17jz, finding MJD$_{\rm peak}$ = 58321.9.


Figure~\ref{fig:lum_comp} shows the luminosity evolution of ASASSN-17jz compared to our comparison sample. SNe are indicated in shades of gold, TDEs in shades of blue, and ANTs in shades of red. All light curves are in rest-frame days relative to peak light, except for ASASSN-14li and ASASSN-15oi, which are in rest-frame days relative to discovery, as the peak was not observed for those two transients.

ASASSN-17jz peaks at a luminosity higher than every object in  our comparison sample except for ASASSN-15lh, and in general is similar in luminosity to the ANTs. The ANTs tend to be more luminous than the TDEs and SNe, which is perhaps unsurprising given that many of the ANTs are interpreted as AGN activity in addition to transient emission. Although the ANTs are more luminous objects, there are no significant differences in decline rate between the three different classes of objects, showing that light-curve shape alone cannot distinguish between these transients.


Comparing the luminosity evolution to the other ANTs specifically, we find that the light-curve shapes of ASASSN-15lh, PS16dtm, and ASASSN-18jd are noticeably different from that of ASASSN-17jz, with the first showing a steeper initial decline followed by a second peak and the second two exhibiting a slower decline with some short-term variation not seen in ASASSN-17jz. ZTF18aajupnt has a similar light-curve shape, but is nearly two orders of magnitude dimmer than ASASSN-17jz. CSS100217 is the only ANT that is a good match in both luminosity and decline rate, though it unfortunately was observed with \swift{} for only 40~days and 4 epochs near peak. There were several epochs of \swift{} UVOT data taken of CSS100217 several hundred days after peak brightness, but these epochs showed little-to-no evolution and were comparable in flux to the pretransient {\it GALEX} photometry of the host, so we used them as an estimate of the host flux. Thus, while CSS100217 is a good match to ASASSN-17jz near peak, it likely faded more rapidly, and does not exhibit the long-term UV emission of ASASSN-17jz.

Figure~\ref{fig:temp_comp} shows the temperature evolution of ASASSN-17jz and our comparison sample. The two SLSNe-II in our comparison sample stand out quite clearly from the rest of the objects in that they exhibit declining temperatures, while SN~2010jl, the TDEs, and the ANTs all exhibit relatively flat temperature evolution, after some early evolution in some cases. There is a clear delineation between the three types of objects, with the SNe being the coolest with temperatures in the range $5000\lesssim T \lesssim20,000$~K, TDEs being the hottest with temperatures in the range $20,000\lesssim T \lesssim50,000$~K, and ANTs falling in the middle, with temperatures in the range $9000 \lesssim T \lesssim 30,000$~K.

ASASSN-17jz exhibits a roughly constant temperature of $T \approx 12,500$~K throughout the course of the event, which places it roughly in the middle of the ANTs. CSS100217 and ASASSN-15lh are the objects with temperatures most similar to that of ASASSN-17jz, though ASASSN-15lh also shows an early decline followed by a rise between 50 and 100~days after peak brightness before leveling off. Interestingly, though ASASSN-18jd and PS16dtm have similar luminosities, they exhibit quite different temperatures, with ASASSN-18jd being more similar to the TDE sample and PS16dtm being more similar to the SN sample.


Finally, in Figure~\ref{fig:rad_comp} we show the radius evolution of ASASSN-17jz and our comparison sample. As with temperature, there is a clear difference between SNe and TDEs in radius evolution, with TDEs generally exhibiting steadily declining radii after peak, and SNe exhibiting increasing or roughly constant radii. The ANTs in our comparison sample generally exhibit radius evolution similar to the SNe, both in size and in decline rate. The exception is ZTF18aajupnt, which has a blackbody radius more similar to those of TDEs.

\edit1{From} our sample of ANTs, ASASSN-17jz is most comparable in radius evolution to ASASSN-15lh, with the two objects looking quite similar in both size and decline rate. CSS100217 and ASASSN-18jd both decline similarly to ASASSN-17jz, though they are larger and smaller in size, respectively. Unlike the other ANTs, PS16dtm shows a steadily \emph{increasing} radius, and ZTF18aajupnt is an order of magnitude smaller in radius compared to the other ANTs.

Looking at all aspects of the blackbody evolution, most of the ANTs in our comparison sample differ from ASASSN-17jz in at least one aspect. ZTF18aajupnt is quite different in all three properties; ASASSN-18jd and PS16dtm differ in radius and temperature evolution; and ASASSN-15lh has an early decline in its temperature and luminosity that is not seen in ASASSN-17jz, though the two objects look similar at later times. The only object that resembles ASASSN-17jz in all three of its luminosity, temperature, and radius is CSS100217. In the discovery paper for this object, \citet{drake11} claimed it was likely an ``extremely luminous'' SN~IIn occurring in the vicinity of an AGN. This explanation is also plausible for ASASSN-17jz, and we discuss this and other plausible origin scenarios in Section~\ref{sec:disc}.
\\

\section{X-ray Analysis}
\label{sec:xray_anal}

\begin{figure}
\centering
\includegraphics[width=0.425\textwidth]{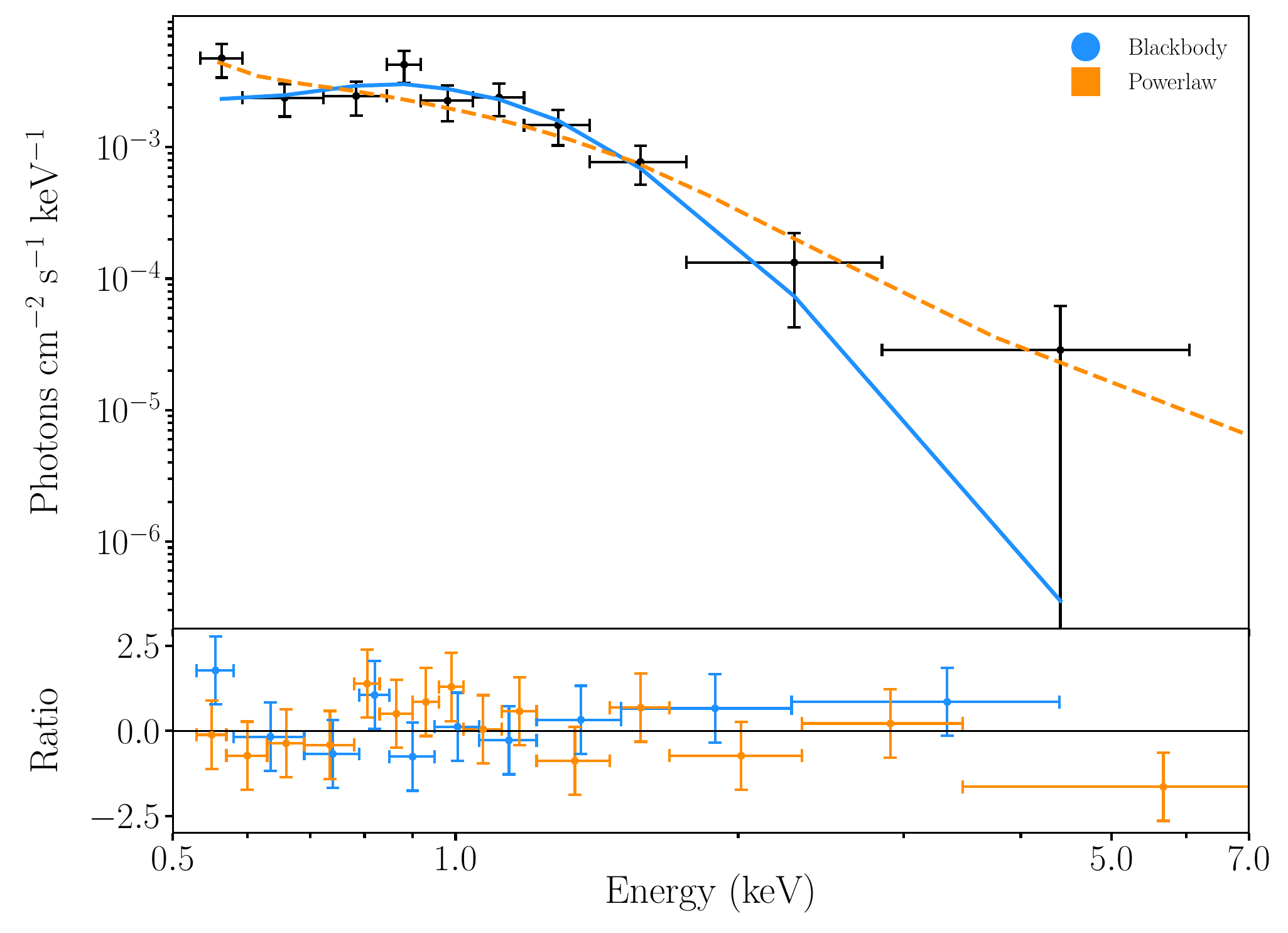}
\caption{X-ray spectrum from merged \swift{} XRT observations taken between 1081 and 1083 days after peak light (black) compared to absorbed blackbody (blue solid line) and absorbed power-law (orange dashed line) models. The bottom panel shows the residuals for the two models.}
\label{fig:xray_spec}
\end{figure}

For the first $\sim150$~days of \swift{} XRT observations, we do not detect any X-ray emission from ASASSN-17jz. In the epoch near 200~days after peak brightness there is a detection in X-rays, but  400~days after peak we again do not detect any X-ray emission. However, in our last two epochs of observation, roughly 1080~days after peak, X-rays are clearly detected. To better understand the nature of this late-time X-ray emission, we  merged the late-time observations and extracted a low S/N spectrum. We then fit this spectrum with both an absorbed blackbody component and an absorbed power-law component, with both models redshifted to the host and fit with $N_H$ frozen to the Galactic column density along the line of sight, $N_H=1.67\times10^{20}$~cm$^{-2}$ \citep{hi4pi16}. We show the extracted spectrum and the blackbody and power-law components in Figure~\ref{fig:xray_spec}.


The best-fit blackbody temperature and radius are $kT_{\rm BB}=0.8\pm0.03$~keV and $R_{\rm BB}=2.1^{+0.9}_{-0.7}\times10^{10}$~cm. This temperature is consistent with but on the high end for temperatures found in the 70~month \swift{} AGN catalog \citep{ricci17}. For the power-law model, we find a best-fit power-law index of $\Gamma=3.4\pm0.6$, which is significantly softer than what is commonly seen in \edit1{most} AGNs \citep[$\Gamma \approx 1.7$;][]{auchettl17}, \edit1{but is in the range commonly seen in narrow-line Seyfert 1 (NLSy1) AGNs \citep{boller96}}. Neither model clearly fits the data better, as the blackbody and power-law models have reduced chi-squared values of $\chi_{\nu,{\rm BB}}^{2}=0.7$ and $\chi_{\nu,{\rm PL}}^{2}=0.8$, respectively.

\begin{figure}
\centering
\includegraphics[width=0.425\textwidth]{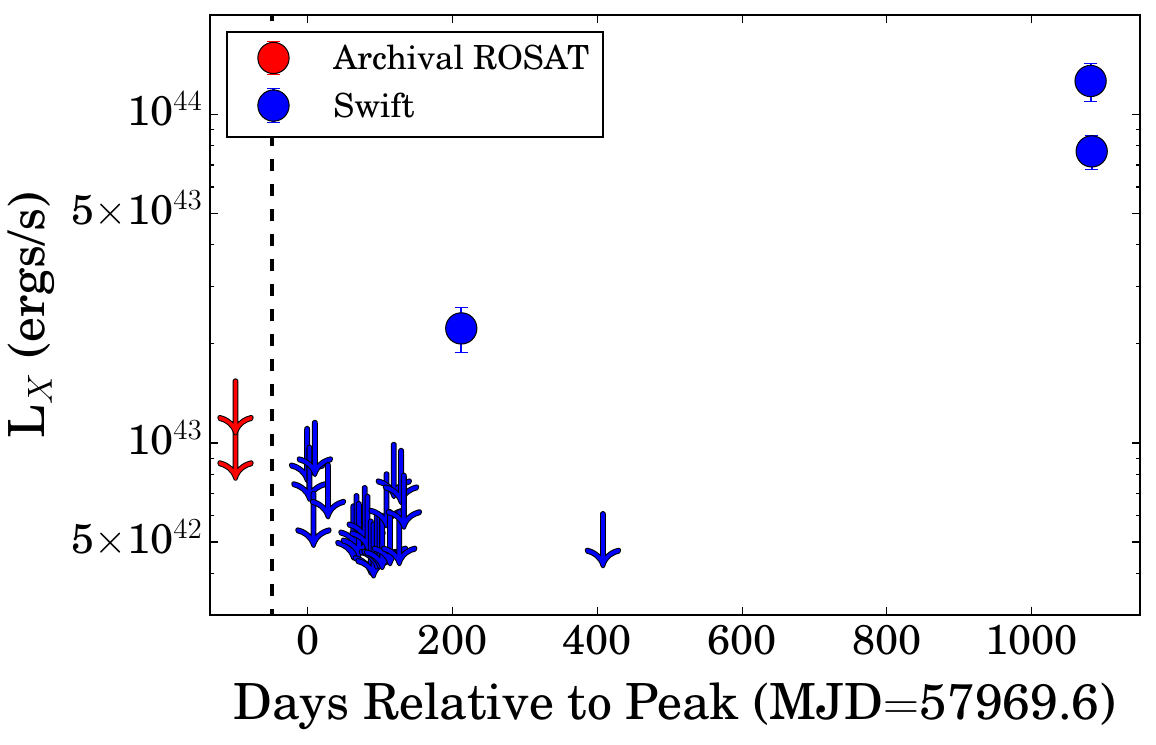}
\caption{X-ray light curve calculated using the best-fit power-law model shown in Figure~\ref{fig:xray_spec} to estimate fluxes in each epoch. $3\sigma$ upper limits are shown as downward-facing arrows. The red arrows to the left of the vertical dashed line show pre-outburst limits from {\it ROSAT} data, which were taken in the early 2000s but are shown at $t=-100$~days to improve readability.}
\label{fig:xray_lc}
\end{figure}

\begin{figure}
\centering
\includegraphics[width=0.425\textwidth]{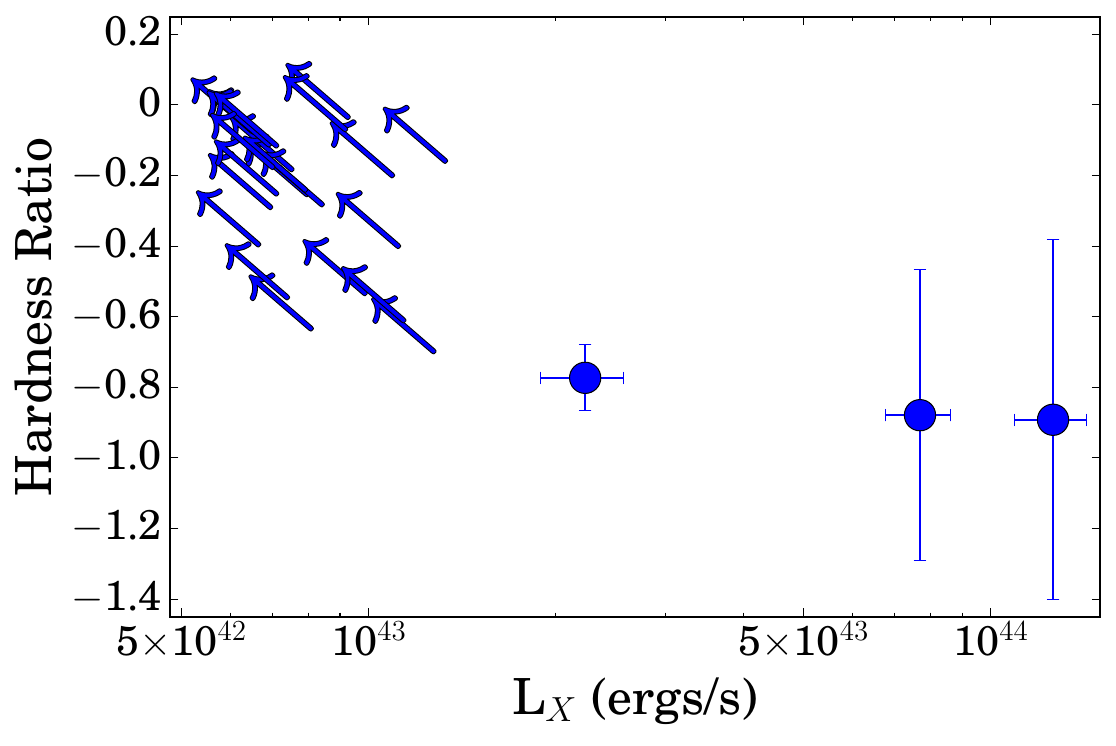}
\caption{X-ray hardness ratio compared to the X-ray luminosity. $3\sigma$ upper limits are shown as arrows.}
\label{fig:xray_hr}
\end{figure}

We used the best-fit power law to derive fluxes, flux upper limits, and their corresponding luminosities. We used the same model to derive upper limits for the pre-outburst data from {\it ROSAT}. The full X-ray light curve is shown in Figure~\ref{fig:xray_lc}. We also derived the hardness ratios for each \swift{} epoch, and compare the hardness ratio to the luminosity in Figure~\ref{fig:xray_hr}. 

The late-time X-ray brightening can be clearly seen in Figure~\ref{fig:xray_lc}, with a luminosity at least an order of magnitude above the pre-outburst limits from {\it ROSAT}. Some theoretical models of TDE emission predict that TDEs would exhibit late-time X-ray brightening, if the formation of the accretion disk is delayed \citep[e.g.,][]{gezari17} or if intervening material absorbing X-ray emission from the disk at early times becomes optically thin \citep[e.g.,][]{metzger15}. The latter seems unlikely since a remarkable peculiarity of all TDEs is that none have shown variable X-ray absorption \citep{auchettl17}. The late-time brightening of ASASSN-17jz is likely not consistent with either of these pictures. First, the timescale of the late-time brightening is significantly longer than predicted by these TDE models, which suggest that the late-time X-ray brightening occurs roughly a year after peak \citep[e.g.,][]{metzger15}. Observed cases of late-time X-ray brightening in TDEs have been consistent with this picture, with the X-rays typically brightening $\sim200$--300 days after peak and fading back to prepeak levels roughly a year after that \citep[e.g.,][]{holoien18a,hinkle21a}. Second, as Figure~\ref{fig:xray_hr} shows, the X-ray emission from ASASSN-17jz becomes softer as the luminosity increases, but X-ray emission from TDEs has typically maintained a constant hardness ratio as the luminosity changes \citep{auchettl17}, \edit1{with only a few exceptions occurring in cases with higher Eddington ratios \citep[e.g.,][]{wevers20}. In combination, the timing of the late-time X-ray brightening and the softening of the spectrum disfavor a TDE origin.}


TDEs are not the only transients that can produce late-time X-ray emission. SNe, especially those in dense circumstellar environments, can too. In this scenario, the shock interaction between the SN ejecta and the dense CSM around the SN produces bright X-ray emission \citep{chevalier06,murase11}. At early times, the optical/UV emission dominates, but the X-ray emission may peak at late times, on the order of 10--60 times the shock-breakout timescale \citep{svirski12}. While X-rays have been detected from SNe at early times (e.g., SN~2008D, \citealt{chevalier08}; SN~2006jc, \citealt{immler08}; SN~2009ip,  \citealt{margutti14}; SN~2020bvc, \citealt{izzo20}), there is now a growing number of SNe that show X-ray emission at later times (i.e., $\sim 100$~days or more after peak optical brightness). This includes SN~2005ip \citep{smith17}, SN~2014C \citep{margutti17, brethauer20}, SN~2010jl \citep{chandra15}, and SN~2016coi \citep{terreran19}. However, the X-ray emission seen from these SNe mostly occurs closer to optical maximum and peaks at luminosities of $<10^{42}$~erg~s$^{-1}$  before decaying \citep[e.g., SN~2010jl; ][]{chandra15}. This is an order of magnitude lower in luminosity than ASASSN-17jz, and the X-rays from \jz{} peak over a thousand days after the optical peak. It seems unlikely that what we are observing arises from late-time X-ray emission from an SN. 


\begin{figure*}
\centering
\includegraphics[width=0.95\textwidth]{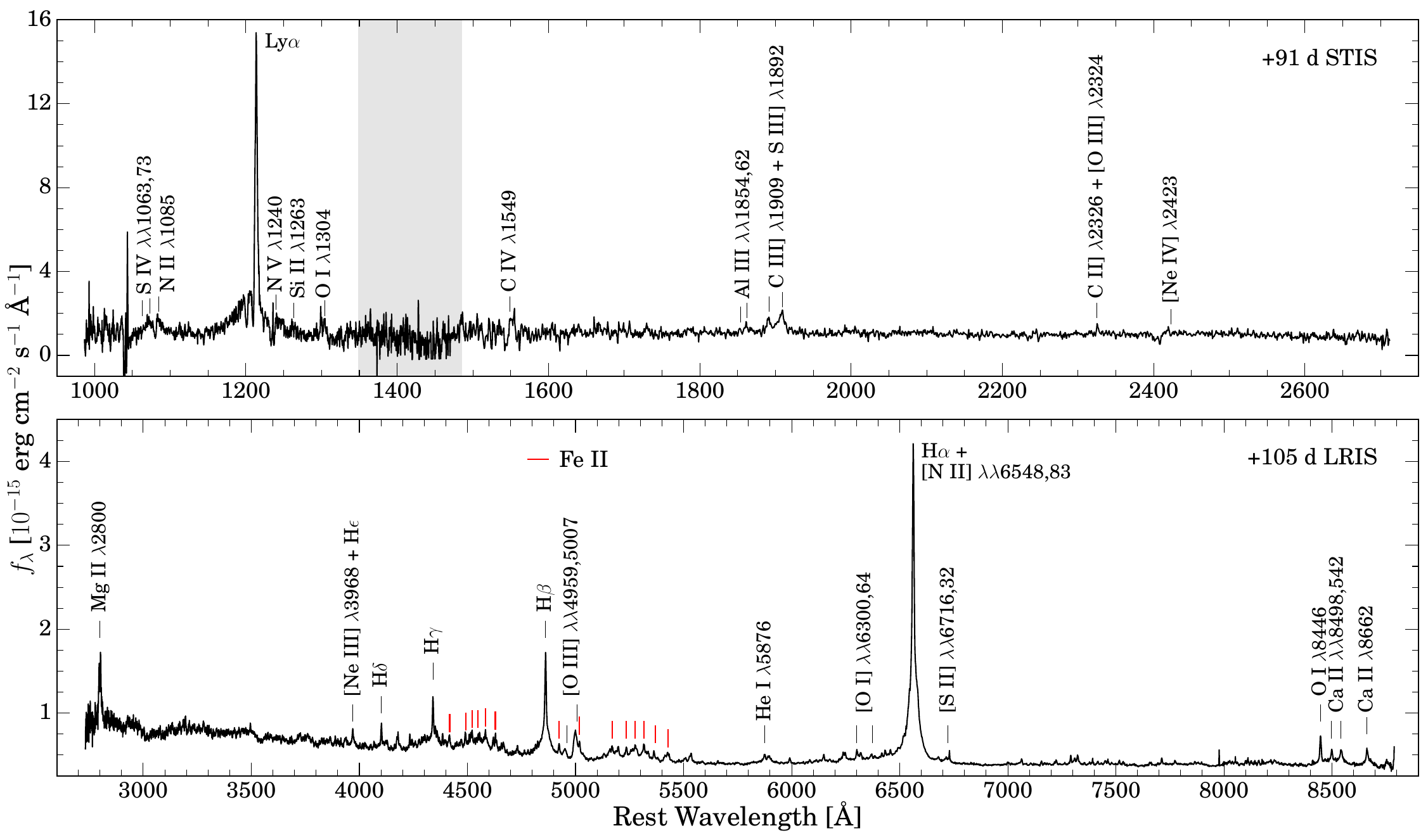}
\caption{+91~d STIS UV spectrum (top) and +105~d LRIS optical spectrum (bottom). Prominent spectral features are identified. In the top panel, the gray-shaded region is meant to highlight the overlap region of the far-UV and near-UV channels on STIS, where the noise is much higher. In the bottom panel, the red ticks indicate \ion{Fe}{2} multiplet lines.}
\label{fig:lines}
\end{figure*}

Finally, we examine the possibility that the X-ray emission from ASASSN-17jz could be due to a change in AGN activity from the host galaxy. \edit1{\citet{auchettl17} found that AGNs often exhibit softer X-ray emission as their X-ray luminosities increase, unlike the majority of TDEs.} Thus, \edit1{the X-ray behavior exhibited by ASASSN-17jz is consistent with an NLSy1 AGN that has become more luminous}. If we consider the blackbody model instead, the blackbody temperature and radius are also consistent with the properties seen in other AGNs with X-ray emission detected by \swift{} \citep{ricci17}. 

We thus conclude that the late-time X-ray brightening seen in ASASSN-17jz is highly likely to be associated with increased AGN activity in the host galaxy, rather than with late-time transient emission. We do not have the data to definitively say whether this increased activity was caused by a transient occurring near the AGN or it coincidentally brightened $\sim 3$~yr following the peak of the transient emission. However, it is notable that no X-ray emission is detected from the host in pre-outburst data spanning several years, and that the change in X-ray behavior is only seen after ASASSN-17jz occurred.


\section{Spectroscopic Analysis}
\label{sec:spec_anal}

\subsection{Emission-Line Evolution}

Figure~\ref{fig:lines} shows examples of the UV and optical spectra and identifies most of the prominent spectral features present throughout the evolution of the transient. We observe many lines which are prominent in the spectra of AGNs, namely \ion{S}{4}~\llambda1063, 1073, \ion{N}{2}~$\lambda$1085, \ion{N}{5}~$\lambda$1240, \ion{Si}{2}~$\lambda$1263, \ion{O}{1}~$\lambda$1304, \ion{C}{4}~$\lambda$1549, \ion{Al}{3}~\llambda1854, 1862, \ion{S}{3}]~$\lambda$1892, \ion{C}{3}]~$\lambda$1909, [\ion{O}{3}]~$\lambda$2324, \ion{C}{2}]~$\lambda$2326, \ion{Mg}{2}~$\lambda$2800, [\ion{Ne}{3}]~$\lambda$3968, [\ion{O}{3}]~\llambda4959, 5007, [\ion{O}{2}]~\llambda6300, 6364, [\ion{S}{2}]~\llambda6716, 6731, and \ion{O}{1}~$\lambda$8446. Hydrogen lines (Balmer lines + Ly$\alpha$) are also clearly present. Additionally, the optical spectrum shows prominent \ion{Fe}{2} emission as well as the near-infrared (NIR) calcium triplet (CaT) \ion{Ca}{2}~$\lambda\lambda$8498, 8542, 8662. These features are not present in all AGNs, but they appear in spectra of \edit1{NLSy1} nuclei \citep{osterbrock85,persson88}. Many of these lines, especially the hydrogen lines, \ion{Fe}{2}, and CaT, are also associated with Type IIn SNe \citep[e.g.,][]{filippenko97}.


\begin{figure*}
\centering
\includegraphics[width=0.95\textwidth]{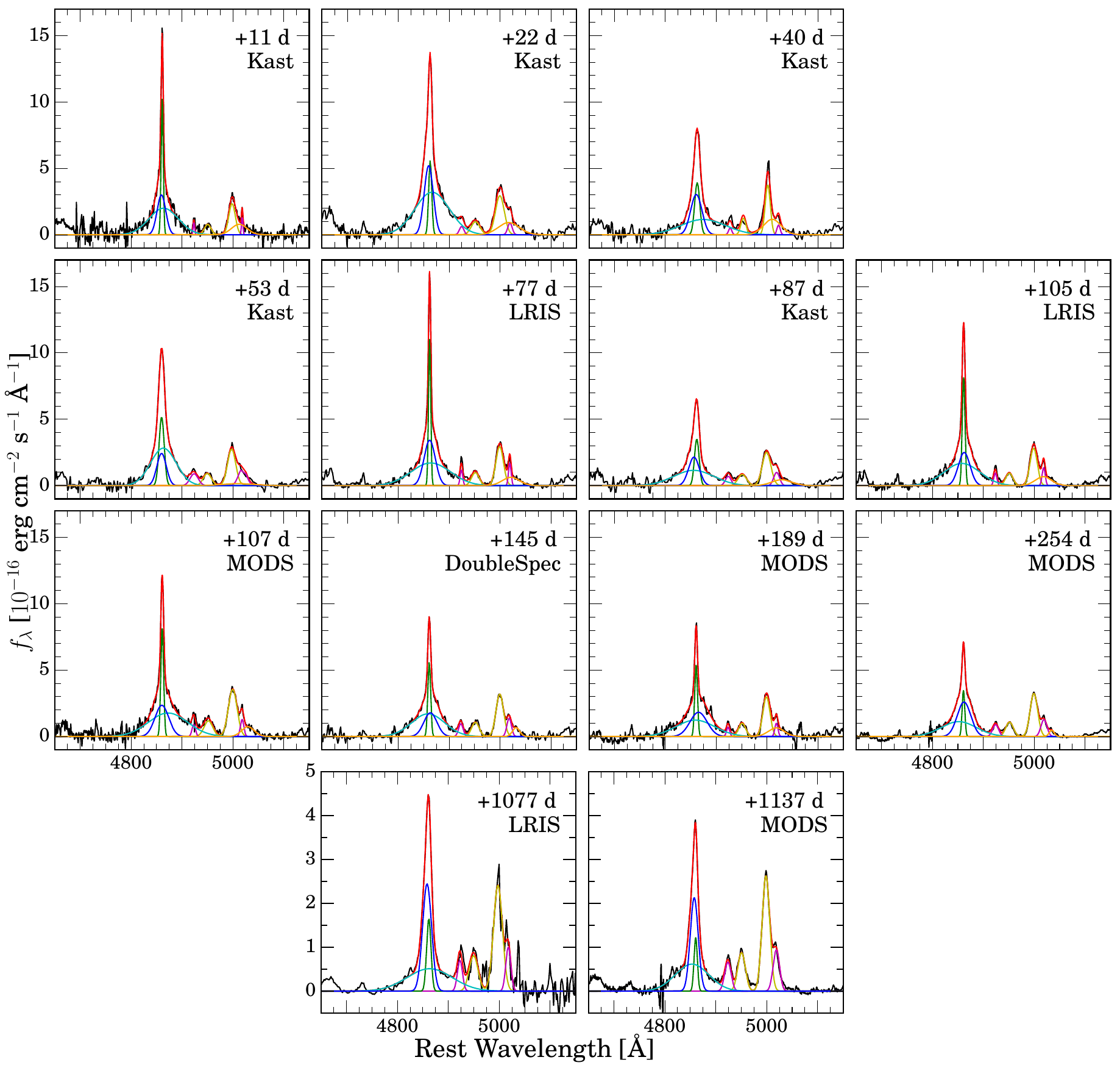}
\caption{Evolution of the \hbeta+[\ion{O}{3}] profile in the spectra (excluding the low-resolution LT spectra). The spectra are labeled with their epoch and instrument, and the lines are color-coded as follows: black --- continuum-subtracted spectrum; green --- narrow \hbeta; blue --- intermediate \hbeta; cyan --- broad \hbeta; yellow --- [\ion{O}{3}] \llambda 4959, 5007; magenta --- \ion{Fe}{2} $\lambda$4923 and $\lambda$5018; orange --- unidentified ``red wing'' feature; red --- combined fit profile. Note that the ordinate for the two late-time spectra has been changed to reflect the significantly lower fluxes of the \hbeta{} features. As described in the text, the line fits for the two late-time spectra do not include the red wing.}
\label{fig:hbeta_spec}
\end{figure*}

To determine the physical mechanisms driving the lines, we attempted to track how the features evolve over time. Owing to the CaT being in the NIR, not enough of our spectra had coverage of the feature, and thus we cannot say how the CaT evolved over time. Unfortunately, owing to the relatively high redshift of this object, a telluric absorption band is coincident with \halpha. As most of our spectra were not corrected for telluric features, it is also difficult to model the evolution of \halpha. This leads us to focus on the evolution of one particular region of the spectra ranging from 4600 to 5100~\AA, encompassing the emission from \hbeta~$\lambda$4861, [\ion{O}{3}]~\llambda4959, 5007, and \ion{Fe}{2}~$\lambda$4923 and $\lambda$5018. Figure \ref{fig:hbeta_spec} shows the continuum-subtracted region of each of the spectra with ``good'' spectral resolution --- specifically, this refers to all except the low-resolution LT spectra, the Kast spectrum from 2017/08/17 where the red and blue parts of the spectrum did not overlap and part of the \hbeta+[\ion{O}{3}] profile is consequently not covered, and the Kast spectrum from 2017/10/19, which had relatively low-resolution.


\begin{figure*}
\centering
\includegraphics[width=0.86\textwidth]{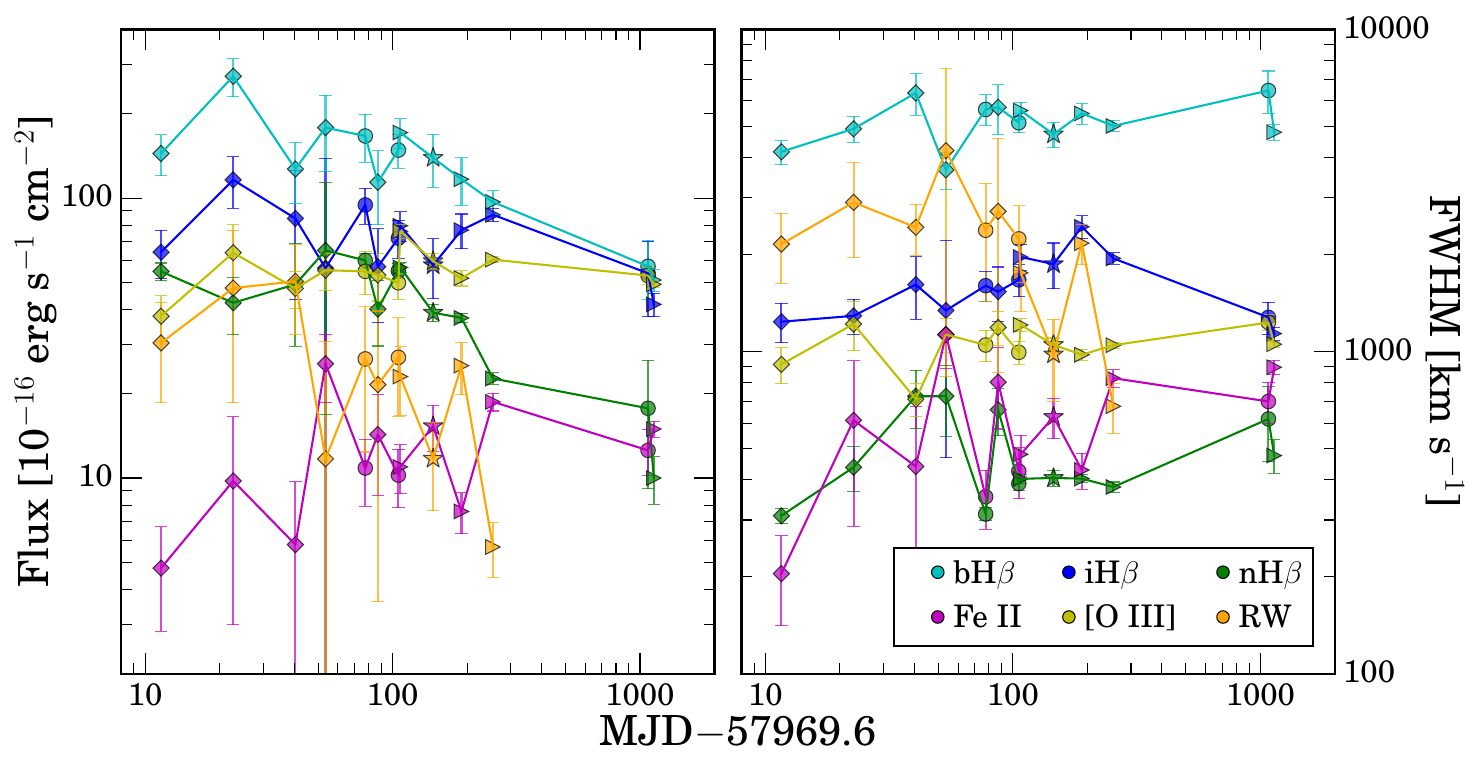}
\caption{Evolution of the flux (left) and FWHM (right) of the fitted lines in Figure~\ref{fig:hbeta_spec}. Color-coding is the same as in Figure~\ref{fig:hbeta_spec} (b\hbeta{} --- broad \hbeta, i\hbeta{} --- intermediate \hbeta, n\hbeta{} --- narrow \hbeta). The symbols of the points in both panels reflect the telescope associated with the data. Note the nearly power-law decay of the flux of the broad and narrow \hbeta{} features. By contrast, the other features do not show coherent evolution, with the possible exception of the red wing. Because the red wing is not included in the fits of the late-time spectra, it is not shown.
}
\label{fig:hbeta_fits}
\end{figure*}

Owing to the complexity of the region, we attempt to track the evolution of the individual features by creating a composite profile consisting of eight Gaussians. These can be divided into three \hbeta{} features with narrow, intermediate, and broad widths (hereafter referred to as narrow, intermediate, and broad \hbeta), two \ion{O}{3} lines, two \ion{Fe}{2} lines, and one extra Gaussian. The three \hbeta{} features were free to vary in centroid, flux, and width, but the two [\ion{O}{3}] lines were constrained such that the offset from zero velocity and the line widths were made to match, and the flux of $\lambda$4959 was set to $\sim 0.33$ times the flux of $\lambda$5007. The two \ion{Fe}{2} lines were similarly constrained, with $\lambda$4923 set to $\sim \edit1{0.70}$ times the flux of $\lambda$5018 \citep[see][for model flux ratios of \ion{Fe}{2} multiplets]{kovacevic10}. The final feature, the extra Gaussian, is introduced because of the peculiar, broad, red wing of \ion{O}{3}~$\lambda$5007, where there is clearly excess flux above that of a single Gaussian. The physical origin of this feature is unclear, so we refer to it as the ``red wing.'' As explained below, we did not include the red wing while fitting the two late-time spectra in Figure~\ref{fig:hbeta_spec} (+1077~d LRIS and +1137~d MODS). 

In Figure \ref{fig:hbeta_spec}, we illustrate the evolution of this spectroscopic region over time along with the model Gaussian profiles. Figure \ref{fig:hbeta_fits} shows the evolution of the fluxes and FWHMs of the features. Of all the features, the one with the most coherent evolution is broad \hbeta, which drops in flux by a factor of $\sim 4$ from its peak brightness at +22~d to its minimum at +1137~d. Interestingly, this feature also appears to stay nearly fixed in FWHM, in contrast to what is often seen in reverberation-mapping studies of AGNs, where line width increases with decreasing luminosity \citep[e.g.,][]{peterson04,denney09}. The narrow \hbeta{} feature evolves in a similar way, dropping by a factor of $\sim 3$--4 over the course of the observations. The flux and FWHM of the narrow \hbeta{} does seem to fluctuate at early times, though this is likely due to the lower spectral resolution of the +22~d and +53~d Kast spectra.

By contrast, the [\ion{O}{3}], \ion{Fe}{2}, and intermediate \hbeta{} features do not coherently evolve over time. [\ion{O}{3}] especially does not vary significantly over the course of our observations. \ion{Fe}{2} fluctuates in flux and FWHM apparently randomly, though these features are relatively weak compared to the others and the flux and FWHM measurements have large uncertainties, so this evolution is also consistent with little-to-no actual variation. The intermediate \hbeta{} feature is somewhat more complicated. While there is no coherent change in its flux, the FWHM appears to slowly rise, hit a maximum at +189~d, and then drop to its original value at late times. It is unclear whether this is physical. Regardless, this evolution is clearly distinct from the coherent flux decay and consistent widths of the narrow and broad \hbeta~features.


\begin{figure}
\centering
{\includegraphics[width=0.5\textwidth]{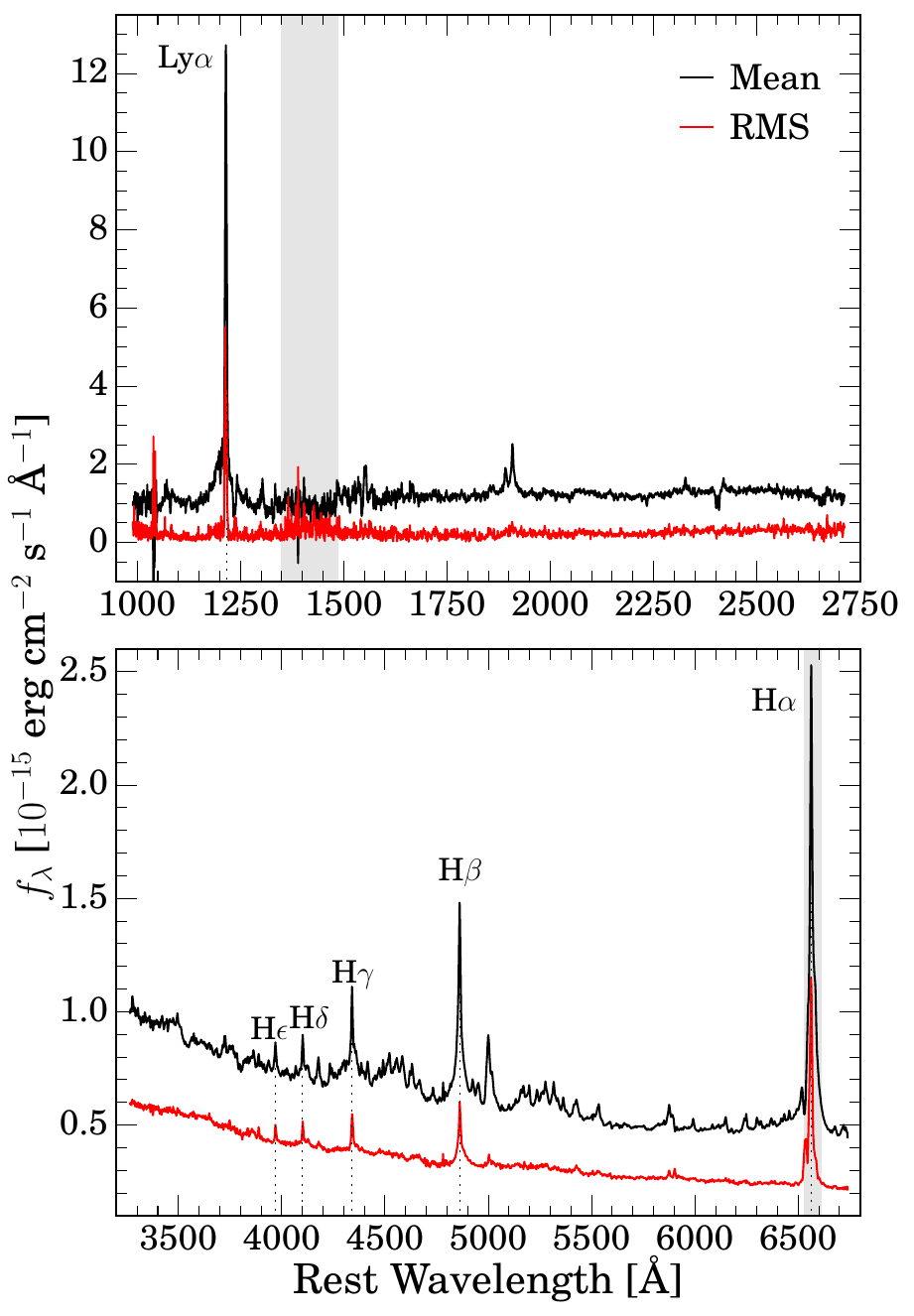}}
\caption{Mean (black) and RMS (red) UV (top) and optical (bottom) spectra. Prominent spectral features of the RMS spectra are highlighted. The emission lines associated with AGN activity (\ion{Fe}{2}, various forbidden lines, etc.) are not strong in the RMS spectra, implying that they are not likely associated with the transient.}
\label{fig:RMS}
\end{figure}

The evolution of the red wing is also complicated. If we examine the +137~d MODS spectrum and compare it to the +1137~d MODS spectrum (the key detail here is that these spectra were taken using the same instrument) in Figure~\ref{fig:hbeta_spec}, there is clearly a broad red wing to the [\ion{O}{3}]~$\lambda$5007+\ion{Fe}{2}~$\lambda$5018 profile in the early spectrum, while there is clearly no such feature in the late spectrum. Because of this, we do not include the red-wing component in our fits of the other features for the two very late-time spectra. Looking at the evolution of the feature in the early-time spectra in Figures~\ref{fig:hbeta_spec} and \ref{fig:hbeta_fits}, it is hard to discern how the feature is evolving over time. It appears to generally become weaker and narrower over time, though there is quite a bit of fluctuation in both the flux and the FWHM from epoch to epoch. Its centroid also varies between 5006 and 5034~\AA. One problem we are likely encountering is that the resolutions of the spectra are not uniform, which can change the apparent profiles of [\ion{O}{3}]~$\lambda$5007 and \ion{Fe}{2}~$\lambda$5018, and consequently the fit of the red wing. However, as the feature does seem to evolve somewhat coherently over time, and disappears at late times, we conclude it is likely transient in nature.

Based on this analysis, we believe the narrow and broad \hbeta{} and the red wing are associated with the transient (\jz{}), whereas the intermediate \hbeta, [\ion{O}{3}], and \ion{Fe}{2} are more likely to be associated with the host galaxy's AGN.

Our UV spectra of \jz{} do not show prominent changes like the optical spectra. This is at least partially due to the UV spectra being observed over a much smaller time window than the optical spectra (46~days vs. 1126~days). The most prominent feature in the spectra is Ly$\alpha$, and there are no obvious changes to it over time.

In addition to our analysis of the individual lines in our spectra, we analyze the RMS UV and optical spectra to see which features show the most prominent evolution. Figure \ref{fig:RMS} shows the UV and optical RMS spectra, as well as the mean spectra from which the RMS spectra were constructed. While there are many prominent spectral lines in the mean spectra, including those discussed above, the RMS spectra show very few, almost exclusively hydrogen lines (Ly$\alpha$ + Balmer series). As the RMS spectrum highlights those features that are evolving continuously while suppressing those features that vary randomly or are constant, this is further evidence that the broad and narrow hydrogen features are associated with the transient, and that the other spectral lines, like the many forbidden lines, \ion{Fe}{2}, etc., are not. It is also worth noting the prominence of Ly$\alpha$ in the RMS spectrum. This suggests that Ly$\alpha$ is subtly evolving over time, likely fading along with the narrow Balmer features. 

\subsection{Comparison with Other Optical Transients}

\begin{figure}
\centering
\includegraphics[width=0.47\textwidth]{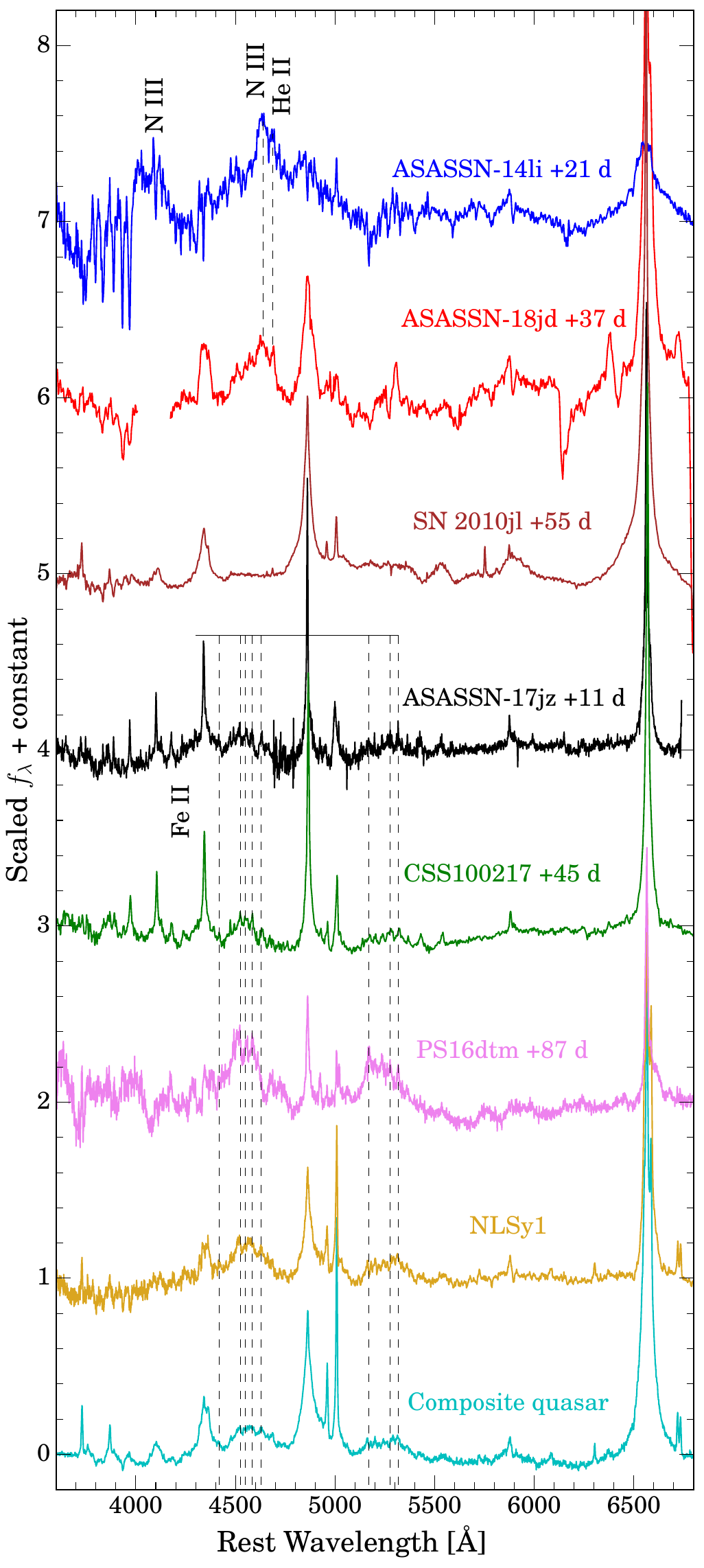}
\caption{Comparison of an optical spectrum of \jz{} to other objects. Prominent spectral features, except for the Balmer series and various forbidden emission lines, are labeled. }
\label{fig:spec_comp}
\end{figure}

In Figure~\ref{fig:spec_comp} we compare an optical spectrum of ASASSN-17jz to those of a TDE \citep[ASASSN-14li,][]{holoien16a}, a conventional Type IIn SN \citep[SN~2010jl,][]{jencson16}, a composite AGN spectrum \citep{vandenberk01}, a NLSy1 spectrum \citep[SDSS J011929.06--000839.7][]{williams02}, and the ambiguous nuclear transients CSS100217 \citep{drake11}, ASASSN-18jd \citep{neustadt20}, and PS16dtm \citep{blanchard17}.


It is immediately apparent that the spectra of \jz{} and CSS100217 are very similar. Indeed, both transients are similarly ambiguous in that they are nuclear transients with AGN-like spectra. Interestingly, \citet{drake11} also found that the H emission lines of CSS100217 were best fit by a three-component model. They similarly see little evolution in the intermediate-width component of both \halpha{} and \hbeta, but unlike \jz, they find that the broad component shows an \emph{increasing} flux and FWHM, while the narrow component shows little evolution. However, it is notable that the lines are so similar between the two transients. \citet{drake11} only have spectra covering the first 164~days after peak light, so it is possible that the longer-term evolution of these features might look more similar to that of \jz.

The early-time spectrum of the Type IIn SN~2010jl somewhat resembles that of \jz{}, in that there are prominent Balmer lines, although the lines of SN~2010jl are clearly broader. SN~2010jl also lacks the many forbidden lines ([\ion{O}{1}], [\ion{S}{2}], etc.) and \ion{Fe}{2} that \jz{} shares with transients like CSS100217 and PS16dtm, although this could simply be due to the latter transients being near their host galaxies' AGNs. SN~2010jl also has much stronger \ion{He}{1}~$\lambda$5876. \citet{jencson16} used a combination of Gaussian and Lorentzian profiles to model their Balmer features, making it difficult to compare the evolution of those components to the evolution we infer for the H$\beta$ features in ASASSN-17jz using our triple Gaussian model. However, they found that the intermediate-width component of H$\alpha$ seen in SN~2010jl narrowed as the transient faded, similar to the broad and narrow H$\beta$ components seen in ASASSN-17jz.


\begin{figure}
\centering
\includegraphics[width=0.47\textwidth]{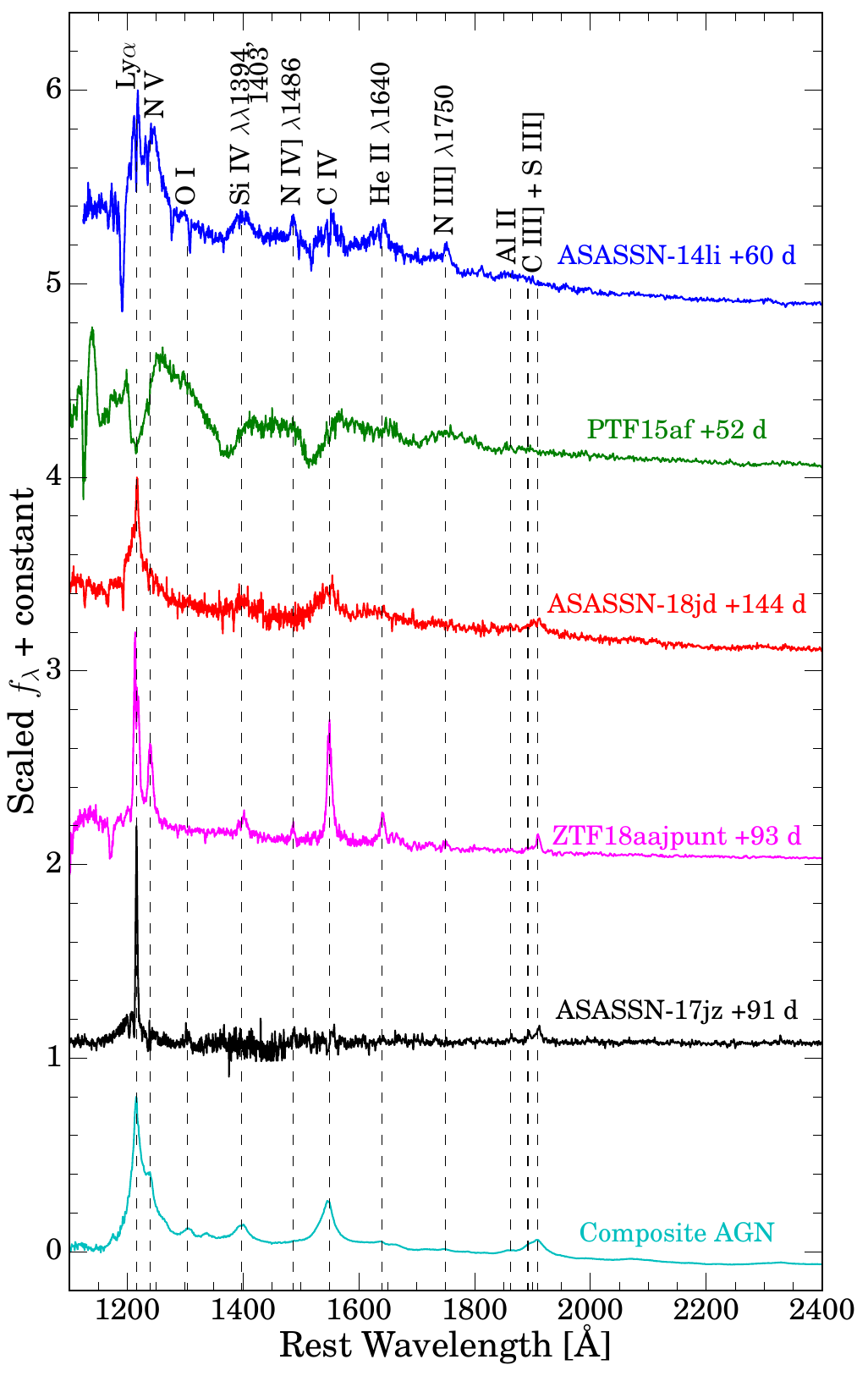}
\caption{Comparison of a UV spectrum of \jz{} to that of other objects. Prominent spectral features are labeled. Note that many of the labeled spectral features, especially those with wavelengths given, are \textit{not} present in the spectrum of \jz{}, and only prominent in the other transients and the AGN spectrum.}
\label{fig:uvspec_comp}
\end{figure}


\begin{figure*}
\centering
\includegraphics[width=0.85\textwidth]{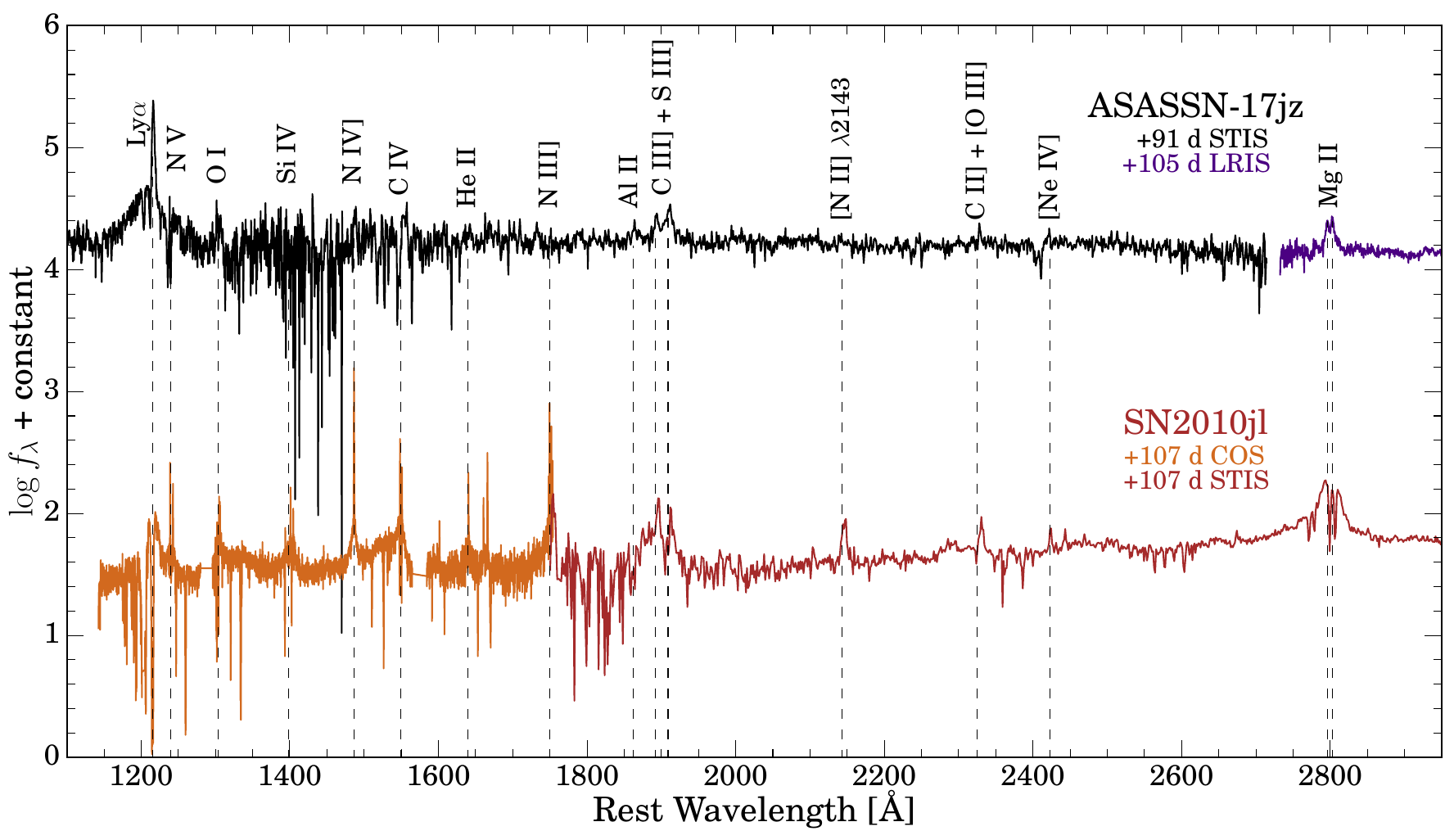}
\caption{Comparison of a UV spectrum of \jz{} to SN~2010jl. Prominent spectral features are labeled. Note that many of the labeled spectral features, especially those with wavelengths given, are \textit{not} present in the spectrum of \jz{}, and only prominent in SN~2010jl.  The spectrum of SN~2010jl is shown apart from the other spectra in Figure \ref{fig:uvspec_comp} only because of the large difference in scale between the narrow and broad features, making it difficult to plot in linear units.}
\label{fig:uvspec_10jl}
\end{figure*}

It is also apparent that \jz{} is quite distinct from the spectra of ASASSN-14li and ASASSN-18jd, both of which show relatively prominent \ion{He}{2} and \ion{N}{3} lines in their spectra. ASASSN-14li exhibits significantly broader H emission as well as broad \ion{He}{2}~$\lambda$4686 emission that is not present in \jz, and it does not have the \ion{Fe}{2} features. While the complex of lines blueward of \ion{He}{2}~$\lambda$4686 in ASASSN-18jd resembles the \ion{Fe}{2} complex seen in the spectrum of \jz{}, \citet{neustadt20} showed that those lines were likely not \ion{Fe}{2} owing to the spacing of the lines and the lack of a comparable complex redward of [\ion{O}{3}]. In any case, strong \ion{He}{2} and \ion{N}{3} are usually indicators of a TDE, and \jz{} does not have these. However, this is not necessarily an indicator that \jz{} is not a TDE, as there are TDEs that do not show these features \citep[e.g., PS18kh/AT2018zr,][]{holoien19b,hung19,velzen21}. PS16dtm, proposed to be a TDE in an existing AGN by \citet{blanchard17}, does resemble \jz{} in terms of spectral features, especially the \ion{Fe}{2} complex and strong Balmer features. However, PS16dtm has much stronger and evolving \ion{Fe}{2} emission \citep[see,][]{blanchard17}, whereas the \ion{Fe}{2} emission of \jz{} is fainter and does not coherently evolve. 

Finally, the optical spectrum of \jz{} shows many common features with the composite quasar spectrum and the NLSy1 spectra. The main differences are that \jz{} has narrower Balmer features and a relatively low [\ion{O}{3}]/\hbeta{} intensity ratio compared to the quasar spectrum.  These differences are also the key differences between NLSy1s and most Type 1 AGNs \citep{osterbrock85}. This implies that many of the emission features in \jz{} could be consistent with being generated by an underlying NLSy1 AGN.

As there are many features in the optical spectra that are similar to those of several other types of transients, we perform a similar comparison with the UV spectra to determine if they may be more constraining. In Figure~\ref{fig:uvspec_comp}, we compare a UV spectrum of \jz{} to that of other objects: the TDEs ASASSN-14li \citep{brown17a} and PTF15af \citep{blagorodnova17}, the ambiguous nuclear transients ASASSN-18jd \citep{neustadt20} and ZTF18aajupnt \citep{frederick20}, and a composite AGN spectrum \citep{vandenberk01}.  In Figure \ref{fig:uvspec_10jl}, we compare a UV spectrum of \jz{} to that of SN~2010jl \citep{fransson14}.  We plot these spectra separately from those in Figure \ref{fig:uvspec_comp} only because of the large difference in scale between the narrow and broad features of SN~2010jl, making it difficult to plot in linear units.  None of the comparison sample are particularly good matches to \jz: all the spectral features of \jz{} are very weak compared to Ly$\alpha$, while the other objects all show a diversity of spectral emission and absorption lines of varying strength. The UV spectra of \jz{} thus imply that the UV emission is dominated by a type of transient other than those shown in the figure.


\edit1{\section{Radio Analysis}}
\label{sec:radio_analysis}

\edit1{It is difficult to determine from the available data whether the emission detected with the VLA was emitted by the transient, by the host galaxy, or by both the host and the transient. We infer a rest-frame luminosity $>9\times10^{28}$~\lunits{} from our VLA observations. This emission surpasses that of the brightest Type IIn SNe to date (e.g., SN~1988Z and SN~1986J) at their peak luminosities \citep{chandra17}. When fitting the flux  across the 8--12~GHz band in images made every 256~MHz, we find a rather flat spectral index ($\alpha=0.03\pm0.16$). This behavior indicates that the emission most likely corresponds instead to an AGN-like host. }

\edit1{The average brightness temperature that we obtain at 10~GHz is $<100$~K. This matches better with a thermal origin, which is further at odds with a radio SN. Moreover, the nondetection at 5~GHz is difficult to explain unless we consider that the emission is resolved at milliarcsec scales and/or has a combination of both free-free and synchrotron emission and absorption. The free-free opacity we calculate using the flux density at 10~GHz considering $\alpha=-2.1$ is 0.7, indicating that any free-free emission would be optically thin. On the other hand, if we use the X-band spectral index ($\alpha=-0.03$), we find that the free-free opacity is $\sim0$, which is explained by optically-thin synchrotron emission suppressed by free-free absorption. The radio emission is likely a combination of both free-free and synchrotron emission and absorption which cannot be quantified with the current data.}

\edit1{Some ANTs from our comparison sample also have been observed in the radio, including radio observations have only been pursued for CSS100217 \citep{drake11}, PS16dtm \citep{blanchard17}, and PS1-10adi \citep{kankare17}. The hosts of PS1-10adi and PS16dtm do not have detected radio counterparts at 1.4\,GHz \citep[see][]{kankare17,blanchard17}. CSS100217's host was reported as not detected at 1.4\,GHz \citep[][]{drake11}, however there is significant emission $>3\sigma$ in an image cutout from the NRAO VLA Sky Survey \citep[NVSS;][]{condon98}. Because of this, the hosts of these transients have been classified as radio-quiet NLSy1 galaxies. The host of ASASSN-17jz is not detected in archival radio images from the Faint Images of the Radio Sky at Twenty-cm survey \citep[FIRST;][]{becker95} or the NVSS.}

\edit1{\citet{drake11} reported three epochs of observations toward CSS100217 at 4.5 and 7.9\,GHz spread within a month, and one epoch at 607.95\,MHz observed in a date between their second and third epochs at higher frequencies. They detected a point source at the position of CSS100217, whose emission at 4.5 and 7.9\,GHz results in a relatively flat spectral index that varied between $-0.20$ and $+0.38$ within one month \citep{drake11}. \citeauthor{drake11} noted that the change in spectral index from being relatively flat to being inverted is difficult to explain with known mechanisms. Indeed, the normal behaviour of transients is to show first an inverted spectrum as the emission rises, changing to a flat spectrum when reaching its peak, and finally becoming steep as the transient emission is turning off. However, the two-point spectral index between the low-frequency observations and those made at higher-frequencies resulted in a steeper index (between $-0.4$ and $-0.5$). The inferred radio luminosity (${L_{R}} \sim 2.3 \times 10^{29}$ erg~s$^{-1}$) seemed brighter than what is expected from other SNe at the same wavelengths known at the time. Giving the compactness of the radio emission, flat spectral indices, and the putative upper limits at 1.4\,GHz from previous surveys, \citeauthor{drake11} argue that the radio emission they observed might have originated from nuclear activity of the central black hole, thus being unrelated to the transient itself.}

\edit1{Based on our analyses of the radio data at the position of ASASSN-17jz and the similarly inconclusive radio emission seen in CSS100217, we find that the most likely possibility is that we are detecting radio emission from the host galaxy AGN, and not from the transient itself.}


\section{Discussion}
\label{sec:disc}

ASASSN-17jz was both highly energetic and extremely long-lasting, and it exhibited a dramatic change in its X-ray emission several years after peak optical light. Its emission shares qualities with several types of transient events, but also differs from all of them in some ways. It is thus one of the most ambiguous of the growing class of ANTs that has been identified in recent years, and if the nature of the transient can be determined, it may be helpful in illuminating the physics behind extreme supernova or accretion scenarios. Here we summarize the results of our analysis in the previous section, and discuss the implications for the possible origin of ASASSN-17jz. 

\subsection{Single Physical Origins}
\label{sec:disc_single}

We first examine whether \jz{} could be the result of only a single type of event: an SN, a TDE, or an AGN outburst.

The relatively smooth rise and early decline after peak seen in ASASSN-17jz's light curve is a common feature of SNe, and \edit2{several} of the emission features in its optical spectra and the shapes of those features have been observed in SNe~IIn \citep[e.g.,][]{filippenko97}. Furthermore, while no single SN emission model we tested was able to power the entire bolometric light curve of ASASSN-17jz, interaction between the SN ejecta and a surrounding CSM wind was able to reproduce the multiband light curves, and a combination of SN+CSM interaction and magnetar spindown is able to reproduce the bolometric light curve. However, these models require extreme physical properties, such as ejecta masses of 40--250~\msun{} and CSM masses of up to 130~\msun, in order to replicate the very high luminosity of the transient. A comparison of the blackbody evolution of ASASSN-17jz to those of SNe shows that while its radius evolution is similar to that of SNe, it is significantly more luminous than most SLSNe, and ASASSN-17jz maintains a relatively constant temperature for several hundred days, while SNe typically become much cooler shortly after peak light. \edit1{While the spectra of \jz{} do share several characteristics with the SN~IIn spectrum, many differ. The SN spectrum has lines not present in \jz, the features they share are narrower in \jz{} than in the SN, and ASASSN-17jz exhibits several forbidden lines and \ion{Fe}{2} lines that are not seen in the SN}. Finally, while it is theoretically possible for an SN to power the late-time UV emission, an SN is unlikely to be able to generate the late-time X-ray brightening seen $\sim1100$ days after peak light. We thus disfavor an SN-only origin for ASASSN-17jz.

The light-curve shape of ASASSN-17jz is also similar to those of TDEs, which often exhibit both smooth evolution and long-term UV emission \citep[e.g.,][]{holoien18a,velzen19b}. TDEs have also been observed to produce brightening X-rays several hundred days after peak light \citep[e.g.,][]{holoien18a,hinkle21a}, have a roughly constant temperature, and often exhibit broad Balmer features in their optical spectra. This, however, is where the similarities between TDEs and ASASSN-17jz end. \edit1{While the early decline of \jz{} was consistent with that of a TDE,} none of the TDE models we tested were able to reproduce either the full multiband or bolometric light curves of \jz. While \jz{} does have a constant blackbody temperature evolution, its temperature is very low ($T\approx10,000$~K) for a TDE, it is much more luminous, and \edit1{while its radius does decrease over time, it does not exhibit the rapid radius decline seen after peak brightness in most TDEs}. The optical spectra of \jz{} do not show the He or Bowen lines commonly seen in TDE spectra, and do exhibit several forbidden lines and \ion{Fe}{2} lines not seen in TDEs. UV spectra of TDEs typically have several emission or absorption features that often evolve shortly after peak light \citep[e.g.,][]{cenko16,brown18,hung20}, while the only strong feature in the UV spectra of ASASSN-17jz is Ly$\alpha$ emission, and there is little evolution over the course of our {\it HST}/STIS observations. \edit1{\jz{} also exhibits \ion{Mg}{2}~$\lambda$2800 emission, which has been absent in UV spectra of TDEs to-date.} Finally, while TDEs can produce a late-time X-ray brightening, the timescale of this brightening is typically a few hundred days, not 1100 days as observed in \jz, and the X-rays from \jz{} become softer as they become more luminous, which is not typical of a TDE \citep[e.g.,][]{auchettl17}. We therefore conclude that a TDE-only origin is also disfavored for ASASSN-17jz.

Finally, we examine the possibility that \jz{} could result from solely an AGN, likely one undergoing some kind of transient accretion event. AGNs typically exhibit some short-term variation in their light curves \citep[e.g.,][]{shappee14}, rather than the smooth evolution seen in \jz. However, \citet{frederick20} report on a subset of peculiar AGN flares occurring in NLSy1s that show smooth, coherent evolution, in contrast to the normally stochastic variability of AGNs. One of the transients, ZTF19aatubsj, is highlighted as being spectroscopically similar to CSS100217, which is itself similar to \jz{}. Despite this, the photometric evolution of ZTF19aatubsj is much slower than that of \jz. It is unclear if these are similar events or distinct phenomena, but it is notable that there is some precedent for smooth light-curve evolution in NLSy1 transients. The other transients included by \citet{frederick20} are classified as belonging to the class of transients discussed by \citet{trakhtenbrot19a} that have prominent \ion{He}{2}/Bowen features, or are classified as TDEs. Because \jz{} does not show any Bowen features, it is thus appropriate to say that if \jz{} is related to the transients described by \citet{frederick20}, it is likely similar to ZTF19aatubsj, rather than to the other events discussed in that manuscript.

There are also differences between the features seen in the spectra of \jz{} and those of composite quasar spectra, particularly in the UV, where typical quasar spectra exhibit several lines not seen in the UV spectra of \jz. We also observe little variation in the spectral lines of \jz, with only the broad and narrow Balmer components clearly evolving over time. These differences from typical AGNs are not, however, disqualifying of a possible AGN origin for \jz. The differences between the spectral features of \jz{} and the composite quasar spectra are similar to the differences between the composite quasar and NLSy1s. The lack of changes in the prominent AGN spectral features, like the [\ion{O}{3}] and \ion{Fe}{2} lines, does not mean that the AGN is not itself changing. Physical models of AGNs place [\ion{O}{3}] in the narrow-line \edit1{region}, which is several light-years away from the central SMBH and the broad-line region \citep{antonucci93}. The \ion{Fe}{2} is thought to be associated with the dusty torus \citep{marziani01,popovic04}, which in most models is also several light-years away from the central SMBH. Thus, it might take several years for changes in the narrow lines and \ion{Fe}{2} to reflect the changes seen in the continuum and broad lines. 

Finally, the X-ray emission exhibited by \jz{} at late times \edit1{has a power-law index of $\Gamma=3.4$, which is typical of NLSy1 AGNs \citep[e.g.,][]{boller96}.} Further, \citet{auchettl17} found that X-ray emission from AGNs becomes softer as it brightens, which is the behavior exhibited by \jz. This, combined with the dissimilarities between the X-ray emission of \jz{} and X-rays that would be expected from TDEs and SNe, lead us to conclude that the X-ray emission almost certainly originates from an \edit1{NLSy1} AGN in \host.

\edit1{Based on all of these properties, \jz{} appears to be a unique event that differs in some way from all of the individual transient comparison sources. However, \emph{if we had to select a single physical origin} for \jz,} it was likely a transient accretion event in an AGN in \host. 

\edit1{In light of the possibility that \jz{} was the result of a transient accretion event in an AGN, we also consider whether this could be the result of an SMBH binary (SMBHB) system. In systems with subparsec separations, the SMBHs can carve out a cavity in a circumbinary accretion disk via tidal torques \citep[e.g.,][]{gold19}. Individual, smaller accretion disks can also form around each black hole \citep[e.g.,][]{ryan17,gold19}. Interaction between the accretion stream(s) in the SMBHB and the cavity can then cause outbursts on timescales similar to that of \jz, and should result in periodic or semiperiodic flaring events \citep[e.g.,][]{komossa06}. Apart from repeated flaring behavior, such systems can have offset line centers or even multiple velocity components in their spectra due to there being multiple accretion streams. The lack of variability going back $\sim12$~yr prior to \jz{} in archival data and the lack of such spectroscopic signatures argue against an SMBHB interpetation. However, we cannot rule this possibility out completely, as the signatures can be hidden at certain viewing angles or BH separations. Further observation of \jz{} and its host could be helpful in testing the possibility of an SMBHB system causing the event.}

\subsection{Multiple Physical Origins}
\label{sec:disc_multi}

While neither SN emission nor TDE emission are solely able to replicate the emission seen from \jz, and the X-rays likely originate from an underlying AGN, it is also possible that the emission from ASASSN-17jz has multiple physical origins, with a transient event occurring in an AGN. These are the types of scenarios commonly invoked to explain the observations of other ANTs, and we discuss these possibilities here. In these scenarios, we assume that the late-time emission is driven by the underlying AGN, while the emission up to the first few hundred days after peak is dominated by a transient event.

By far the most similar event to ASASSN-17jz in both photometric and spectroscopic properties is CSS100217, which \citet{drake11} suggested was the result of an SN~IIn occurring in or near the NLSy1 AGN of its host galaxy. \edit1{In their discovery paper for the transient PS1-10adi, \citet{kankare17} identified CSS100217 as a similar event, and also consider an SN occurring in an existing AGN a plausible explanation for these transients.} SNe~IIn result from SNe which are surrounded by dense CSM, and models of SN ejecta interacting with both a CSM shell and a CSM wind are able to replicate the rise and early decline of both the multiband and bolometric light curves of \jz. The blackbody evolution of \jz{} is consistent with that of CSS100217 in all three aspects, and while the constant temperature evolution and high luminosity of \jz{} are not similar to those of SNe~IIn like SN~2010jl, this could perhaps be explained by the underlying AGN contributing to the emission. The spectra are extremely similar to those of CSS100217 as well. Unfortunately, UV spectra were not obtained of CSS100217 at early times, so we cannot comment on how similar the UV spectra of \jz{} might be to that event transients.

CSS100217 was also analyzed by \citet{frederick20}, and they conclude that the event was not an SN, but was instead a peculiar AGN flare. Their reasoning was somewhat flawed, however, in that it is explicitly stated by \citet{frederick20} that (among other reasons) CSS100217 is likely not an SN owing to the lack of P~Cygni profiles in the spectra. Such profiles are not always present in the spectra of Type IIn SNe \citep[e.g.,][]{filippenko97}, and so this detail cannot be used to reject CSS100217 as an SN~IIn. \edit1{Further, their analysis is more focused on whether a SN alone can explain the observed emission from CSS100217, rather than whether some of the observed features of that flare could be due to an SN while others are caused by the underlying AGN. For example, they point out that while SNe~IIn can exhibit \ion{Fe}{2} lines in their spectra at late times, CSS100217 exhibited these features throughout the flare. \citet{frederick20} use this as a piece of evidence against the SN interpretation, but do not consider the possibility that these lines are caused by an underlying AGN, and that an SN could produce some of the other observed features. Thus, while we agree that CSS100217 (and \jz) were not likely caused solely by an SN, we consider an SN+AGN scenario a viable interpretation for both events.}

Given the similarity of the early-time light curve to those of SNe~IIn and the similarity in all aspects to CSS100217, which can plausibly be explained as an SN~IIn in a NLSy1 host, we conclude that an SN~IIn in a NLSy1 host galaxy accompanied by late-time activity from the AGN is a very plausible explanation for ASASSN-17jz.

We also consider the possibility of a TDE occurring in a host galaxy with AGN activity. This scenario has been explored to some degree in theoretical work, which has found that the emission from such events could be nonthermal and differ substantially from ``typical'' TDEs and AGNs \citep[e.g.,][]{chan19}. However, this work is based on highly restrictive simulations, and the full region of parameter space remains largely unexplored. \edit1{Interestingly, \citet{kankare17} also consider a TDE occurring in an AGN to be a plausible scenario for PS1-10adi and CSS100217, based on the idea that the unbound material from the disrupted star could interact with broad-line region clouds in a way similar to that of SN ejecta interacting with CSM. They note, however, that there are several unexplained points for this case that would require additional study.} We thus base this analysis largely on comparison with the transient PS16dtm, which had compelling evidence that it was a TDE occurring in an AGN \citep{blanchard17} and provides us with observable characteristics to compare with those of ASASSN-17jz.

As with the SN models, TDE models are able to replicate the early multiband and bolometric light curves of ASASSN-17jz, with perhaps more reasonable model parameters than the SN models. However, there are many more differences between PS16dtm and ASASSN-17jz than there are between CS100217 and ASASSN-17jz. The blackbody luminosity of PS16dtm shows short-term variability and a slower decline than that of \jz; the temperature of PS16dtm, while roughly constant, is lower than that of \jz; and the radius of PS16dtm increases over time, compared to the decreasing radius of \jz. The two objects' spectra also show larger differences. The spectra of PS16dtm clearly evolve and show much stronger \ion{Fe}{2} emission than those of \jz{}. The spectra of \jz{} are also consistent with being a combination of SN~IIn and AGN spectra with little resemblance to TDE spectra. TDEs typically have much broader Balmer lines that fade and become narrower over time. \edit1{Many} TDEs also exhibit lines of helium and/or lines attributed to Bowen fluorescence \citep[e.g.,][]{velzen20,velzen21}, while \jz{} lacks any He or Bowen features. 

Finally, while late-time X-ray brightening is predicted and seen in some TDEs \citep[e.g.,][]{metzger15,gezari17,holoien18a,wevers19,hinkle21b}, many TDEs do not show any strong X-ray emission \citep[e.g.,][]{holoien14a,holoien18a,holoien20a}. As discussed above, late-time X-ray brightening in TDEs typically occurs roughly a year after peak optical light, and the hardness ratio has remained largely consistent throughout the luminosity changes \citep{auchettl17}. The X-ray emission from \jz{} does not exhibit this behavior, and is more consistent with AGN emission. The X-ray emission we observe from \jz{} could be consistent with a scenario where an X-ray-faint TDE occurs around an existing AGN, which becomes brighter in X-rays at late times. However, it is typically assumed that X-ray emission is obscured from TDEs in which X-ray emission is not detected \citep[e.g.,][]{auchettl17}, and it seems unlikely that X-ray emission from a TDE could be obscured while late-time X-ray emission from an AGN around the same SMBH is not.

Thus, while we cannot completely rule out a scenario where the observed emission from \jz{} is the result of the combination of a TDE and AGN activity, we conclude that this scenario is less likely than that of a SN+AGN discussed above.
\\

Ultimately, it seems likely that ASASSN-17jz was either the result of transient accretion activity in an NLSy1 AGN, or an SN~IIn occurring in an NLSy1 host. Neither scenario is significantly favored over the other, but given the evidence of an SN being partially responsible for the emission (e.g., light-curve fits, blackbody fits, and similarities with CSS100217), we conclude that the SN+AGN scenario is the most likely. 

As the AGN emission becomes dominant at late times in this scenario, it is reasonable to ask whether the increased AGN activity is a consequence of the SN. This would require the SN to have occurred close enough to the SMBH to affect the accretion flow, which is unlikely but possible. Our constraints on the position of the transient and lack of a secondary source seen in images from \HST\ imply that any non-AGN transient would need to be very close to the host nucleus, making this picture plausible. 

\edit1{The likelihood and effects of transients occurring near enough to AGNs to affect the accretion disk has been explored theoretically and in simulations \citep[e.g.,][]{rozycka95,chan19,grishin21,moranchel21}. These studies have found that the transients can affect the accretion flow, particularly in the cases of TDE debris streams colliding with the accretion disk \citep[e.g.,][]{chan19} or SNe occurring within the disk itself \citep[e.g.,][]{grishin21}. Most of these theoretical studies have focused more on the effects of the transients on the AGN, rather than the observable properties of the events, but \citet{grishin21} predict that an SN occurring within an AGN accretion disk would produce a much faster event than we observed in the case of ASASSN-17jz. This is because they treat the material of the accretion disk itself as the ``CSM'' in an SN$+$CSM scenario, and this material has different properties from a stellar CSM. The scenario we suggest, where a more traditional, luminous SN IIn occurs near or in the disk, may therefore produce a flare like the one we observe, rather than what is predicted by \citet{grishin21}.}

\edit1{There have also been cases of observed transients that have been claimed to be TDEs or SNe occurring in and around AGNs and affecting the accretion flow and/or broad-line region of the AGN \citep[e.g.,][]{merloni15,blanchard17,smith18}. The idea that ASASSN-17jz was an SN that affected the accretion of its host AGN would thus be rare, but not a unique case of a transient occurring in and affecting an AGN. We do note that the timescale of the changes would be unique among these objects.} We have no definitive proof, however, of a causal relationship between a potential early SN and the increased AGN emission at late times. Longer-term monitoring of this system, particularly in the UV and X-rays, may be able to shed further light on this possibility. 

Regardless of whether ASASSN-17jz was the result of an SN in an AGN or solely an AGN, the increased AGN activity implies that the accretion flow must have substantially changed in some way. ASASSN-17jz and other, similar nuclear transients thus provide us with an opportunity to study how large-scale changes to accretion disks can occur, how the accretion flow can change as a result, and how long it takes the accretion disk to stabilize again.
\\

\acknowledgments

The authors thank Y. Chen, J. Mauerhan, C. Melis, C. C. Steidel, and R. L. Theios for obtaining follow-up data. \edit1{We thank Lizelke Klindt and Fiona Harrison for contributing to observations made with the Palomar 200-inch telescope.} U.C. Berkeley students Nick Choksi, Edward Falcon, Romain Hardy, Goni Halevy, Emily Ma, Yukei Murakami, Jackson Sipple, Costas Soler, Samantha Stegman, James Sunseri, Sergiy Vasylyev, and Jeremy Wayland contributed to observations with the Lick Nickel 1~m telescope. We are grateful to S. Gomez for assistance with running \texttt{MOSFiT} supernova models. We thank the \swift{} PI, the Observation Duty Scientists, and the science planners for promptly approving and executing our \swift{} observations. The Las Cumbres Observatory and its staff are gratefully acknowledged for its continuing assistance with the ASAS-SN project. We thank the staffs of the various observatories where data were obtained for their assistance.

ASAS-SN is supported by the Gordon and Betty Moore Foundation through grant GBMF5490 to the Ohio State University and National Science Foundation (NSF) grant AST-1515927. Development of ASAS-SN has been supported by NSF grant AST-0908816, the Mt. Cuba Astronomical Foundation, the Center for Cosmology and AstroParticle Physics at the Ohio State University, the Chinese Academy of Sciences South America Center for Astronomy (CASSACA), the Villum Foundation, and George Skestos.

Support for T.W.-S.H. was provided by NASA through the NASA Hubble Fellowship grant HST-HF2-51458.001-A awarded by the Space Telescope Science Institute (STScI), which is operated by the Association of Universities for Research in Astronomy, Inc., for NASA, under contract NAS5-26555. P.J.V. is supported by the National Science Foundation Graduate Research Fellowship Program Under Grant No. DGE-1343012. C.R.-C. acknowledges financial support from the Chinese Academy of Sciences (CAS), through the CAS South America Center for Astronomy (CASSACA) and CONICYT grant CAS16013. K.Z.S., C.S.K., and T.A.T. are supported by NSF grants AST-1515876, AST-1515927, and AST-1814440. B.J.S., C.S.K., and K.Z.S. are supported by NSF grant AST-1907570/AST-1908952. B.J.S. is also supported by NSF grants AST-1920392 and AST-1911074. Support for J.L.P. and C.R.-C. is provided in part by ANID through the Fondecyt regular grant 1191038 and through the Millennium Science Initiative grant ICN12009, awarded to The Millennium Institute of Astrophysics, MAS. T.A.T. is supported in part by NASA grant 80NSSC20K0531. A.V.F.'s supernova research group at U.C. Berkeley has been supported by the TABASGO Foundation, Gary and Cynthia Bengier (T.d.J. was a Bengier Postdoctoral Fellow), the Christopher R. Redlich Fund, the Miller Institute for Basic Research in Science (A.V.F. is a Miller Senior Fellow), NASA/{\it HST} grant GO-15166 from STScI, and Google (K.D.Z. was a Google/Lick Predoctoral Fellow). \edit1{M.I. acknowledges the support from the National Research Foundation (NRF) grants  2020R1A2C3011091 and  2021M3F7A1084525, supervised by the Ministry of Science and ICT (MSIT) of Korea. H.D.J. is supported by the NRF grant 2022R1C1C2013543 funded by the MSIT of Korea. M.A.T. acknowledges support from the DOE CSGF through grant DE-SC0019323.} Research by S.V. is supported by NSF grants AST-1813176 and AST-2008108.

Parts of this research were supported by the Australian Research Council Centre of Excellence for All Sky Astrophysics in 3 Dimensions (ASTRO 3D), through project number CE170100013.

This material is based upon work supported by the U.S. Department of Energy, Office of Science, Office of Advanced Scientific Computing Research, Department of Energy Computational Science Graduate Fellowship under Award Number DE-FG02-97ER25308.

This research is based on observations made with the NASA/ESA \emph{Hubble Space Telescope} obtained from STScI, which is operated by the Association of Universities for Research in Astronomy, Inc., under NASA contract NAS 5–26555. These observations are associated with programs HST-GO-14781, HST-GO-15166, and HST-GO-15312.

The National Radio Astronomy Observatory is a facility of the NSF operated under cooperative agreement by Associated Universities, Inc.

Research at Lick Observatory is partially supported by a generous gift from Google. A major upgrade of the Kast spectrograph on the Shane 3~m telescope at Lick Observatory was made 
possible through generous gifts from the Heising-Simons Foundation as well as William and Marina Kast. KAIT and its ongoing operation were made possible by donations from Sun Microsystems, Inc., the Hewlett-Packard Company, AutoScope Corporation, Lick Observatory, the NSF, the University of California, the Sylvia \& Jim Katzman Foundation, and the TABASGO Foundation. 

The European VLBI Network is a joint facility of independent European, African, Asian, and North American radio astronomy institutes. Scientific results from data presented in this publication are derived from the following EVN project code(s): RR011. e-VLBI research infrastructure in Europe is supported by the European Union's Seventh Framework Programme (FP7/2007-2013) under grant agreement number RI-261525 NEXPReS. 

The LBT is an international collaboration among institutions in the United States, Italy, and Germany. LBT Corporation partners are The University of Arizona on behalf of the Arizona Board of Regents; Istituto Nazionale di Astrofisica, Italy; LBT Beteiligungsgesellschaft, Germany, representing the Max-Planck Society, The Leibniz Institute for Astrophysics Potsdam, and Heidelberg University; The Ohio State University, representing OSU, University of Notre Dame, University of Minnesota, and University of Virginia.

This research uses data obtained through the Telescope Access Program (TAP).

Observations obtained with the Hale telescope at Palomar Observatory were obtained as part of an agreement between the National Astronomical Observatories, Chinese Academy of Sciences, and the California Institute of Technology. 

Some of the data presented herein were obtained at the W.~M. Keck Observatory, which is operated as a scientific partnership among the California Institute of Technology, the University of California, and NASA; the observatory was made possible by the generous financial support of the W.~M. Keck Foundation. We acknowledge the Keck target-of-opportunity program for the spectrum obtained on 22 Oct. 2017. 

This research has made use of the Keck Observatory Archive (KOA), which is operated by the W.~M. Keck Observatory and the NASA Exoplanet Science Institute (NExScI), under contract with NASA.

\software{FAST (Kriek et al. 2009), IRAF (Tody 1986, Tody 1993), \edit1{LPipe (Perley 2019),} HEAsoft (Arnaud 1996), XSPEC (v12.9.1; Arnaud 1996), MOSFiT (Guillochon et al. 2018)}

\bibliography{bibliography.bib}{}
\bibliographystyle{aasjournal}

\appendix

\setcounter{table}{0}
\renewcommand{\thetable}{A\arabic{table}}


\begin{deluxetable}{cccccc}[h!]
\tabletypesize{\footnotesize}
\tablecaption{Spectroscopic Observations of ASASSN-17jz}
\tablehead{
\colhead{Date} &
\colhead{Telescope} &
\colhead{Instrument} &
\colhead{Grating} &
\colhead{Slit} &
\colhead{Exposure Time}}
\startdata
2017 July 31.92 & LT 2-m & SPRAT & VPH 600 l/mm & 1\farcs{8} & 1$\times$600s \\
2017 July 31.93 & LT 2-m & SPRAT & VPH 600 l/mm & 1\farcs{8} & 1$\times$500s \\
2017 August 06.35 & Palomar 200-in & Doublespec & 600/4000, 316/7500 & 1\farcs{5} & 1$\times$900s, 1$\times$900s \\
2017 August 10.91 & LT 2-m & SPRAT & VPH 600 l/mm & 1\farcs{8} & 1$\times$500s \\
2017 August 13.90 & LT 2-m & SPRAT & VPH 600 l/mm & 1\farcs{8} & 1$\times$500s \\
2017 August 16.23 & Shane 3-m & Kast & 600/4310, 830/8460 & 2\farcs{0} & 1$\times$3600s, 2$\times$1200s \\
2017 August 16.91 & LT 2-m & SPRAT & VPH 600 l/mm & 1\farcs{8} & 1$\times$500s \\
2017 August 17.31 & Shane 3-m & Kast & 600/4310, 600/7500 & 2\farcs{0} & 1$\times$1860s, 3$\times$600s \\
2017 August 18.29 & Keck 10-m & DEIMOS & 1200G, 600ZD & 1\farcs{0} & 3$\times$200s, 1$\times$180s \\
2017 August 20.30 & Palomar 200-in & Doublespec & 300/3990, 316/7500 & 1\farcs{5} & 1$\times$600s, 1$\times$600s \\
2017 August 20.90 & LT 2-m & SPRAT & VPH 600 l/mm & 1\farcs{8} & 1$\times$500s \\
2017 August 28.29 & Shane 3-m & Kast & 600/4310, 300/7500 & 2\farcs{0} & 1$\times$2160s, 3$\times$700s \\
2017 September 02.86 & LT 2-m & SPRAT & VPH 600 l/mm & 1\farcs{8} & 1$\times$600s \\
2017 September 14.12 & Palomar 200-in & Doublespec & 600/4000, 316/7500 & 1\farcs{5} & 1$\times$900s, 1$\times$900s \\
2017 September 14.20 & Shane 3-m & Kast & 600/4310, 300/7500 & 2\farcs{0} & 1$\times$2160s, 3$\times$700s \\
2017 September 15.88 & LT 2-m & SPRAT & VPH 600 l/mm & 1\farcs{8} & 1$\times$600s \\
2017 September 19.87 & LT 2-m & SPRAT & VPH 600 l/mm & 1\farcs{8} & 1$\times$600s \\
2017 September 27.18 & Shane 3-m & Kast & 600/4310, 300/7500 & 2\farcs{0} & 1$\times$2160s, 3$\times$700s \\
2017 September 27.85 & LT 2-m & SPRAT & VPH 600 l/mm & 1\farcs{8} & 1$\times$600s \\
2017 October 07.87 & LT 2-m & SPRAT & VPH 600 l/mm & 1\farcs{8} & 1$\times$600s \\
2017 October 19.20 & Shane 3-m & Kast & 600/4310, 300/7500 & 2\farcs{0} & 1$\times$2460s, 3$\times$800s \\
2017 October 21.20 & Keck 10-m & LRIS & 400/3400, 600/5000 & 1\farcs{0} & 1$\times$900s \\
2017 October 22.84 & LT 2-m & SPRAT & VPH 600 l/mm & 1\farcs{8} & 1$\times$600s \\
2017 October 30.12 & Shane 3-m & Kast & 600/4310, 300/7500 & 2\farcs{0} & 1$\times$2460s, 3$\times$800s \\
2017 November 01.83 & LT 2-m & SPRAT & VPH 600 l/mm & 1\farcs{8} & 1$\times$600s \\
2017 November 18.18 & Keck 10-m & LRIS & 600/4000, 400/8500 & 1\farcs{0} & 1$\times$1800s \\
2017 November 20.07 & LBT 8.4-m & MODS & Dual & 1\farcs{0} & 1$\times$300s+1$\times$260s \\
2017 December 28.53 & Palomar 200-in & Doublespec & 300/3990, 316/7500 & 1\farcs{5} & 1$\times$600s, 1$\times$600s \\
2018 January 03.31 & LT 2-m & SPRAT & VPH 600 l/mm & 1\farcs{8} & 1$\times$600s \\ 
2018 February 10.51 & LBT 8.4-m & MODS & Dual & 1\farcs{0} & 5$\times$1200s \\ 
2018 April 16.37 & LBT 8.4-m & MODS & Dual & 1\farcs{0} & 3$\times$1200s \\
2018 September 15.25 & Keck 10-m & DEIMOS & 600ZD & 1\farcs{0} & 7$\times$600s \\
2020 July 17.31 & Keck 10-m & LRIS & 400/3400, 400/8500 & 1\farcs{0} & 9$\times$1200s+1$\times$350s, 9$\times$1200s \\
2020 September 15.17 & LBT 8.4-m & MODS & Dual & 1\farcs{0} & 6$\times$1200s \\
\enddata 
\tablecomments{Date, telescope, instrument, grating, slit width, and exposure time for each of the optical spectroscopic observations obtained of ASASSN-17jz for the initial classification of the transient and as part of our follow-up campaign. For instruments with blue and red channels, the blue and red gratings and exposure times are separated by a comma, with blue listed first.} 
\label{tab:spec_details} 
\end{deluxetable}


\begin{deluxetable}{ccccc}[h!]
\tabletypesize{\footnotesize}
\tablecaption{\emph{HST} STIS Spectroscopic Observations of ASASSN-17jz}
\tablehead{
\colhead{Date} &
\colhead{Detector} &
\colhead{Grating} &
\colhead{Slit} &
\colhead{Total Exposure Time}}
\startdata
2017 September 09.62 & FUV/NUV-MAMA & G140L, G230L & 52\farcs{0}$\times$0\farcs{2} & 2460s, 1854s \\
2017 October 05.14 & FUV/NUV-MAMA & G140L, G230L & 52\farcs{0}$\times$0\farcs{2} & 3420s, 2920s \\
2017 October 19.03 & FUV/NUV-MAMA & G140L, G230L & 52\farcs{0}$\times$0\farcs{2} & 5809s, 3954s \\
2017 November 04.58 & FUV/NUV-MAMA & G140L, G230L & 52\farcs{0}$\times$0\farcs{2} & 5809s, 3954s \\
\enddata 
\tablecomments{Date, instrument, grating, slit, and total exposure time for each of the UV spectroscopic observations obtained of ASASSN-17jz with \emph{HST} STIS as part of our follow-up campaign. The gratings and exposure times used for the FUV and NUV detectors are separated by a comma, with FUV listed first.} 
\label{tab:uv_spec_details} 
\end{deluxetable}



\end{document}